\documentclass[10pt]{iopart}
\pdfoutput=1

\usepackage{iopams}
\usepackage[mathcal]{euscript}
\usepackage{amsgen,amsfonts,amssymb,amsbsy}

\usepackage{bm}
\usepackage[utf8]{inputenc}
\usepackage[T1]{fontenc}
\usepackage{ae,aecompl}

\usepackage{graphics}
\usepackage{graphicx}
\usepackage{float}
\usepackage{stackrel}

\usepackage[usenames,dvipsnames]{xcolor}

\usepackage{color}
\definecolor{darkgreen}{rgb}{0,0.6,0}
\definecolor{darkblue}{rgb}{0,0,0.6}
\definecolor{darkred}{rgb}{0.6,0,0}
\definecolor{darkpurple}{rgb}{0.5,0,0.5}

\usepackage{hyperref}

\hypersetup{
bookmarksopen=true,
pdftitle=Current statistics and depinning transition for a one-dimensional Langevin process in the weak-noise limit,
pdfauthor=Nicolás Tizón-Escamilla and Vivien Lecomte and Eric Bertin, 
pdftoolbar=false, 
pdfstartview={FitH},		
pdfmenubar=true,			
pdfhighlight=/O,			
colorlinks=true,			
urlcolor=darkblue,
citecolor=darkblue,		
linkcolor=darkpurple	
}

\usepackage[left=1.0in,right=1.0in,top=1.0in,bottom=1.0in,includeheadfoot]{geometry}

\newcommand{\tf}{{t_{\rm{f}}}}

\newcommand{\WW}{\mathbb W}

\newcommand{\fla}{f_{\mkern-1.5mu \lambda}}
\newcommand{\dd}{\!{\rm{d}}}
\newcommand{\ee}{{\rm{e}}}
\newcommand{\cc}{{\rm{c}}}

\renewcommand{\epsilon}{\varepsilon}
\newcommand{\Ss}{{\rm{s}}}
\newcommand{\mm}{{\rm{m}}}

\newcommand{\deltaf}{\delta{\mkern-2mu}f}

\providecommand{\eqref}[1]{\eref{#1}}
\providecommand{\text}[1]{{\rm{#1}}}
\providecommand{\operatorname}[1]{{\,\rm{#1}}}

\newcommand{\eff}{{\mkern+1.mu\text{eff}\mkern-1.5mu}}

\providecommand{\mathds}[1]{\mathbf{#1}}

\newenvironment{align}
 {\begin{eqnarray}}
 {\end{eqnarray}\ignorespacesafterend}

\newenvironment{casesl}
    {
     \arraycolsep=1.1pt
     \left\{
       \begin{array}{ll}
    }
    { 
       \end{array}
     \right.
    }

\begin{document}

\title[]{Current statistics and depinning transition for a one-dimensional Langevin process in the weak-noise limit}

\author{Nicol\'as Tiz\'on-Escamilla$^{1,2}$, Vivien Lecomte$^2$ and Eric Bertin$^2$}

\address{$^1$
Aix Marseille Universit\'e, Universit\'e de Toulon, CNRS, CPT, Turing Center for Living Systems, Marseille, France
}
\address{$^2$
Universit\'e~Grenoble Alpes, CNRS, LIPhy, F-38000 Grenoble, France
}

\begin{abstract}
We consider a particle with a Langevin dynamics driven by a uniform non-conservative force, in a one-dimensional potential with periodic boundary conditions. 
We are interested in the properties of the system for atypical values of the time-integral of a generalized particle current.
To study these, we bias the dynamics, at trajectory level, by a parameter conjugated to the current, within the large-deviation formalism.
We investigate, in the weak-noise limit, the phase diagram spanned by the physical driving force and the parameter defining the biased process.
We focus in particular on the depinning transition in this two-dimensional phase diagram.
In the absence of trajectory bias, the depinning transition as a function of the force is characterized by the standard exponent $\frac{1}{2}$. We show that for any non-zero bias, the depinning transition is characterized by an inverse logarithmic behavior as a function of either the bias or the force, close to the critical lines. We also report a scaling exponent $\frac{1}{3}$ for the current when considering the depinning transition in terms of the bias, fixing the non-conservative force to its critical value in the absence of bias.
Then, focusing on the time-integrated particle current, we study the thermal rounding effects in the zero-current phase when the tilted potential exhibits a local minimum. We derive in this case the Arrhenius scaling, in the small noise limit, of both the particle current and the scaled cumulant generating function. This derivation of the Arrhenius scaling relies on the determination of the left eigenvector of the biased Fokker-Planck operator, to exponential order in the low-noise limit. An effective Poissonian statistics of the integrated current emerges in this limit.
\end{abstract}

\tableofcontents

\newpage

\section{Introduction}
\label{sec:introduction}

The dynamics of an equilibrium system can be described in many cases as a Langevin dynamics \cite{vankampen_stochastic_2007,risken_fokker-planck_1996}. This dynamics can be driven out of equilibrium by introducing a non-conservative force, that generates a current and breaks the microscopic time reversibility, or detailed balance.
Another way to bias the equilibrium statistics is by conditioning the statistics of full trajectories to have a given average value of a time-integrated observable, like the average particle current, the average activity, or the average energy dissipation \cite{jack_large_2010,chetrite_nonequilibrium_2013,chetrite_nonequilibrium_2015,tizon-escamilla_effective_2019,tociu_how_2019}.
Such an approach relies on the full statistics of trajectories \cite{onsager53a,onsager53b} instead of the single-time probability distribution of microscopic configurations.
When trajectories are considered in the limit of long time intervals, this approach can be conveniently formulated using the large deviation framework \cite{ellis_entropy_1985,ellis_overview_1995,ellis_theory_1999,bertini15a,touchette_large_2009}, which is particularly instrumental to evaluate the statistics of time-integrated observables~\cite{bertini15a,touchette_large_2009,derrida07a,bertini_current_2005,bertini06a,hurtado14a,derrida_exact_1998,lebowitz_gallavotticohen-type_1999,derrida_universal_1999,harris_phase_2017}.
One may guess from physical intuition that biasing the dynamics by imposing a non-zero average current should be similar to imposing a physical driving force that generates a current. 

This correspondence can be put on a firm ground by using methods based on an abstract transformation of the deformed Markov operator to define a probability-conserving effective dynamics \cite{miller_convexity_1961,simon_construction_2009,popkov_asep_2010,jack_large_2010,chetrite_nonequilibrium_2013,chetrite_nonequilibrium_2015,proesmans_large-deviation_2019,derrida_sadhu_2019}.
An illustration of this procedure in the simple case of a particle in a one-dimensional periodic potential with a Langevin dynamics has been given in Ref.~\cite{tsobgni_nyawo_large_2016} (see also our previous contribution~\cite{tizon-escamilla_effective_2019}). 
Such driven one-dimensional problems are of interest since they can be studied experimentally (for instance to study modified fluctuation-dissipation relations~\cite{gomez-solano_experimental_2009})
and provide examples of Brownian ratchets (see~\cite{reimann_brownian_2002} for a review).
In this work, we are interested in the characteristic behavior of the particle for trajectories conditioned to present an atypical value of the time-averaged velocity $v$ on a long time window.
By a change of ensemble at trajectory level, such `microcanonical' (or \emph{conditioned}) ensemble is shown to be equivalent to a \emph{biased} ensemble where trajectories are weighted by $\ee^{\lambda t v}$, where $\lambda$ plays the role of a Lagrange multiplier conjugated to $v$ (see~\cite{chetrite_nonequilibrium_2015}).
In this paper, we focus our analytical study on the biased ensemble only.
In the 1D problem we consider, biasing the dynamics
amounts to an effective Langevin dynamics with a uniform drive in a `renormalized' periodic potential~\cite{tsobgni_nyawo_large_2016,tizon-escamilla_effective_2019}. Hence, perhaps at odds with a naive physical picture, biasing the dynamics by the average current is not simply equivalent to including a uniform driving force in the system, but the potential energy is also modified in a non-trivial way.

Interestingly, this modification of the potential energy has important consequences on the scaling properties of the depinning transition which separates, in the weak-noise limit, the zero-current regime where the particle is trapped in a minimum of the tilted potential, from the non-zero current regime where a propagative motion sets in. 
In the absence of bias ($\lambda=0$) the particle is trapped in a local minimum of the potential unless the drive $f$ is large enough to allow the particle to reach a steady state with non-zero average current $\bar{v}$.
This standard `depinning transition' leads for the average current $\bar{v}$ to a scaling $\bar{v} \sim (f-f_{\cc})^{1/2}$ with the distance to the critical force $f_{\cc}$ \cite{stratonovich_oscillator_1965,strogatz_nonlinear_2001,fisher_collective_1998,brazovskii_pinning_2004}.
If, instead, one fixes $f=0$ and varies the value of $\lambda$, one observes an analogous `dynamical phase transition' (DPT): the bias $\lambda$ towards atypical velocities has to be large enough for the particle to reach a non-zero velocity state \cite{tsobgni_nyawo_large_2016}, but in that case one finds an inverse logarithmic scaling of the current $\bar{v}$ close to the DPT \cite{tizon-escamilla_effective_2019}, in terms of the Lagrange multiplier $\lambda$.
One may then wonder how the two scalings can be matched when considering the full parameter plane $(\lambda,f)$ of the statistical bias $\lambda$ and the physical drive~$f$, rather than the two parameter lines $(\lambda,f=0)$ and $(\lambda=0,f)$.
Remark that, in our system of interest, the transition lines in the $(\lambda,f)$ plane separate zero-current and propagative regimes; for convenience, the corresponding DPTs will also be called  depinning transitions, by extension of the $\lambda=0$ case. 
Note that the relation between the $\lambda=0$ depinning transition and the DPT was also investigated in Ref.~\cite{tsobgni_nyawo_large_2016}.

In this paper, we investigate in the weak-noise limit the phase diagram of the depinning transition of a one-dimensional biased Langevin process corresponding to a particle in a potential, with a non-conservative driving force $f$
and a biasing parameter $\lambda$ conjugated to a generalized current $v$. The transition lines between zero-current and non-zero-current phases in the phase diagram $(\lambda,f)$ are determined.
The scalings of the scaled cumulant generating function (SCGF) $\phi(\lambda,f)$ and of the average current $\bar v$ close to one of the transition lines are obtained. We find in particular that the standard depinning exponent $\frac{1}{2}$ of the depinning transition is only valid for the non-biased dynamics ($\lambda=0$), while any amount of bias ($\lambda\ne 0$) leads to a generic inverse logarithmic scaling of the generalized current $\bar v$ close to the critical line. We also report a scaling exponent $\frac{1}{3}$ for the current when considering the depinning transition in terms of~$\lambda$, right at the critical force $f_{\cc}$. Crossover lines between the different scaling behaviors in the plane $(\lambda, f)$ are briefly discussed.
In addition, we study the thermal rounding effects in the zero-current phase. Denoting by~$\epsilon$ the amplitude of the noise, we obtain an Arrhenius scaling $ \asymp \ee^{-\tilde\Phi(\lambda,f)/\epsilon }$ for the SCGF, as well as for the generalized current $\bar{v}(\lambda,f)$, in the low-noise asymptotics $\epsilon \to 0$. 
A justification for such an Arrhenius scaling was given in Ref.~\cite{tsobgni_nyawo_large_2016} (within the Freidlin--Wentzell--Graham formalism~\cite{graham_weak-noise_1985,freidlin_random_2012}), and
we provide in this paper a method for the explicit determination of the Arrhenius function $\tilde\Phi(\lambda,f)$. We discuss the physical implication of the result in terms of an effectively Poissonian distribution of the integrated current in the small-noise limit.

The study of the SCGF in this problem has a long history, starting with the determination of the entropy current distribution in models of colloidal particles~\cite{speck_distribution_2007,mehl_large_2008,speck_large_2012} and studies of the current distribution~\cite{alexander_rayleigh-ritz_1997,nemoto_variational_2011,nemoto_thermodynamic_2011,tsobgni_nyawo_large_2016} but it is only recently that analytical studies of the dynamical phase transitions at small noise were achieved~\cite{tizon-escamilla_effective_2019,proesmans_large-deviation_2019}.
We refer the reader to Sec.~\ref{sec:crit-lines-separ} for a comparison of our results to previous ones.
A difficulty is that the order of the transition as a function of the Lagrange multiplier $\lambda$ is not obvious (with a logarithmic singularity~\cite{tizon-escamilla_effective_2019}) whenever the potential in which the particle evolves presents metastable states. 
Proesmans and Derrida~\cite{proesmans_large-deviation_2019} have determined analytically the finite-temperature exponential corrections of the form $\ee^{-\tilde\Phi(\lambda,f)/\epsilon }$ for the SCGF in situations where the potential presents no metastable state, using a WKB approach.
In the present paper, following a different approach, we consider the case of a potential with a metastable state, which, as we detail, modifies the nature of the DPTs while also bringing a new host of technical issues that we overcome in order to determine the SCGF associated with the particle current.

\begin{figure}[t]
\begin{center}
  \includegraphics[width=0.49\columnwidth]{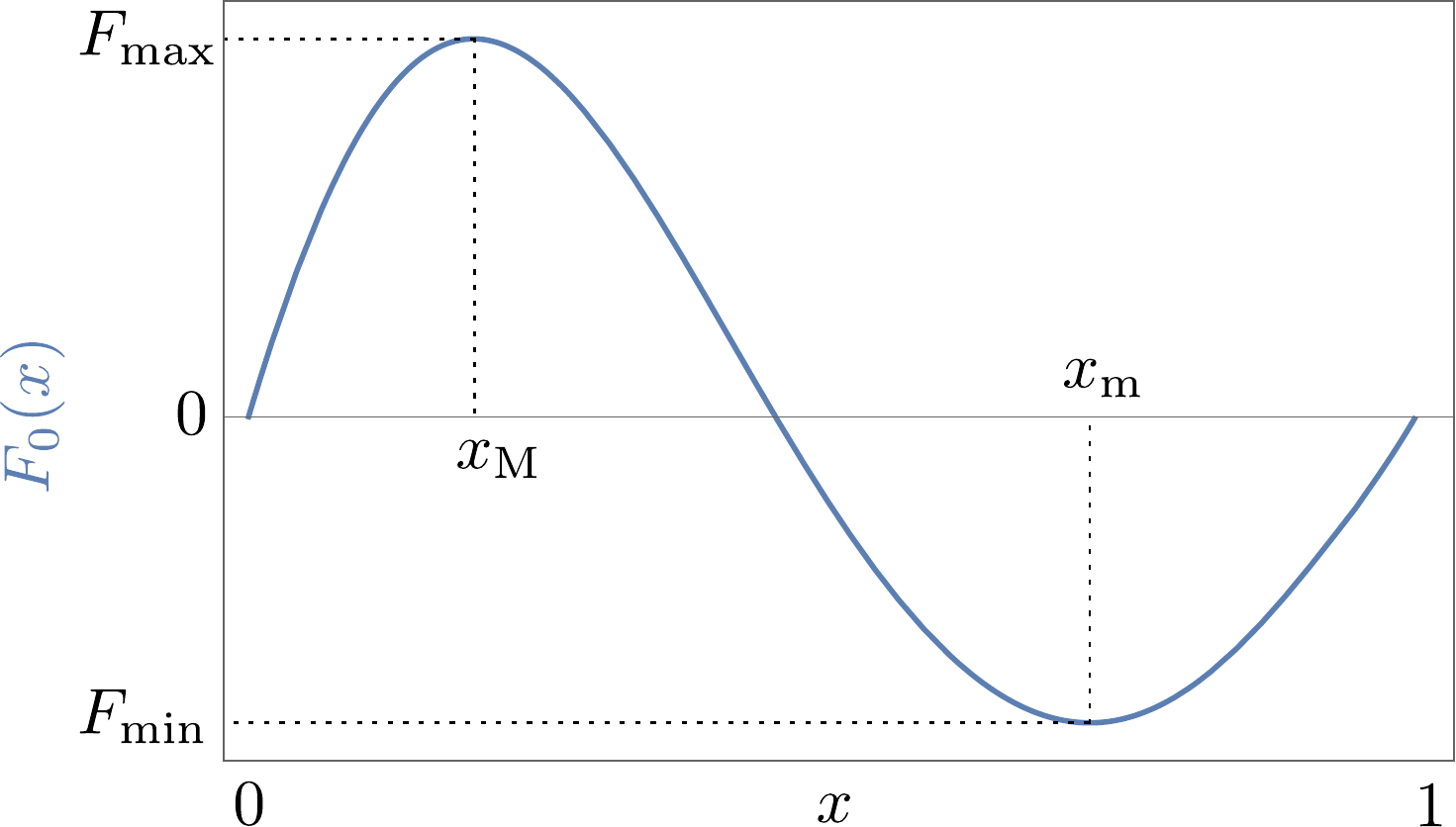}
\end{center}
\caption{Schematic representation of the conservative contribution $F_0(x)=-V'(x)$ to the one-dimensional force $F(x,f)=F_0(x)+f$ that we consider. The potential $V(x)$ and its corresponding force $F_0(x)$ are periodic functions on $[0,1]$. The constant drive $f$ verifies $f=\int_0^1 \dd x \, F(x,f)$.
\label{fig:scheme_F0ofx}
}
\end{figure}

\section{Model and dynamics}

\subsection{Model: a driven particle in a one-dimensional potential with metastability}
\label{sec:model:-driv-part}

Consider a particle of position $x(t)$ at time $t$, evolving on a ring of period $1$, subjected to a continuous force field $F(x)$ and to a Gaussian white noise $\eta(t)$. In the overdamped limit, the Langevin equation describing the evolution of the position is:
\begin{equation} \label{eq:Langevin}
\dot{x}(t) = F(x(t)) + \sqrt{\epsilon} \, \eta(t)
\end{equation}
(the overdot denotes a time derivative). The noise $\eta(t)$ has zero mean and correlation $\langle \eta(t) \eta(t') \rangle=\delta(t-t')$. The spatial period of the force is equal to $1$, i.e., $F(x+1)=F(x)$, and the temperature is $\epsilon/2$.

In our previous study~\cite{tizon-escamilla_effective_2019}, we have studied the statistics of an observable of the type
\begin{equation}\label{eq:generalobservable}
A(\tf) = \int_0^{\tf} \dd t \, h(x(t))\,  + \int_0^{\tf} \dd t\, g(x(t))\, \dot{x}(t)
\end{equation}
with arbitrary functions $g(x)$ and $h(x)$,
and where the stochastic integral involving $g(x)$ is understood in a Stratonovich sense (see e.g.~\cite{oksendal_stochastic_2013}). This will be also the case of all the stochastic integrals described in this paper.
In the following, we will specialize to current-type observables, for which $h(x)=0$, so that the observable $A(\tf)$ reads as $A(\tf) = \int_0^{\tf} \dd t\, g(x(t))\, \dot{x}(t)$.
As discussed in \cite{tizon-escamilla_effective_2019}, the force field $F(x)$ can be decomposed without loss of generality into a conservative part $F_0(x)=-V'(x)$ (where $V(x)$ is a periodic function of $x$ of period $1$) and a homogeneous drive $f$,
\begin{equation} \label{eq:general:force}
F(x,f) = F_0(x) \: + \: f \,.
\end{equation}
By definition, $\int_0^1 \dd x\, F_0(x) =0$, so that
$f = \int_0^1 \dd x\, F(x,f)$.
Since we are specifically interested in this work in the dependence of the dynamics on the non-conservative force $f$, we have made the $f$-dependence explicit and wrote the force as $F(x,f)$.
We assume that the conservative part $F_0(x)$ of the force has a single local maximum $F_{\max}=F_0(x_{\rm M})>0$ and a single local minimum $F_{\min}=F_0(x_{\rm m})<0$ (see Fig.~\ref{fig:scheme_F0ofx}). 
We also assume that $F_0(x)$ is monotonous on the intervals separating the maximum and the minimum (there are two such intervals because of the periodic boundary conditions), so that there exist two stationary points where $F_0(x)=0$.
A typical example of such a force is $F_0(x)=\sin (2\pi x)$. Note that we may further assume that $x_{\rm M}<x_{\rm m}$ as in the purely sinusoidal case, which can be achieved through a simple translation of the  $x$ variable.
We also focus on the case of a positive drive $f\geq 0$ (without loss of generality, since, as seen from Eq.~(\ref{eq:Langevin}), one can change the sign of $f$ through $x\mapsto 1-x$).

In the following, we tackle the general form~(\ref{eq:general:force}) of the force, considering in some illustrations the simple sinusoidal force field
\begin{equation} \label{eq:sinus:force}
F_{\sin}(x,f) = \sin (2\pi x) \: + \: f \,
\end{equation}
for instance to compute explicitly quantities which depend on the details of the potential.
The amplitude of the sine contribution is set to $1$ without loss of generality, through a rescaling of time and temperature.

\subsection{Distribution of an additive observable}
\label{sec:distr-an-addit}

The distribution of the observable $A(\tf)$ in the large-time limit is determined through the SCGF $\varphi_\epsilon(\lambda,f)$ defined as\footnote{Throughout the paper, we use the symbol $\asymp$ for the asymptotic logarithmic equivalence in the case of large deviation scalings, and the symbol $\approx$ for the standard asymptotic equivalence (i.e., when the correct prefactors are included). Following standard practice in physics papers, we use the symbol $\sim$ to denote equivalent (typically power-law) scalings when prefactors are not included explicitly.}
\begin{equation}
  \label{eq:defSCGF}
  \big\langle \ee^{-\frac\lambda\epsilon A(\tf)} \big\rangle  \stackrel[\tf\to\infty]{}{\asymp} \ee^{\tf\,\varphi_\epsilon(\lambda,f)}
\end{equation}
where the average is taken over trajectories of duration $\tf$. 
The parameter $\lambda$ is a statistical bias acting at the level of trajectories, by reweighting the probability of trajectories associated with the Langevin equation~(\ref{eq:Langevin}). The standard Langevin dynamics is recovered in the case $\lambda=0$.
Up to a rescaling of the parameter $\lambda$, one sets without loss of generality $\int_0^1 \dd x\, g(x) =1$, which will prove useful to lighten notations\footnote{%
We easily see that $A(\tf)=\int_0^\tf\dd t\, g(x(t))\,\dot x(t)$ is split into a sum of two terms: one that is equal to  $\int_0^1 \dd x\, g(x)$ times the number of turns (counted algebraically) that $x(t)$ performs on $[0,\tf]$; and one that can be bounded by $\max_{[0,1]} g$.
Thus, if $\int_0^1 \dd x\, g(x) =0$, the observable $A(\tf)$ is not extensive in time at large $\tf$ and the relevant large-deviation scaling to consider is different from~(\ref{eq:defSCGF}).
}. 
In the large $\tf$ limit, the conditioned trajectory ensemble (where $A(\tf)/\tf=a$ is fixed) and the canonical trajectory ensemble (weighted by $\ee^{-\frac\lambda\epsilon A(\tf)}$) become equivalent, for an appropriate value of $\lambda$ (see for instance~\cite{garrahan_first-order_2009,touchette_large_2009,chetrite_nonequilibrium_2013,chetrite_nonequilibrium_2015}).
Also, the distribution of $A$ verifies a large-deviation principle $P[A(\tf)=a\,\tf]\asymp \ee^{\tf\,\Pi_\epsilon(a)}$ as $\tf\to\infty$. If $\Pi_\epsilon$ is a concave function of $a$, then it is related to $\varphi_\epsilon$ via a Legendre transform.
In the present paper, we focus our study on $\varphi_\epsilon(\lambda)$.

\smallskip

In \cite{tizon-escamilla_effective_2019}, we studied in particular how a biased conservative dynamics (i.e., $\lambda \ne 0$, but $f=0$) can be mapped to non-conservative, unbiased dynamics (i.e., $\lambda = 0$ and $f\ne 0$) \cite{tsobgni_nyawo_large_2016}, a case which is intuitively easier to grasp.
In the present work, we instead leave aside this mapping, and rather consider the interplay of the statistical bias $\lambda$ and the physical driving $f$ on the same dynamics. In other words, we investigate the $(\lambda,f)$ phase diagram of the model, and thus emphasize the dependence of relevant quantities on the parameters $\lambda$ and $f$.

For the original dynamics, at zero temperature and in the absence of bias ($\lambda=0$), the behavior of the average velocity
\begin{equation}
 \bar{V}=\lim_{\tf \to \infty} \frac{1}{\tf} \, \int_0^\tf \dd t \, \dot{x}(t)
 \end{equation}
as a function of the driving force~$f$ is very simple:  
there is a critical value of the force $f_{\cc}=-F_{\text{min}}>0$ such that for $f<f_\cc$, the particle is trapped in a stationary point $x_\Ss$ satisfying $F(x_\Ss,f)=0$ in the noiseless situation, and can only escape the trap from thermal activation; for $f>f_\cc$, there are no stationary points, the particle is no longer trapped and can thus move freely across the system. 
This phenomenology is well known~\cite{stratonovich_oscillator_1965,strogatz_nonlinear_2001} and is one of the simplest instances of the depinning transition (see e.g.~\cite{fisher_collective_1998,brazovskii_pinning_2004} for reviews).
The average velocity of the particle behaves as $|f-f_\cc|^{1/2}$ for $f>f_\cc$ and is governed by the so-called mean-field depinning exponent $\beta=\frac 12$.
See also Ref.~\cite{tsobgni_nyawo_large_2016} for a discussion on the role of such depinning behavior for the DPT as a function of $\lambda$.

To characterize more generally this transition for the biased process, we are specifically interested in the behavior of the average value
\begin{equation}\label{eq:gen-velocity}
\bar{v}(\lambda,f) = \lim_{\tf \to \infty} \frac{1}{\tf} \, \big\langle A(\tf) \big\rangle_\lambda 
\qquad
\text{with}
\quad
\langle\,\cdot\,\rangle_\lambda = \frac{\big\langle \ee^{-\frac\lambda\epsilon A(\tf)} \,\cdot\, \big\rangle}{\big\langle \ee^{-\frac\lambda\epsilon A(\tf)} \big\rangle }
\end{equation}
as a function of the two relevant parameters $(\lambda,f)$, in the small noise asymptotics $\epsilon\to 0$.
At a qualitative level, both parameters $\lambda$ and $f$ can be understood as bias parameters that impose a nonzero value of $\bar{v}$. However, we have shown in~\cite{tizon-escamilla_effective_2019} that both parameters have quite different effects at a quantitative level.
Note that the notation $\bar{v}$ is chosen here because, for $g(x)=1$, the observable $\bar{v}$ is simply the average current, or average velocity of the particle. In the following, we thus call $\bar{v}$ the generalized average current.
Also, in the Legendre transformation between $\varphi_\epsilon(\lambda,f)$ and $\Pi_\epsilon(a)$, the quantity $\bar v$ is precisely such that $\varphi_\epsilon(\lambda,f)= \Pi_\epsilon(\bar v) -\frac\lambda\epsilon \bar v $ so that understanding the dependency of $\bar v$ on $\lambda$ allows one to reconstitute the behavior of the distribution $P[A(\tf)=a\,\tf]$ from the knowledge of $\varphi_\epsilon(\lambda,f)$.

We now turn to the determination of the critical lines separating the zero-current and non-zero-current phases in the phase diagram $(\lambda,f)$.

\section{Critical lines separating fixed points and propagative trajectories}
\label{sec:crit-lines-separ}

It the zero-temperature limit, a DPT separates zero-current and non-zero-current phases; it was investigated numerically 
using an eigenvector decomposition in~\cite{mehl_large_2008} and~\cite{speck_large_2012}, with the observation of a ``kink'' or a ``wedge'' shape of the SCGF.
The current distribution was investigated in a systematic way in~\cite{tsobgni_nyawo_large_2016} where, at low temperature, a ``kink'' behavior was also obtained numerically using a Fourier--Bloch decomposition.
In~\cite{tizon-escamilla_effective_2019}, for the non-driven dynamics ($f=0$) we derived an analytical characterization of the singularity, showing that it presents a continuous but logarithmic behavior as a function of $\lambda$ close
to the transition point.
In~\cite{proesmans_large-deviation_2019} the transition was studied analytically in the regime where the tilted potential presents not metastable state.
In this section and the next one,  we fully characterize the phase diagram in the $(\lambda,f)$ plane and identify the critical behavior close to singularities.

The determination of the critical lines that we now present relies on results derived in
\cite{tizon-escamilla_effective_2019,proesmans_large-deviation_2019}, that we briefly recall below.
In the small-noise limit, the path-integral description of trajectorial average in Eq.~(\ref{eq:defSCGF}) is dominated by an `optimal trajectory'.
It has been shown that depending on $(\lambda,f)$, the optimal trajectory can be either a fixed point or a propagative trajectory, and the criterion allowing one to distinguish between these two possibilities was also determined in \cite{tizon-escamilla_effective_2019}.
In the limit $\epsilon\to 0$, the SCGF $\varphi_\epsilon(\lambda,f)$ behaves as 
\begin{equation}
  \label{eq:defphi0noise}
  \varphi_\epsilon(\lambda,f)  
  \ \stackrel[\epsilon\to 0]{}{\sim} \
  \frac 1\epsilon \phi(\lambda,f)
\end{equation}
where $\phi(\lambda,f)$ does not depend on $\epsilon$.
Optimal trajectories are then propagative if and only if the following criterion is satisfied \cite{tizon-escamilla_effective_2019}:
\begin{equation} \label{eq:propag:cdtn}
|\lambda-f| > \int_0^1 \dd x \, \sqrt{2\mathcal{V}_{\max}(f)+F(x,f)^2}
\end{equation}
where $\mathcal{V}_{\max}(f)=\max_x[-\frac{1}{2}F(x,f)^2]$.
The assumptions $h(x)=0$ and $\int_0^1 \dd x \, g(x) =1$ have been used to simplify the criterion
given in~\cite{tizon-escamilla_effective_2019}.
When Eq.~(\ref{eq:propag:cdtn}) holds, the SCGF $\phi(\lambda,f)$ is solution of the equation
\begin{equation} \label{eq:SCGF}
\int_0^1 \dd x \, \sqrt{2\phi(\lambda,f)+F(x,f)^2} = |\lambda-f| \,.
\end{equation}
Otherwise, if Eq.~(\ref{eq:propag:cdtn}) is not satisfied, the optimal trajectories are fixed points and $\phi(\lambda,f)=\mathcal{V}_{\max}(f)$. The domain of validity of Eq.~(\ref{eq:propag:cdtn}) is limited by two lines
$\lambda_{\cc}^{-}(f)$ and $\lambda_{\cc}^{+}(f)$, respectively defined by
\begin{eqnarray}
\label{eq:lambdacm:def}
\lambda_{\cc}^{-}(f) &=& f - \int_0^1 \dd x \, \sqrt{2\mathcal{V}_{\max}(f)+F(x,f)^2} \,,\\
\label{eq:lambdacp:def}
\lambda_{\cc}^{+}(f) &=& f + \int_0^1 \dd x \, \sqrt{2\mathcal{V}_{\max}(f)+F(x,f)^2}\,.
\end{eqnarray}
\begin{figure}[t]
\begin{center}
  \includegraphics[height=0.4\columnwidth]{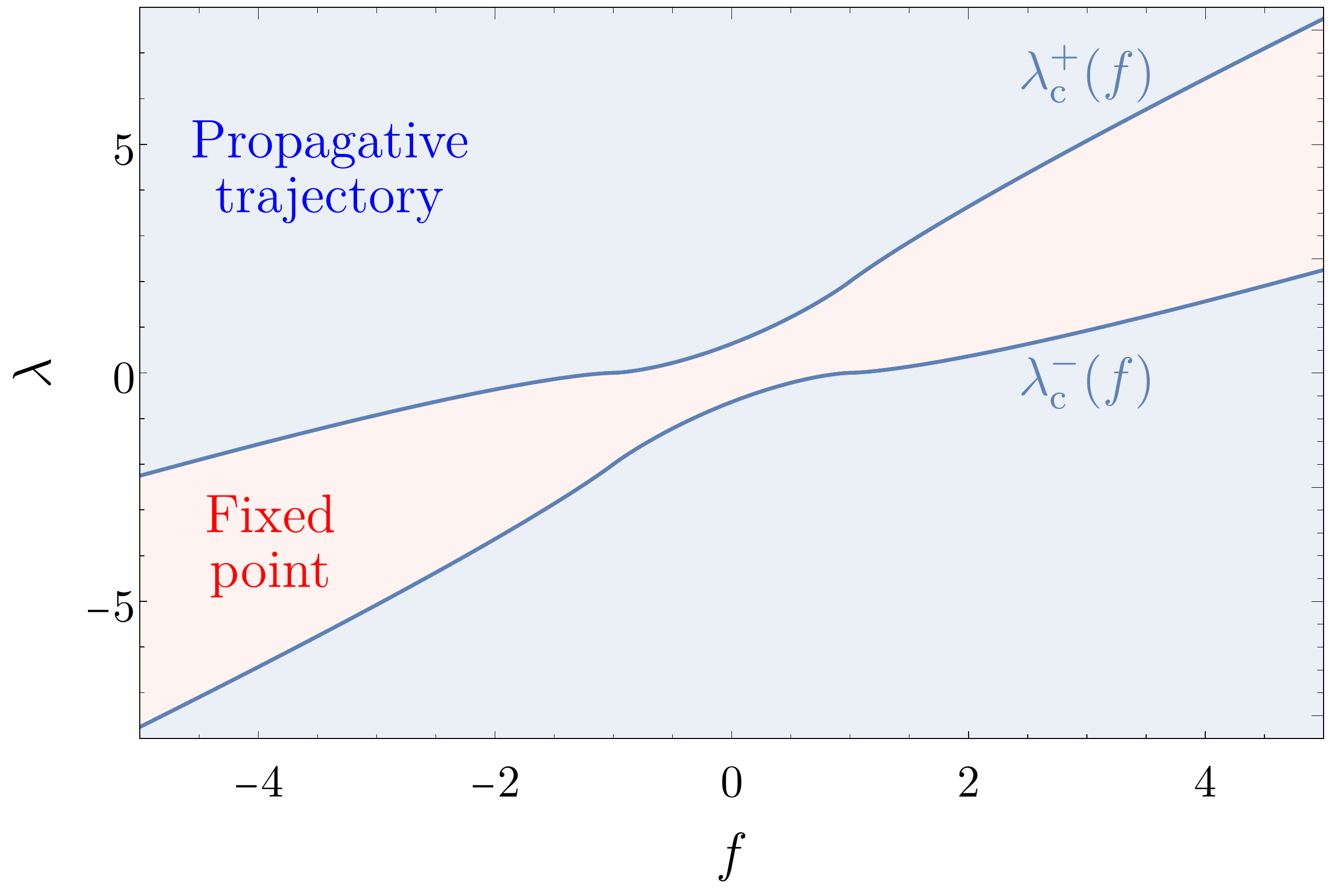}
\end{center}
\caption{
Phase diagram in the plane $(\lambda,f)$ for the sinusoidal force field $F_{\sin}(x,f)$ of Eq.~(\ref{eq:sinus:force}). The two types of optimal trajectories are a fixed point ({\textbf{red}}), associated with a zero current, or a propagative trajectory ({\textbf{blue}}), associated with a non-zero current. The critical lines $\lambda_\cc^\pm(f)$ separate these propagative and zero-current phases.
The order of the transitions along these lines is non trivial and discussed in Sec.~\ref{sec:scaling-phi-bar}.
\label{fig:phasediagram}
}
\end{figure}%
In Fig.~\ref{fig:phasediagram} we show a representation of the corresponding phase diagram in the plane $(\lambda,f)$ for the sinusoidal force field $F_{\sin}(x,f)$.
As a consistency check, one sees that the critical value $f_{\cc}=-F_{\min}>0$ of the force is such that $\lambda_{\cc}^{-}(f_\cc)=0$.
Indeed $f_\cc$ is the minimum force for which $F(x,f)\geq 0$ for all $x\in[0,1]$, hence $\mathcal V_{\max}(f_\cc)=0$ and
\begin{equation} \label{eq:critical:force}
\int_0^1 \dd x \, \sqrt{2\mathcal{V}_{\max}(f_{\cc})+F(x,f_{\cc})^2} = \int_0^1\dd x\, F(x,f_\cc)=f_{\cc}
\ ,
\end{equation}
leading to $\lambda_{\cc}^{-}(f_\cc)=0$.
By symmetry, the condition $\lambda_{\cc}^{+}(f)=0$ yields the negative critical force $-F_{\max}$ that is not in the range of force we consider.
In~\cite{tsobgni_nyawo_large_2016,tizon-escamilla_effective_2019}, for the sine force~\eqref{eq:sinus:force}, the values $\lambda_{\cc}^{\pm}(0)=\pm\frac{2}{\pi}$ have been determined. Here we wish to determine the values of $\lambda_{\cc}^{-}(f)$ and $\lambda_{\cc}^{+}(f)$ for any $f>0$ (the values for $f<0$ are obtained by a simple symmetry, as discussed in Sec.~\ref{sec:model:-driv-part}). We thus need to evaluate the integral appearing in Eq.~(\ref{eq:propag:cdtn}), which requires to distinguish the cases $0\le f\le f_{\cc}$ and $f>f_{\cc}$.

\subsection{Critical lines for $0\le f\le f_{\cc}$}

When $0\le f\le f_{\cc}$, there exist crossing points such that $F(x,f)=0$ and one thus has $\mathcal{V}_{\max}=0$.
The integral appearing in Eqs.~(\ref{eq:lambdacm:def}) and~(\ref{eq:lambdacp:def}) then reads
\begin{equation}
\int_0^1 \dd x \, \sqrt{2\mathcal{V}_{\max}(f)+F(x,f)^2} = \int_0^1 \dd x \, |F(x,f)|
= \int_0^1 \dd x \, |F_0(x)+f|\,,
\end{equation}
which leads to
\begin{equation} \label{eq:lambdac:fltfc}
\lambda_{\cc}^{\pm}(f) = f \pm \int_0^1 \dd x \, |F(x,f)|\,.
\end{equation}
In the specific case where $F_{\sin}(x,f) = \sin (2\pi x) + f$, we get the explicit result
\begin{equation}\label{eq:lambdac:fltfc:sin}
\lambda_{\cc}^{\pm}(f) = f \pm \frac{2}{\pi} \left( \sqrt{1-f^2}
+ f \, {\rm arcsin} f \right).
\end{equation}
We can check here explicitly that $f_{\cc}=-F_{\min}=1$  is the solution of $\lambda_{\cc}^{-}(f_\cc)=0$.
In addition, one can use the explicit form (\ref{eq:lambdac:fltfc:sin}) of $\lambda_{\cc}^{-}(f)$ to evaluate its asymptotic behavior for $f \to 0$ and $f \to 1^{-}$,
\begin{eqnarray}
&& \lambda_{\cc}^{-}(f) = -\frac{2}{\pi} + f - \frac{1}{\pi} f^2 + o(f^{2})\, , \qquad f \to 0 \,,\\
&& \lambda_{\cc}^{-}(1-\deltaf) = -\frac{4\sqrt{2}}{3\pi} \, (\deltaf)^{3/2} + o(\deltaf^{3/2})\, , \qquad
\deltaf \to 0^{+} \,.\label{eq:lambdac-_1-}
\end{eqnarray}
The behavior of $\lambda_{\cc}^{+}(f)$ can be obtained from the relation
$\lambda_{\cc}^{+}(f)=2f-\lambda_{\cc}^{-}(f)$ (which expresses that two critical values $\lambda^\pm_\cc$ are related by the Gallavotti--Cohen symmetry discussed in Appendix~\ref{sec:GC-sym}).

\subsection{Critical lines for $f>f_{\cc}$}

When $f>f_{\cc}$, $\mathcal{V}_{\max}(f)$ is nonzero and depends on the details of the force field $F(x,f)$. To go beyond the general expressions
(\ref{eq:lambdacm:def}) and~(\ref{eq:lambdacp:def}), one needs to specify explicitly the force field.
In the case $F(x,f) = F_{\sin}(x,f)= \sin (2\pi x) + f$, for which $f_{\cc}=1$, we get $\mathcal{V}_{\max}(f)=-\frac{1}{2}(f-1)^2$. One thus has
\begin{equation} \label{eq:def:integ}
\int_0^1 \dd x \, \sqrt{2\mathcal{V}_{\max}+F(x)^2} =
\frac{1}{2\pi} \int_0^{2\pi} \dd\theta\, \sqrt{(2f-1) + \sin^2\theta +2f \sin \theta}
\end{equation}
which, after some algebra, yields
\begin{eqnarray}
\lambda_{\cc}^{-}(f) &=& \frac{2}{\pi} \left[f \, {\rm arctan} \sqrt{f-1} - \sqrt{f-1} \right] ,\\
\lambda_{\cc}^{+}(f) &=& 2f - \frac{2}{\pi} \left[f \, {\rm arctan} \sqrt{f-1}-\sqrt{f-1} \right] .
\end{eqnarray}
From these expressions, the asymptotic forms of $\lambda_{\cc}^{\pm}(f)$ for $f \to \infty$ and for $f \to 1^{+}$ are easily determined.
One finds in particular for $\lambda_{\cc}^{-}(f)$
\begin{eqnarray}
&& \lambda_{\cc}^{-}(f) = f - \frac{4}{\pi} f^{1/2} + o(f^{1/2})\, , \qquad f \to \infty \,,\\
&& \lambda_{\cc}^{-}(1+\deltaf) = \frac{4}{3\pi} (\deltaf)^{3/2} + o(\deltaf^{3/2})\, , \qquad
\deltaf \to 0^{+} \,. \label{eq:lambdac-_1+}
\end{eqnarray}
Eqs.~(\ref{eq:lambdac-_1-}) and~(\ref{eq:lambdac-_1+}) show that the behavior of the critical line $\lambda_\cc^-(f)$ is not symmetric in the two limits $f\to 1^+$ or $f\to 1^-$: the power-law behavior $\propto \vert \deltaf\vert^{3/2}$ is the same but the prefactors are different.
Note last that the behavior of $\lambda_{\cc}^{+}(f)$ is easily obtained from the Gallavotti--Cohen symmetry
$\lambda_{\cc}^{+}(f)=2f-\lambda_{\cc}^{-}(f)$ (see Appendix~\ref{sec:GC-sym}).


\section{Scaling of $\phi$ and ${\bar v}$ close to the critical line $\lambda_{\cc}^{-}(f)$}
\label{sec:scaling-phi-bar}

For a generic force $F(x,f)$,
we now specialize, for simplicity, to the vicinity of the lower critical line $\lambda_{\cc}^{-}(f)$. Results for the other critical line $\lambda_{\cc}^{+}(f)$ can be obtained in a similar way using the Gallavotti--Cohen symmetry (see Appendix~\ref{sec:GC-sym}).
To determine the SCGF $\phi(\lambda,f)$, we need to solve the equation
\begin{equation} \label{eq:solve:phi:v1}
\int_0^1 \dd x \sqrt{2\phi(\lambda,f)+F(x,f)^2} = -\lambda+f
\end{equation}
for $\lambda<\lambda_{\cc}^{-}(f)$. The current ${\bar v}$ is then obtained as~\cite{tizon-escamilla_effective_2019}
\begin{equation} \label{eq:def:barv:phi:deriv}
{\bar v}(\lambda,f) = -\frac{\partial \phi}{\partial \lambda}(\lambda,f)\,.
\end{equation}
We set $\lambda=\lambda_{\cc}^{-}(f)-\delta \lambda$ with $\delta \lambda>0$ and look for the asymptotic behavior of $\phi$ for $\delta \lambda \to 0$.
Using Eq.~(\ref{eq:lambdacm:def}), one rewrites Eq.~(\ref{eq:solve:phi:v1}) as
\begin{equation}
\int_0^1 \dd x\, \left[ \sqrt{2\phi(\lambda,f)+F(x,f)^2}
- \sqrt{2\mathcal{V}_{\max}(f)+F(x,f)^2} \right] = \delta \lambda
\end{equation}
which is reexpressed using the relation $\sqrt{a}-\sqrt{b} = (a-b)/(\sqrt{a}+\sqrt{b})$ as
\begin{equation} \label{eq:solve:phi:v2}
2\big(\phi(\lambda,f)-\mathcal{V}_{\max}(f)\big) \int_0^1  \frac{\dd x}{\sqrt{2\phi(\lambda,f)+F(x,f)^2} + \sqrt{2\mathcal{V}_{\max}(f)+F(x,f)^2}} = \delta \lambda \,.
\end{equation}
To proceed further, it is convenient to deal separately with the cases $0\le f <f_{\cc}$ and $f >f_{\cc}$, as done above for the determination of the critical lines.

\subsection{Transition as a function of $\lambda$ for $0\le f <f_{\cc}$}\label{sect:trans-f01}
\label{sec:trans:fleqfc}

For $0\le f <f_{\cc}$, one has $\mathcal{V}_{\max}(f)=0$, and Eq.~(\ref{eq:solve:phi:v2}) simplifies to
\begin{equation} \label{eq:solve:phi:flefc}
2 \phi \int_0^1  \frac{\dd x}{\sqrt{2\phi+F(x,f)^2} + |F(x,f)|} = \delta \lambda \,,
\end{equation}
where we have dropped the explicit $(\lambda,f)$-dependence of $\phi$ to lighten notations.
For small $\delta \lambda$, $\phi$ is also small, so that one needs to evaluate the asymptotic behavior for small $\phi$ of the integral appearing in Eq.~(\ref{eq:solve:phi:flefc}).

We recall that, as shown on Fig.~\ref{fig:scheme_F0ofx}, we have assumed that the conservative part $F_0(x)$ of the force has a single local maximum and a single local minimum, and that it is monotonous in the two intervals between the maximum and the minimum.
By continuity of the force field, this implies that the force field $F(x,f)=F_0(x)+f$ exhibits two zero-crossing points, i.e., points $x$ such $F(x,f)=0$ in the case $0\le f < f_{\cc}$ considered here (and no zero-crossing points for $f > f_{\cc}$).
Let us denote as $x_1$ and $x_2$ the distinct values of $x$ satisfying $F(x_i,f)=0$ ($i=1$, $2$). We assume that $\partial_x F(x_i,f) \ne 0$ ($i=1$, $2$), so that $F(x,f)$ behaves linearly in the vicinity of each $x_i$.
Defining the intermediate point $x_1^* = \frac{1}{2} (x_1+x_2)$, one can split the interval $[0,1]$ into two subintervals $[0,x_1^*]$ and  $[x_1^*,1]$, each subinterval containing a single zero-crossing point $x_i$.
For later convenience, we define the notations $x_0^*=0$ and $x_2^*=1$, so that we can generically write the subintervals as $[x_{i-1}^*,x_i^*]$, for $i=1$, $2$.

The asymptotic evaluation of the integral in Eq.~(\ref{eq:solve:phi:flefc}) goes as follows. On each subinterval $[x_{i-1}^*,x_i^*]$, the integral
\begin{equation}
\mathcal{I}_i= \int_{x_{i-1}^*}^{x_i^*} \, \frac{\dd x}{\sqrt{2\phi+F(x,f)^2} + |F(x,f)|}
\end{equation}
is logarithmically diverging if $\phi=0$, due to a divergence of the integrand around $x=x_i$. A small, but nonzero $\phi$ thus effectively acts as a cut-off to regularize the diverging integral.
For small $\phi$, the integral is dominated by the vicinity of the crossing point $x_i$ where $F(x_i,f)=0$, and the leading behavior of the integral can be obtained by expanding $F(x,f)$ to first order around $x=x_i$, yielding
\begin{equation}
\mathcal{I}_i
\approx
\int_{x_{i-1}^*}^{x_i^*} \frac{\dd x}{\sqrt{2\phi+F_i'^{\,2\,}(x-x_i)^2} + |F_i'(x-x_i)|}
\end{equation}
where we have introduced the shorthand notation $F_i' = \partial_xF(x_i,f)$.
With the change of variable $x=x_i+\sqrt{\phi} \, y$, we obtain
\begin{equation}
\int_{x_{i-1}^*}^{x_i^*} \frac{\dd x}{\sqrt{2\phi+F_i'^{\,2\,}(x-x_i)^2} + |F_i'(x-x_i)|}
= \int_{(x_{i-1}^*-x_i)/\sqrt{\phi}}^{(x_i^*-x_i)/\sqrt{\phi}} \frac{\dd y}{\sqrt{2+F_i'^{\,2\,}y^2} + |F_i'y|} \,.
\end{equation}
The last integral over the variable $y$ diverges logarithmically, when $\phi \to 0$, at both the lower and upper bounds of the integral.
We are thus led to the following asymptotic estimation, to leading order in $\phi$ when $\phi \to 0$,
\begin{equation}
\int_{(x_{i-1}^*-x_i)/\sqrt{\phi}}^{(x_i^*-x_i)/\sqrt{\phi}} \frac{\dd y}{\sqrt{2+F_i'^{\,2\,}y^2} + |F_i'y|} \approx \frac{1}{2|F_i'|} |\ln\phi|.
\end{equation}
Gathering terms, Eq.~(\ref{eq:solve:phi:flefc}) reads, in this asymptotic regime, as
\begin{equation} \label{eq:solve:phi:flefc:v2}
\phi\, |\!\ln\phi| \left( \frac{1}{|F_1'|} + \frac{1}{|F_2'|} \right)
\approx \delta \lambda \,.
\end{equation}
Inverting the relation to get the leading order expression of $\phi$ in terms of $\delta\lambda$, one gets
\begin{equation} \label{eq:phi:scaling:gen}
\phi\big(\lambda_{\cc}^{-}(f)-\delta\lambda,f\big) \approx \frac{\tilde\phi(f) \, \delta \lambda}{|\ln\delta \lambda|} \,, \qquad
\tilde\phi(f) =  \frac{|F_1' F_2'|}{|F_1'|+|F_2'|}\,,
\end{equation}
where we recall that $F_i' = \partial_xF(x_i,f)$ depends on $f$.
From Eq.~(\ref{eq:def:barv:phi:deriv}), the average generalized current ${\bar v}$ then reads as
\begin{equation} \label{eq:barv:scaling:gen}
{\bar v}\big(\lambda_{\cc}^{-}(f)-\delta\lambda,f\big) \approx \frac{\tilde\phi(f)}{|\ln\delta \lambda|} \,.
\end{equation}
The depinning transition (which is a DPT) as a function of $\lambda$ for $0\le f < f_{\cc}$ thus behaves with a logarithmic divergence,
at odds with standard DPTs, which are first- or second-order~\cite{bodineau05a,bertini_current_2005,PhysRevLett.98.195702,lecomte_thermodynamic_2007,jack_dynamical_2020}.
This shows that the logarithmic, almost discontinuous DPT as a function of $\lambda$ found in~\cite{tizon-escamilla_effective_2019} for $f=0$ is actually present in the whole range of force $0\leq f<f_{\cc}$.

In the specific case of a sinusoidal force field $F_{\sin}(x,f) = \sin (2\pi x) + f$,
one has $|F_1'|=|F_2'|=\pi\sqrt{1-f^2}$, resulting in
\begin{eqnarray}
\label{eq:phi:scaling:sinus}
\phi\big(\lambda_{\cc}^{-}(f)-\delta\lambda,f\big) &\approx& \frac{\pi}{2} \sqrt{1-f^2}\, \frac{\delta \lambda}{|\ln \delta \lambda|} \,, \\
\label{eq:barv:scaling:sinus}
{\bar v}\big(\lambda_{\cc}^{-}(f)-\delta\lambda,f\big) &\approx&  \frac{\pi}{2}\, \frac{\sqrt{1-f^2}}{|\ln \delta \lambda|} \,.
\end{eqnarray}
Note that one generically expects, for $f=f_{\cc}-\deltaf$ (with $\deltaf>0$ and small) the scaling $\tilde\phi(f) \sim (\deltaf)^{1/2}$, as found on the example of the sine force field~(\ref{eq:sinus:force}).

\subsection{Transition as a function of $\lambda$ for $f>f_{\cc}$}
\label{sec:trans:fgeqfc}

We now turn to the case $f>f_{\cc}$, for which $F(x,f)>0$ for all $x$.
The force field $F(x,f)$ has a single minimum $F_{\mm}(f)$ at a point $x_{\rm m}$
and $\mathcal{V}_{\max}(f) = -\frac{1}{2}F_{\mm}(f)^2 < 0$.
Writing $\delta\phi = \phi(\lambda,f)-\mathcal{V}_{\max}(f)$, Eq.~(\ref{eq:solve:phi:v2}) turns into
\begin{equation} \label{eq:solve:phi:v3}
2\delta\phi \int_0^1  \frac{\dd x}{\sqrt{2\delta\phi+F(x,f)^2-F_{\mm}(f)^2} + \sqrt{F(x,f)^2-F_{\mm}(f)^2}} = \delta \lambda \,,
\end{equation}
where we recall that $\delta \lambda>0$ is defined by $\lambda=\lambda_{\cc}^{-}(f)-\delta \lambda$.
For $\delta\phi\to 0$, the integral in Eq.~(\ref{eq:solve:phi:v3}) is dominated by the divergence of the integrand close to $x=x_{\rm m}$. 
Expanding $F(x,f)$ in the vicinity of $x_{\rm m}$, we get to leading order
$F(x,f)^2-F_{\mm}^2 = F_{\mm}\, F_0''\,(x-x_{\rm m})^2$, with the shorthand notation
$F_0''=\partial_{x}^{2} F_0(x_{\rm m},f)$, and where we have dropped the explicit $f$-dependence of $F_{\mm}$ on $f$ to lighten notations. Note that $F_0''>0$ because it corresponds to a minimum of the force field.
In the small $\delta\phi$ limit, Eq.~(\ref{eq:solve:phi:v3}) can thus be rewritten as
\begin{equation} \label{eq:solve:phi:v3bis}
2\delta\phi \int_0^1  \frac{\dd x}{\sqrt{2\delta\phi + F_{\mm} F_0''(x-x_{\rm m})^2} + \sqrt{F_{\mm} F_0''}\,|x-x_{\rm m}|} = \delta \lambda \,.
\end{equation}
Using the change of variable $x=x_{\rm m}+\sqrt{\delta\phi} \, y$, we get in the limit $\delta\phi\to 0$,
\begin{equation}
\int_{-x_{\rm m}/\sqrt{\delta\phi}}^{(1-x_{\rm m})/\sqrt{\delta\phi}} \frac{\dd y}{\sqrt{2+
F_{\mm} F_0'' y^2} + \sqrt{F_{\mm} F_0''}|y|}
\approx \frac{1}{2\sqrt{F_{\mm} F_0''}}\, |\!\ln\delta\phi|.
\end{equation}
It follows that 
\begin{equation} \label{eq:phi:scaling:gen2}
\phi\big(\lambda_{\cc}^{-}(f)-\delta\lambda,f\big) \approx \frac{\tilde{\phi}(f) \, \delta \lambda}{|\!\ln\delta \lambda|} \,, \qquad
\tilde{\phi}(f) = \sqrt{F_{\mm} F_0''}\,,
\end{equation}
where we recall that both $F_{\mm}$ and $F_0''$ depend on $f$.
From Eq.~(\ref{eq:def:barv:phi:deriv}), the average generalized current ${\bar v}$ then reads, for $\delta\lambda\to 0^{+}$, as
\begin{equation} \label{eq:barv:scaling:gen2}
{\bar v}\big(\lambda_{\cc}^{-}(f)-\delta\lambda,f\big) \approx \frac{\tilde{\phi}(f)}{|\!\ln\delta \lambda|} \,.
\end{equation}
In the specific case of a sinusoidal force field $F_{\sin}(x,f) = \sin (2\pi x) + f$,
one has $F_{\mm}=f-1$ and $F_0''=4\pi^2$, so that $\bar\phi(f)=\pi\sqrt{f-1}$, leading for $\delta\lambda\to 0^{+}$ to
\begin{eqnarray}
\label{eq:phi:scaling:sinus2}
\phi\big(\lambda_{\cc}^{-}(f)-\delta\lambda,f\big) &\approx& 2\pi \sqrt{f-1}\, \frac{\delta \lambda}{|\!\ln \delta \lambda|} \,, \\
\label{eq:barv:scaling:sinus2}
{\bar v}\big(\lambda_{\cc}^{-}(f)-\delta\lambda,f\big) &\approx& \frac{2\pi \sqrt{f-1}}{|\!\ln \delta \lambda|}\,.
\end{eqnarray}
We thus recover a logarithmic depinning behavior (or DPT) very similar to that of the case $0\leq f<f_{\cc}$, but with different prefactors.
Note that the range of force $f>f_\cc$ was also studied in Ref.~\cite{proesmans_large-deviation_2019} but the critical behavior of the phase transition was not determined.

\subsection{Transition as a function of $\lambda$ at the critical force $f=f_{\cc}$}
\label{sec:trans:feqfc}

We now investigate the criticality of the SCGF for $f=f_{\cc}$. As a first indication, one sees that the leading term found for the SCGF $\phi$ and for the generalized average current ${\bar v}$ vanish when $f \to f_{\cc}$, see Eqs.~(\ref{eq:phi:scaling:sinus}) and~(\ref{eq:barv:scaling:sinus}) for $f<f_{\cc}$
and Eqs.~(\ref{eq:phi:scaling:sinus2}) and~(\ref{eq:barv:scaling:sinus2}) for $f>f_{\cc}$. This suggests that a different scaling behavior may take place at the critical force $f_{\cc}$.
We thus consider the case $f=f_{\cc}$ and $\lambda=\lambda_{\cc}^{-}(f_{\cc})-\delta\lambda=-\delta\lambda$ with $\delta\lambda>0$ (we recall that $\lambda_{\cc}^{-}(f_{\cc})=0$). Thus, according to Eq.~(\ref{eq:solve:phi:v2}), we obtain
\begin{equation} \label{eq:solve:phi:fc1}
2\phi \int_0^1  \frac{\dd x\,}{\sqrt{2\phi+(F_0(x)+f_{\cc})^2} + |F_0(x)+f_{\cc}|} = \delta \lambda
\end{equation}
where we have used $\mathcal{V}_{\max}(f_{\cc})=0$. 
For $f=f_{\cc}$, the force field $F(x,f_{\cc})$ now has a single zero-crossing point at $x=x_{\rm m}$, which is also the minimum of $F_0(x)$. We can then follow the same initial steps as in Sec.~\ref{sec:trans:fleqfc}. 
Expanding $F_0(x)$ around $x=x_{\rm m}$, we get to leading order
$F_0(x)+f_{\cc} = \frac{F_0''}{2}(x-x_{\rm m})^2$, with the shorthand notation
$F_0''=F_0''(x_{\rm m})$ (note that $F_0''>0$).
To determine $\phi=\phi(-\delta\lambda,f_{\cc})$, we thus have to solve the equation
\begin{equation} \label{eq:solve:phi:fc2}
4\phi \int_0^1  \frac{\dd x\,}{\sqrt{8\phi+F_0''^{\,2\,}(x-x_{\rm m})^4} + F_0''^{\,}(x-x_{\rm m})^2} \approx \delta \lambda \,.
\end{equation}
Performing the change of variable $x=x_{\rm m}+(\phi^{1/4}/\sqrt{F_0''})y$ and taking the limit $\phi \to 0$ in the bounds of the integral, we eventually get
\begin{equation}
 \frac{8C}{\sqrt{F_0''}} \, \phi^{3/4} \approx \delta \lambda \,,
\qquad \quad \text{with} \quad C = \int_0^{\infty} \frac{\dd y}{\sqrt{y^4+8}+y^2} 
=  \frac{\Gamma \big(\frac{1}{4}\big)^2}{6\times 2^{3/4} \sqrt{\pi}}\,.
\end{equation}
It follows that
\begin{equation}
\phi \approx \frac{F_0''^{\,2/3}}{(8C)^{4/3}}\, \delta \lambda^{4/3}\,,
\qquad \quad \text{and} \quad
{\bar v} \approx \frac{4}{3} \frac{F_0''^{\,2/3}}{(8C)^{4/3}} \, \delta \lambda^{1/3}\,.
\end{equation}
In the specific case of a sinusoidal force field $F_{\sin}(x,f)=\sin{2\pi x}+f$, both the SCGF and the average generalized velocity take the form:
\begin{equation}
\phi \approx \left(\frac{\pi}{4C}\right)^{4/3}  \delta \lambda^{4/3}\,,
\qquad \quad \text{and} \quad
{\bar v} \approx \frac{4}{3} \left(\frac{\pi}{4C}\right)^{4/3} \delta \lambda^{1/3}\,.
\end{equation}
We thus get a non-standard exponent $\frac{1}{3}$ for the depinning transition as a function of $\lambda$ right at the critical force $f_{\cc}$.

\subsection{Depinning transition as a function of the force $f$ for fixed $\lambda$}

We have analyzed above the DPT in the $(\lambda,f)$-plane, close to the critical line $\lambda_{\cc}^{-}(f)$, by varying the parameter $\lambda$ at a fixed driving force $f$. This was motivated in part by technical considerations, because the asymptotic evaluation of $\phi$ is more easily carried out when varying $\lambda$ at fixed $f$. In addition, evaluating the generalized average current ${\bar v}$ requires to differentiate $\phi$ with respect to $\lambda$ at fixed $f$, and thus to know explicitly the $\lambda$-dependence of the SCGF $\phi$.

While looking at the transition as a function of $\lambda$ is standard in the context of DPT, it may be more natural to consider the transition as a function of the physical drive $f$ (at fixed $\lambda$) when thinking in terms of a depinning transition.
We thus discuss in this section how the above results can be recast in terms of a depinning transition as a function of the drive $f$. Yet, before coming to this point, we start by showing how the standard depinning transition
can be reformulated in the present framework.

\subsubsection{Standard depinning transition around $f=f_{\cc}$ for $\lambda =0$.} \label{sec:std:depinning}

We first show how the standard depinning transition, corresponding to the unbiased dynamics $\lambda =0$, can be obtained from the analysis of Eq.~(\ref{eq:solve:phi:v1}) in the vicinity of the critical force $f_{\cc}$.
Although we are interested in the behavior at $\lambda =0$, the current is obtained by evaluating $\partial_{\lambda}\phi$, hence we need to determine $\phi(\lambda,f)$ in the vicinity of $\lambda=0$, and not only $\phi(0,f)$. We thus consider an arbitrarily small value of $\lambda$.
Using Eq.~(\ref{eq:SCGF}) under the above assumptions, the equation determining $\phi=\phi(\lambda,f)$ is given by
\begin{equation} \label{eq:phi:depin}
\int_0^1 \dd x \, \sqrt{2\phi+F(x,f)^2} = f - \lambda\,.
\end{equation}
For $\lambda=0$, one has $\phi(0,f)=0$ for all $f>f_{\cc}$, as can be seen from Eq.~(\ref{eq:phi:depin}) since $\int_0^1 \dd x \, \sqrt{F(x,f)^2} = f$ if $F(x,f)\ge 0$ for all $x$ (which is true when $f\ge f_{\cc}$).
Subtracting the latter equation to Eq.~(\ref{eq:phi:depin}) and using again the relation $\sqrt{a}-\sqrt{b} = (a-b)/(\sqrt{a}+\sqrt{b})$ we get 
\begin{equation} \label{eq:solve:phi:v21}
2\phi \int_0^1  \frac{\dd x\,}{\sqrt{2\phi+(F_0(x)+f_{\cc}+\deltaf)^2} + |F_0(x)+f_{\cc}+\deltaf|} = -\lambda\,,
\end{equation}
where we have focused on the case $f=f_\cc+\deltaf$. Note that Eq.~(\ref{eq:solve:phi:v21}) differs from Eq.~(\ref{eq:solve:phi:v2}), because we now vary $f$ at fixed $\lambda$ instead of varying $\lambda$ at fixed $f$.
Setting $\phi=0$ in the denominator of the integrand in Eq.~(\ref{eq:solve:phi:v21}) does not induce any divergence of the integral, so we can safely make this replacement to determine the leading order behavior of $\phi$ in terms of $\lambda$ and $\deltaf$.
We then get
\begin{equation} \label{eq:solve:phi:v22}
\phi \int_0^1  \frac{\dd x\,}{F_0(x)+f_{\cc}+\deltaf} = -\lambda.
\end{equation}
Note that we have dropped the absolute value because the integrand is positive.
For small $\deltaf$, the last integral is dominated by the vicinity of $x=x_{\rm m}$ where $F_0(x)$ is minimum, because $F_0(x_{\rm m})+f_{\cc}=0$.
Expanding the integrand to leading order, one finds
\begin{equation} \label{eq:solve:phi:v23}
\phi \int_0^1  \frac{\dd x\,}{\frac{1}{2} \, F_0''\,(x-x_{\rm m})^2+\deltaf} \approx -\lambda\,,
\end{equation}
with $F_0''=F_0''(x_{\rm m})$.
The integral in Eq.~(\ref{eq:solve:phi:v23}) diverges when $\deltaf \to 0$, and its asymptotic behavior can be extracted using the change of variable
$x=x_{\rm m}+\sqrt{2\deltaf/F_0''}\, y$, leading to
\begin{equation}
\sqrt{\frac{2}{F_0''}}\, \frac{\phi}{\sqrt{\deltaf}} \int_{-\infty}^{\infty} \frac{\dd y}{y^2+1} \approx -\lambda\,.
\end{equation}
We thus obtain the following asymptotic behavior of the SCGF $\phi$ and of the generalized current $\bar{v}$ for $\lambda \to 0$ and $\deltaf \to 0$ (in this order),
\begin{equation}
\phi \approx  -\pi \sqrt{\frac{F_0''}{2}} \lambda\, \sqrt{\deltaf}
\end{equation}
and, using Eq.~(\ref{eq:def:barv:phi:deriv}),
\begin{equation}
\bar{v} \approx \pi \sqrt{\frac{F_0''}{2}} \, \sqrt{\deltaf} \,.
\end{equation}
So we recover with this method the standard scaling exponent $\frac{1}{2}$ of the 1D depinning transition, usually obtained from the study of the deterministic dynamics close to the saddle node~\cite{stratonovich_oscillator_1965,strogatz_nonlinear_2001,fisher_collective_1998,brazovskii_pinning_2004,tsobgni_nyawo_large_2016}.

\subsubsection{Crossovers between scaling behaviors around $(\lambda=0,f=f_{\cc})$.}

We have seen above that different scaling behaviors appear in the vicinity of the critical point $(\lambda=0,f=f_{\cc})$. Up to now, we have varied the parameters along specific lines in the $(\lambda,f)$-plane. It is of interest to look at the crossovers between the different types of scaling in the full $(\lambda,f)$-plane.

A simple and heuristic way to study these crossovers is to compare the orders of magnitude of the different scalings obtained.
For instance, for $f=f_{\cc}+\deltaf>f_{\cc}$ and $\lambda = -\delta \lambda<0$, one can compare (omitting prefactors) the behavior $(\delta\lambda)^{1/3}$ obtained for ${\bar v}$ at $f=f_{\cc}$ to the term $(\deltaf)^{1/2}$ obtained at $\lambda=0$.
Balancing the two terms yields a crossover line $\delta \lambda \sim (\deltaf)^{3/2}$, or equivalently $\deltaf \sim (\delta \lambda)^{2/3}$.
One thus expects that for $\deltaf \ll (\delta \lambda)^{2/3}$ (with both $\deltaf$ and $\delta\lambda$ small), $\bar{v} \sim (\delta \lambda)^{1/3}$, while in the opposite regime $\deltaf \gg (\delta \lambda)^{2/3}$, $\bar{v} \sim (\deltaf)^{1/2}$.

A similar reasoning can be performed for $f=f_{\cc}-\deltaf<f_{\cc}$ and $\lambda = \lambda_{\cc}^{-}(f)-\delta \lambda$ ($\delta \lambda>0$). For small enough  $\delta \lambda$, the results of Sec.~\ref{sec:trans:fleqfc} apply, and $\bar{v} \sim (\deltaf)^{1/2} |\ln \delta \lambda|^{-1}$, while for $\deltaf =0$, $\bar{v} \sim (\delta\lambda)^{1/3}$. Equating the two scalings, one finds for the crossover line the scaling behavior
\begin{equation}
\deltaf \sim (\delta \lambda)^{2/3} (\ln \delta \lambda)^2.
\end{equation}
Hence for $\deltaf \ll (\delta \lambda)^{2/3} (\ln \delta \lambda)^2$, one expects $\bar{v} \sim (\delta \lambda)^{1/3}$, while in the opposite case $\bar{v} \sim (\deltaf)^{1/2} |\ln \delta \lambda|$.

\subsubsection{Depinning transition as a function of the force $f$ for arbitrary $\lambda$.}

We have recovered above in Sec.~\ref{sec:std:depinning} that for $\lambda=0$, the depinning transition as a function of the force $f$ is characterized by the standard exponent $\frac{1}{2}$, in the sense that $\bar{v} \sim (f-f_{\cc})^{1/2}$.
For $\lambda \ne 0$, we now try to recast the results obtained above on the depinning transition as a function of $\lambda$ in terms of the transition as a function of $\deltaf$. This can be done as follows.
We know from Secs.~\ref{sec:trans:fleqfc} and \ref{sec:trans:fgeqfc} that $\bar{v}(\lambda_{\cc}^{-}(f)-\delta\lambda,f) \approx \tilde{\phi}(f)\, |\!\ln\delta\lambda|^{-1}$ for a small $\delta\lambda>0$.

Now considering instead a variation of $f$, one can evaluate
$\bar{v}(\lambda_{\cc}^{-}(f),f+\deltaf)$ as
\begin{equation}
\bar{v}(\lambda_{\cc}^{-}(f),f+\deltaf) = \bar{v}(\lambda_{\cc}^{-}(f+\deltaf)-\delta\lambda,f+\deltaf)
\end{equation}
provided that
\begin{equation}
\delta\lambda = \frac{\dd\lambda_{\cc}^{-}}{\dd f} \,\deltaf\,.
\end{equation}
As long as $\dd\lambda_{\cc}^{-}/\,\dd f \ne 0$, which is true for all $f \ne f_{\cc}$ ($f\ge 0$), we thus have that
\begin{equation}
\bar{v}(\lambda_{\cc}^{-}(f),f+\deltaf) \approx \frac{\tilde{\phi}(f+\deltaf)}{|\!\ln\delta\lambda|} \approx \frac{\tilde{\phi}(f)}{|\!\ln\deltaf|} \,,
\end{equation}
neglecting the factor $\dd\lambda_{\cc}^{-}/\,\dd f$ in the logarithm for $\deltaf \to 0$.


\section{Finite-temperature effects on the dynamical phase transition}

\subsection{Description of the thermal rounding}
\label{sec:thermal-rounding}

The results we have obtained so far are describing the leading order in the small-noise asymptotics $\epsilon\to 0$.
In particular, the SCGF $\varphi_\epsilon(\lambda,f)$ behaves as $\frac 1\epsilon \phi(\lambda,f)$, with $\phi(\lambda,f)\neq 0$ in the non-zero current regime ($\lambda>\lambda_\cc^+(f)$ or $\lambda<\lambda_\cc^-(f)$),
while the zero-current regime $\lambda_\cc^-(f)\leq\lambda\leq\lambda_\cc^+(f)$ is characterized by $\phi(\lambda,f)=0$.
This implies that the large deviation function $\Pi_\epsilon(a,f)$ behaves as $\frac 1\epsilon \pi(a,f)$ as $\epsilon\to 0$, with $\pi(a,f)$ presenting a cusp at the generalized average current $\bar v$.
This singular behavior means that even the expected Gaussian fluctuations around $\bar v$ are not described by the dominant order in $\epsilon$ we have computed.

To understand how a small but finite $\epsilon$ can round or amend the observed singularities, one thus needs to determine how $\varphi_\epsilon(\lambda,f)$ scales with $\epsilon$ in the zero-current regime beyond the minimum order $\frac 1\epsilon \phi(\lambda,f)=0$. 
In this case we will assume $g(x)=1$, which implies that the generalized current $\bar v$ is the standard particle current or average velocity (see Eq. (\ref{eq:gen-velocity}) and below).
A natural approach consists then in considering
the biased Fokker--Planck operator
\begin{equation}\label{eq:fp-operator}
\mathbb{W}_\lambda\cdot=
-\partial_x\big((F(x)-\lambda)\cdot\big)
+\frac{\lambda}{\epsilon}\left(\frac{\lambda}{2}-F(x)\right)
+\frac{1}{2}\epsilon\partial_x^2\cdot
\ ,
\end{equation}
whose largest eigenvalue is $\varphi_\epsilon(\lambda,f)$.
The first two terms are responsible for its dominant behavior $\frac 1 \epsilon \phi(\lambda,f)$, while the last term, which accounts for diffusion, describes corrections around it (as generically in the WKB or Freidlin--Wentzell--Graham approaches~\cite{graham_weak-noise_1985,freidlin_random_2012}; see~\cite{tsobgni_nyawo_large_2016,tizon-escamilla_effective_2019} for the problem at hand).
A customary route to follow consists in implementing a perturbation expansion, considering $\frac{1}{2}\epsilon\partial_x^2\cdot$ as a perturbation term in~(\ref{eq:fp-operator}).
The eigenvalue $\varphi_\epsilon(\lambda,f)$ is associated with left and right eigenvectors of $\mathbb W_\lambda$, that we denote $L(x)$ and $R(x)$ and expand as
\begin{align}
L(x)&=\exp{\left[-\frac{U_L(x)}{\epsilon}+U_{L0}(x)+\epsilon U_{L1}(x)+\mathcal{O}(\epsilon^2)\right]}\,,
\\
R(x)&=\exp{\left[-\frac{U_R(x)}{\epsilon}+U_{R0}(x)+\epsilon U_{R1}(x)+\mathcal{O}(\epsilon^2)\right]}\,.
\end{align}
In turn, this corresponds to an expansion of the SCGF as
\begin{equation}
\varphi_\epsilon(\lambda,f)=\frac{\phi(\lambda,f)}{\epsilon}+\phi_0(\lambda,f)+\epsilon\phi_1(\lambda,f)+\mathcal{O}(\epsilon^2)\,.
\label{eq:expvarphiepsilon}
\end{equation}

However, in the zero-current regime, such a perturbative approach fails: it would yield $\varphi_\epsilon(\lambda,f)=0$ at all orders in powers of $\epsilon$, while in fact the SCGF behaves as
\begin{equation}
  \label{eq:defPhi}
  \varphi_\epsilon(\lambda,f) \stackrel[\epsilon\to 0]{}{\asymp} \ee^{-\frac 1\epsilon \tilde \Phi(\lambda,f)}
  \qquad
  \textnormal{for}
  \quad
  \lambda_\cc^-(f)<\lambda<\lambda_\cc^+(f) 
  \,,
\end{equation}
as we will show. This behavior is non-analytic as a function of $\epsilon$ and cannot be described by an expansion of the form~(\ref{eq:expvarphiepsilon}).
Physically, it corresponds to the fact that, in the zero-current regime, the dynamics is governed by metastable states: the time scale defining the current is coming from an escape mechanism whose rate takes an Arrhenius form with an exponential behavior leading to Eq.~(\ref{eq:defPhi}). 
Nyawo and Touchette in Ref.~\cite{tsobgni_nyawo_large_2016} 
justified such behavior of the SCGF (within the Freidlin--Wentzell--Graham approach~\cite{graham_weak-noise_1985,freidlin_random_2012}), and studied it numerically.
The Arrhenius function $\tilde \Phi(\lambda,f)$ was determined analytically by Proesmans and Derrida in~\cite{proesmans_large-deviation_2019} in the special situation where the dynamics presents no metastable state, using a WKB approach and asymptotic matching.
We focus in this paper on the more complex case where the dynamics presents a metastable state, which renders the analysis more intricate while also inducing a different behavior of the Arrhenius function $\tilde\Phi(\lambda,f)$.

There are many ways to derive an Arrhenius scaling in general: in a path-integral approach, one would have to enumerate and sum over the infinite number of possible trajectories going from a metastable state to another (see for instance~§2.2 of Chap.~7 in~\cite{coleman1988aspects}).
In our case, a path-integral formulation is available~\cite{tizon-escamilla_effective_2019} but the instantons are not obvious to determine.
From an operator viewpoint, if the dynamics is reversible (i.e., if the operator~(\ref{eq:fp-operator}) can be made Hermitian), one finds a Boltzmann-like ground state but this is not the case here.
If the operator~(\ref{eq:fp-operator}) was probability-preserving, we could find its ground state, even in the absence of reversibility (see for instance~\cite{risken_fokker-planck_1996,le_doussal_creep_1995}) but when $\lambda\neq 0$ the operator does not preserve probability.
We thus have to resort to a different approach, that we now describe, and that also allows us to determine some aspects of the behavior of the SCGF that go beyond the exponential behavior of Eq.~(\ref{eq:defPhi}).

\subsection{Strategy to obtain the Arrhenius scaling in the zero-current phase}
\label{sec:strat-obta-arrh}

We will base our analysis on the observation that the biased evolution operator can be made probability-preserving after an appropriate transformation consisting in a shift and a change of basis~\cite{miller_convexity_1961,simon_construction_2009,popkov_asep_2010,jack_large_2010}.
Defining a diagonal operator whose elements are the components of left eigenvector of~(\ref{eq:fp-operator}), one performs a similarity transformation that describes
an `effective dynamics' (or `auxiliary dynamics') which is {asymptotically equivalent} at large times to the biased dynamics and to the conditioned dynamics~\cite{jack_large_2010,chetrite_nonequilibrium_2013,chetrite_nonequilibrium_2015} after proper normalization.
As is well-known~\cite{tsobgni_nyawo_large_2016,tizon-escamilla_effective_2019} and recalled in more details below, the effective dynamics takes place in a tilted potential $U^\eff_\lambda$ decomposed as
\begin{equation}
  \label{eq:defUeffVeff}
  U^\eff_\lambda (x) = V_\lambda(x) - \fla\, x
\end{equation}
where $V_\lambda$ is periodic [i.e., $V_\lambda(x)=V_\lambda(x+1)$] and $\fla$ is a uniform force.
The determination of $U^\eff_\lambda$ is not an easy task, and cannot be done at all orders in $\epsilon$ in general
(see Ref.~\cite{tsobgni_nyawo_large_2016} for a perturbative study using a Fourier--Bloch decomposition).

 The derivation of the required order is  detailed in Sec.~\ref{sec:left-right-eigenv-domin}.
If $U^\eff_\lambda (x)$ is known, then, although the dynamics is non-reversible (if $\fla\neq 0$), the steady state can be written explicitly~\cite{risken_fokker-planck_1996}. 
Recalling that we focus on the case $g(x)=1$, it is then known~\cite{le_doussal_creep_1995,scheidl_mobility_1995} that for the Langevin dynamics of a particle in an arbitrary tilted potential
$U^\eff_\lambda (x) = V_\lambda(x) - \fla\, x$ (with $V_\lambda(x)$ a periodic potential), the average velocity $\langle\dot x\rangle_\eff$ is obtained from
\begin{align}
  \label{eq:oneoverV}
  \frac{1}{\langle\dot x\rangle_\eff}
&  =
  \frac 2 \epsilon
  \int_0^\infty \dd z \:
  \ee^{-\frac{2}{\epsilon} \fla z} \,G(z)
\qquad\text{with}
\\
  \label{eq:oneoverVg}
  G(z)
& =
  \int_0^1 \dd x\: \ee^{\frac 2 \epsilon\big[V_\lambda(x+z)-V_\lambda(x)\big]}
\,.
\end{align}
Such expression is valid when $\fla > 0$ (and can be adapted to the case $\fla < 0$). We refer to Appendix~\ref{sec:deriv-diff-coeff} for a self-contained derivation and for other expressions valid for any $\fla$.
Using the fact that $\partial_\lambda \varphi_\epsilon(\lambda,f) = - \frac 1\epsilon \langle\dot x\rangle_\eff$ together with the saddle-point asymptotics
\begin{align}
  G(z)
& 
  \stackrel[\epsilon\to 0]{}\asymp
  \ee^{\frac 2 \epsilon\big[V_\lambda(X(z)+z)-V_\lambda(X(z))\big]}
  \qquad\text{with}
\\
X(z)
&
  \stackrel[\phantom{\epsilon\to 0}]{}=
  \stackrel[0\leq x\leq 1]{}{\operatorname{argmax}} \big[V_\lambda(x+z)-V_\lambda(x)\big]
\ ,
\label{eq:defXz}
\end{align}
one finds that the SCGF takes the following exponential form:
\begin{align}
  \label{def:Phinottilde}
  \partial_\lambda \varphi_\epsilon(\lambda,f)
&
  \stackrel[\epsilon\to 0]{}\asymp
  \ee^{- \frac 1\epsilon \Phi(\lambda,f)}
\\
  \quad\;\,
  \Phi(\lambda,f)
&
  \stackrel[\phantom{\epsilon\to 0}]{}=
  - 2
   \min_{z\geq 0}
   \Big[
     -V_\lambda(X(z)+z) + V_\lambda(X(z))
     + \fla\, z
   \Big]
\ .
\label{eq:exprPhilambda}
\end{align}
Note that the two optimization principles~(\ref{eq:defXz}) and~(\ref{eq:exprPhilambda}) are not done on the same intervals.
In these expressions, the asymptotic equivalents denoted by $\stackrel[\epsilon\to 0]{}\asymp$  are logarithmic: for instance Eq.~\eqref{def:Phinottilde} means that
 $\log |\partial_\lambda \varphi_\epsilon(\lambda,f)|\approx- \frac 1\epsilon \Phi(\lambda,f) $
as $\epsilon\to 0$.
Going beyond (i.e.~obtaining the prefactor of the exponential) would require to integrate the fluctuations around the saddle-point, which is not immediate, since the function $V_\lambda$ present singularities.
It would also require to determine the finite-$\epsilon$ corrections to the potential $V_\lambda$.
As announced in Eq.~(\ref{eq:defPhi}),
the SCGF also presents an Arrhenius scaling of the form $\ee^{- \frac 1\epsilon \tilde \Phi(\lambda,f)}$,
but, as we discuss later in Sec.~\ref{sec:interpr-form-rate}, the two Arrhenius functions $\Phi$ and $\tilde \Phi$ are not the same.
This is due to a constant contribution in the SCGF $\varphi_\epsilon(\lambda,f)$, that vanishes when differentiating w.r.t.~$\lambda$.
We will first determine $\Phi(\lambda,f)$ from the optimization principles~(\ref{eq:defXz}) and~(\ref{eq:exprPhilambda}), and then infer the value of $\tilde\Phi(\lambda,f)$.

We are thus able to determine the SCGF $\varphi_\epsilon(\lambda,f)$ if we know the effective tilted potential $U^\eff_\lambda (x) = V_\lambda(x) - \fla\, x$.
We thus need to evaluate the effective tilted potential $U^\eff_\lambda (x)$, which requires the determination of the left eigenvector $L(x)$ of the operator $\WW _\lambda$ defined in Eq.~(\ref{eq:fp-operator}).
In the following, we first recall in Sec.~\ref{sec:relat-betw-effect} the relation between $U^\eff_\lambda$ and the left eigenvector $L(x)$.
We then compute in Secs.~\ref{sec:left-right-eigenv-domin}
the left eigenvector $L(x)$ at minimal order in $\epsilon$. We present in Sec.~\ref{sec:determ-therm-effects} the calculation of the Arrhenius function $\Phi(\lambda,f)$ from~\eqref{eq:exprPhilambda}, leading to the determination of the SCGF $\varphi_\epsilon(\lambda,f)$ in the scaling form of Eq.~(\ref{eq:defPhi}), before commenting further on its physical interpretation in Sec.~\ref{sec:interpr-form-rate}.
We finally discuss in Sec.~\ref{sec:disc-lambd-0} the non-trivial interchange of the limits $\lambda\to 0$ and $\epsilon\to 0$ in the equilibrium case $f=0$.

\smallskip
Note that, as shown in Appendix~\ref{sec:GC-sym}, the SCGF verifies a Gallavotti--Cohen-type symmetry~\cite{gallavotti_dynamical_1995-1,gallavotti_dynamical_1995,kurchan_fluctuation_1998,lebowitz_gallavotticohen-type_1999} of the form  $  \varphi_\epsilon(\lambda,f) =   \varphi_\epsilon(2f-\lambda,f) $.
This means that the SCGF is symmetric around the point $\lambda=f$; we thus restrict our study to the case $\lambda < f$ (where the average current is positive) without loss of generality.

\subsection{Relation between the effective dynamics and the dominant eigenvectors}
\label{sec:relat-betw-effect}

We recall here the  generic result providing the relation between the potential $U^\eff_\lambda$ of the effective dynamics and the dominant eigenvectors of the tilted operator.
One defines $\langle L|$ as the left eigenvector of $\WW _\lambda$ corresponding to the maximal eigenvalue $\varphi_\epsilon(\lambda,f)$.
As in Refs.~\cite{simon_construction_2009,popkov_asep_2010,jack_large_2010}, one constructs a diagonal operator $\hat L$ whose elements are the components of $\langle L|$. One directly checks that
\begin{equation}
  \label{eq:defWstar}
  \langle -| \WW _\lambda^\eff = 0
\quad
\text{with}
\quad
  \WW _\lambda^\eff = \hat L \WW _\lambda \hat L^{-1} - \varphi_\epsilon(\lambda,f) \mathds{1} \,.
\end{equation}
This means that $\WW _\lambda^\eff$ is probability-preserving. As generically derived in~\cite{jack_large_2010,chetrite_nonequilibrium_2013,chetrite_nonequilibrium_2015} and detailed in~\cite{tizon-escamilla_effective_2019} for our example at hand, if one writes  $L(x)=\ee^{-\frac 1\epsilon U_L(x)}$,
one obtains by direct computation that $\WW _\lambda^\eff$  is a Fokker--Planck evolution operator:
\begin{equation}
 \WW _\lambda^\eff \cdot
  =-\partial_x\Big[\big(F(x,f)-\lambda  g(x)-U'_L(x)\big) \cdot\Big]
  +\frac{1}{2}\epsilon\partial^2_x\cdot\,\,.
\label{eq:Wefflambda-final}
\end{equation}
It is of course  probability-preserving and corresponds to the evolution of a particle subjected to a force $F_\lambda^\eff(x)=F(x,f)-\lambda  g(x)-{U}'_L(x)$.
The function ${U}_L(x)$ is periodic (as discussed in~\cite{tizon-escamilla_effective_2019}) and depends on $\epsilon$ in general (although one expects that it behaves as $\epsilon^0$ at dominant order for $\epsilon\to 0$). 
In this framework, we can thus write the effective force $F_\lambda^\eff(x)$ as deriving from an effective tilted potential $U_\lambda^\eff(x)$
\begin{align}
  \label{eq:FeffUeff}
  F_\lambda^\eff(x) &= -\partial_x U_\lambda^\eff(x)
\\
  U_\lambda^\eff(x) & = V_\lambda(x) - \fla\, x  
  \quad \text{with} \quad
  V_\lambda(x) = V(x)+U_L(x)
  \quad \text{and} \quad
  \fla= f-\lambda
\,.
\label{eq:decompfinalUeff}
\end{align}

\subsection{Left eigenvector at dominant order in $\epsilon\to 0$}
\label{sec:left-right-eigenv-domin}

To compute the required order in $\epsilon$ of the effective periodic potential $V_\lambda$ entering the expression~(\ref{eq:exprPhilambda}) of $\Phi(\lambda,f)$,
we see from Eqs.~(\ref{eq:defXz})-(\ref{eq:exprPhilambda}) that only the leading order, in $\epsilon^0$, of  $V_\lambda$ 
matters for the optimization principles in~(\ref{eq:defXz}) and~(\ref{eq:exprPhilambda}).
Consequently, from~\eqref{eq:decompfinalUeff}, we only have to determine the dominant behavior of $U_L(x)$.
We derive here the leading-order behavior of the left eigenvector.
For completeness, we also provide in Appendix~\ref{app:left-eigen} the corresponding expression of the right eigenvector, although it will not be used in what follows.
 
The left eigenvalue equation associated with the Fokker--Planck operator~(\ref{eq:fp-operator}) reads:
\begin{equation}\label{eq:left-eigen-eq}
\frac{1}{2}\epsilon L''(x)+(F(x,f)-\lambda)L'(x)+\frac{\lambda}{\epsilon}\left(\frac{\lambda}{2}-F(x,f)\right)L(x)=\varphi_\epsilon L(x)\,.
\end{equation}
As we already mentioned, we focus here on the zero-current phase $\lambda_\cc^-(f)\leq\lambda\leq\lambda_\cc^+(f)$, where $\phi(\lambda,f)=0$, i.e., the contribution to order $\epsilon^{-1}$ to the SCGF $\varphi_\epsilon$ vanishes.
It would be natural to assume the left eigenvector $L(x)$ to take for $\epsilon \to 0$ the asymptotic form $L(x) \propto \ee^{-U_L(x)/\epsilon}$
where $U_L(x)$ is sufficiently regular, i.e., has a continuous derivative.
However, the left eigenspace associated with the eigenvalue $\varphi_\epsilon$ may be of dimension larger than one, and we thus rather write the left eigenvector $L(x)$ as a linear superposition of different solutions of the form $\ee^{-U_L(x)/\epsilon}$.
An important point is that only the linear combination needs to fulfill the spatial periodicity condition, whereas the individual solutions do not need to satisfy this constraint.

Introducing the Ansatz $L(x) = \ee^{-U_L(x)/\epsilon}$ in Eq.~(\ref{eq:left-eigen-eq}), we find at leading order $\epsilon^{-1}$ the equation
\begin{equation}
\frac{1}{2}\left(U_L'(x)\right)^2 -(F(x,f)-\lambda) U_L'(x)+\frac{\lambda^2}{2}-\lambda F(x,f)=0\,,
\label{eq:quad:ULprime}
\end{equation}
leading to the solution
\begin{equation}
U_L'(x)=F(x,f)-\lambda-\sigma_i(x)\, |F(x,f)|\,,
\end{equation}
where $\sigma_i(x)=\pm 1$ is a sign that may a priori depend on $x$.
Yet, assuming continuity of the function $U_L'(x)$ imposes that the function $\sigma_i(x)$ can only change sign at points $x$ such that $F(x,f)=0$.
Eq.~(\ref{eq:quad:ULprime}) thus exhibits \emph{four} continuous solution functions $U_{L,i}'(x)$ (an unexpected result given that Eq.~(\ref{eq:quad:ULprime}) is a second-order ordinary equation $U_L'$), given by
\begin{eqnarray}
 U_{L,1}'(x)=F(x,f)-\lambda-|F(x,f)|\\
 U_{L,2}'(x)=F(x,f)-\lambda+|F(x,f)|\\
 U_{L,3}'(x)=-\lambda\\
 U_{L,4}'(x)=2F(x,f)-\lambda\,.
\end{eqnarray}
\begin{figure}[t]%
\begin{center}
  \includegraphics[height=0.2475\columnwidth]{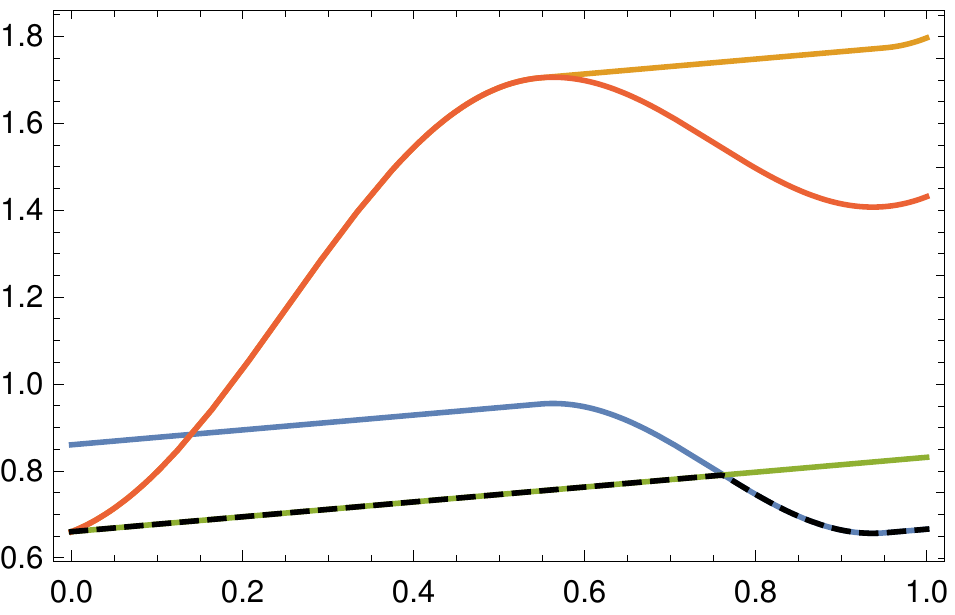}
\quad
  \includegraphics[height=0.25\columnwidth]{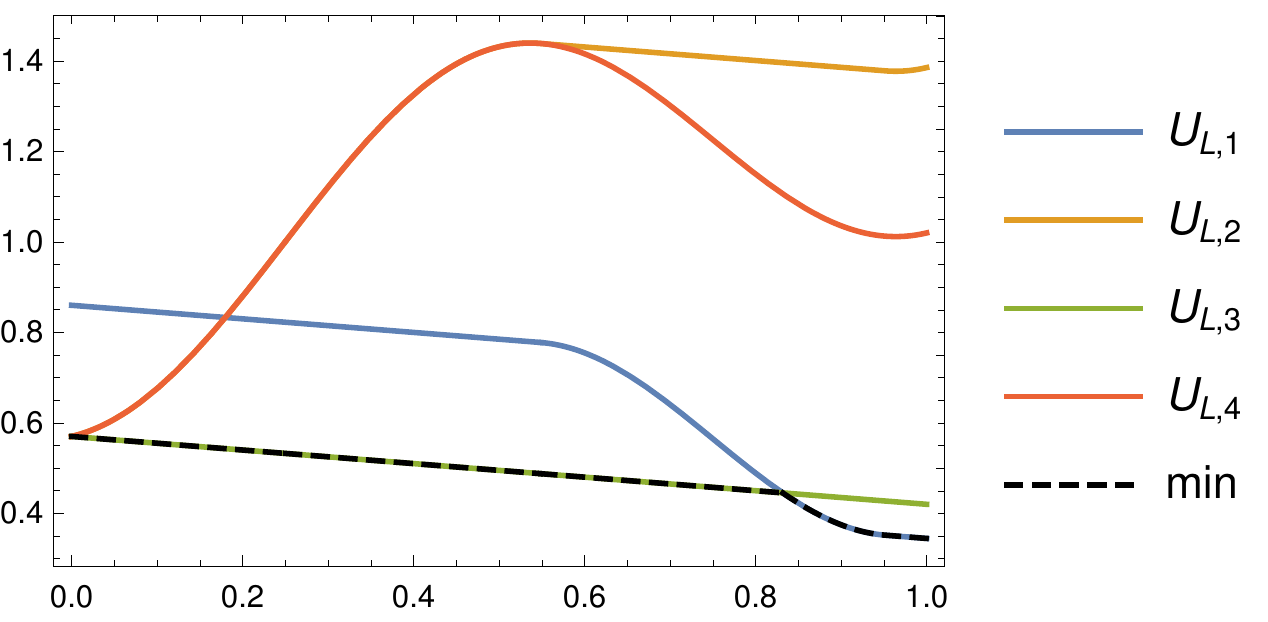}
\end{center}
\caption{
\textbf{(Left)}
The four functions $U_{L,1}$, $\ldots$, $U_{L,4}$ defined in Eqs.~(\ref{eq:eigen-1})-(\ref{eq:eigen-4}) for the optimal values of $k_2$, $k_3$ and $k_4$, in the case $\lambda_\cc^-(f)<\lambda<0$.
The final result $U_L(x)$ of the optimization principle~(\ref{eq:minU1U4}) is plotted in dashed line.
The force is $F_{\sin}(x,f) = \sin (2\pi x) + f$ with $f=0.3$ and $\lambda=-0.17$.
\textbf{(Right)} Same, for the case $0<\lambda\leq f$. The force is the same but now $\lambda=0.15$.
\label{fig:UL1234_min}
}
\end{figure}%
The presence of four solutions instead of two comes from the fact that independent pairs of solutions of Eq.~(\ref{eq:quad:ULprime}) can be connected with sufficient regularity.
By integration we obtain, recalling that $F(x)=-V'(x)+f$,
\begin{eqnarray} \label{eq:eigen-1}
U_{L,1}(x)=-V(x)+(f-\lambda) x-\int_0^x \dd y \, |F(y,f)|+k_1\\ \label{eq:eigen-2}
U_{L,2}(x)=-V(x)+(f-\lambda) x+\int_0^x \dd y \, |F(y,f)|+k_1+k_2\\ \label{eq:eigen-3}
U_{L,3}(x)=-\lambda x+k_1+k_3\\ \label{eq:eigen-4}
U_{L,4}(x)=-2V(x)+ (2f-\lambda) x+k_1+k_4\,,
\end{eqnarray}
where the $k_i$'s are integration constants, and where the potential $V(x)$ satisfies $V(0)=V(1)=0$.
Note that for convenience, we have split the integration constants in the functions $U_{L,2}(x)$, $U_{L,3}(x)$ and $U_{L,4}(x)$ into two separate constants, to set apart the overall constant $k_1$ that is eventually determined by normalization, and should be held fixed but arbitrary when determining the functional form of the left eigenvector $L(x)$.
Given these multiple solutions for $U_L(x)$, the left eigenvector $L(x)$ should in principle be written as a linear combination of all possible solutions:
\begin{equation} \label{eq:linear:combin:L}
L(x)=  \ee^{-U_{L,1}(x)/\epsilon} + \int \dd k_2\, \ee^{-U_{L,2}(x)/\epsilon}
+ \int \dd k_3\, \ee^{-U_{L,3}(x)/\epsilon} + \int \dd k_4 \,\ee^{-U_{L,4}(x)/\epsilon}\,,
\end{equation}
which includes an integration over the undetermined constants $k_2$, $k_3$ and $k_4$.
However, in the limit $\epsilon \to 0$, the integrals over $k_i$ are dominated by their lower bounds $k_i^{\min}$, that remain to be specified.
Hence, $L(x)$ takes for  $\epsilon \to 0$ the form $L(x) \asymp \ee^{-U_L(x)/\epsilon}$ with $U_L(x)$ given by
\begin{equation}
U_L(x)=\min{\left\lbrace U_{L,1}(x),U_{L,2}(x),U_{L,3}(x),U_{L,4}(x)\right\rbrace}\,,
\label{eq:minU1U4}
\end{equation}
where the functions $U_{L,i}(x)$ take the forms given in Eqs.~(\ref{eq:eigen-1})--(\ref{eq:eigen-4}), with the constants $k_2$, $k_3$ and $k_4$ taking their minimal value $k_i^{\min}$ compatible with the constraint of periodicity of the function $U_L(x)$, namely $U_L(0)=U_L(1)$.
Note that we would have obtained the same result as given in Eq.~(\ref{eq:minU1U4}) if we had chosen to fix another integration constant instead of $k_1$.

The solution of the optimization problem~(\ref{eq:minU1U4}) is presented in Appendix~\ref{sec:determ-const-k_is}.
The result reads as follows.
\begin{itemize}
\item \underline{If $\lambda_{\cc}^{-}(f) < \lambda <0$ (see Fig.~\ref{fig:UL1234_min}, left)}\,:  
\begin{equation}
  \label{eq:ULfinalA}
  U_L(x) = k_1 + 
  \begin{casesl}
    \lambda_\cc^-(f)-\lambda-\lambda x & \qquad \text{if}\quad 0<x<x^*_L 
    \\
    -2V(x)+(2f-\lambda) x -\lambda_\cc^+(f) & \qquad \text{if}\quad x^*_L <x < 1\,,
  \end{casesl}
\end{equation}
where $x^*_L$ is determined by the continuity of $U_L(x)$ at $x=x^*_L$, yielding the condition
\begin{equation} \label{eq:xstar}
  V(x^*_L)=\frac{\lambda}{2}+f\,(x^*_L-1)\,.
\end{equation}

\item \underline{If $0<\lambda <f$ (see Fig.~\ref{fig:UL1234_min}, right)}\,: 
\begin{equation}
  \label{eq:app:resULothercase}
  U_L(x) =k_1+
   \begin{casesl}
   -2V(x) + (2f - \lambda) x +  \lambda_\cc^-(f) - \lambda
     & 
    \qquad \text{if} \quad 
    0\leq x \leq x^\dag_L
    \\[2mm]
     - \lambda x +  \lambda_\cc^-(f)
    & 
    \qquad \text{if} \quad 
     x^\dag_L \leq x \leq 1\,,
 \end{casesl}
\end{equation}
where $x^\dag_L$ is determined by imposing $U_L(x)$ to be continuous at $x=x^\dag_L$, i.e.
\begin{equation} \label{eq:app:xdagforULothercase}
  V(x^\dag_L)=f x^\dag_L-\frac{\lambda}{2}\,.
\end{equation}

\end{itemize}

As a consistency check, we remark that in both cases, when $\lambda\to 0$, the function $U_L(x)$ becomes constant (in Eqs.~(\ref{eq:ULfinalA}) and~(\ref{eq:app:resULothercase}), one has $x^*_L\to 1$ and $x^\dag_L\to 0$ respectively).
This result is expected since for $\lambda=0$ the dynamics described by the tilted evolution operator is probability-preserving and its left eigenvector is constant.

Finally, note that the function $U_L(x)$ as given by Eqs.~(\ref{eq:ULfinalA}) and~(\ref{eq:app:resULothercase}) also satisfies Eq.~(\ref{eq:quad:ULprime}), but does not necessarily have a continuous derivative. One may thus wonder if we should have released the constraint that the function $U_L(x)$ should have a continuous derivative when looking for solutions of Eq.~(\ref{eq:quad:ULprime}). 
However, looking from the outset for solutions $U'_L(x)$ of Eq.~(\ref{eq:quad:ULprime}) that would not be continuous yields an infinite set of solutions parametrized by the location of possible discontinuities, out of which it would be very difficult to find the relevant solution. It is thus more convenient to proceed as we have done, by restricting ourselves to functions $U_{L,i}(x)$ having continuous derivatives, and writing $L(x)$ as a linear combination of functions as done in Eq.~(\ref{eq:linear:combin:L}), which yields a well-defined procedure to determine the correct function $U_L(x)$ in the low noise limit.

The computation of the function $U_L(x)$ for the regime $f < \lambda\leq\lambda_\cc^+(f)$, as well as a particularization to the sinusoidal force $F_{\sin}(x,f)$ case can be found in Appendix~\ref{app:left-eigen}.

\subsection{Determination of the thermal effects in the zero-current region}
\label{sec:determ-therm-effects}

Once the left eigenvector is determined, we are now in a position to obtain the Arrhenius scaling in the zero-current regime. Following the approach described in Sec.~\ref{sec:strat-obta-arrh}, to obtain the Arrhenius function $\Phi(\lambda,f)$ we need to solve the two optimization problems~(\ref{eq:defXz}) and (\ref{eq:exprPhilambda}), where $\fla$ is given by Eq.~(\ref{eq:decompfinalUeff}). Again, we restrict our study to the regime $\lambda_\cc^-(f)<\lambda<f$ as a consequence of the Gallavotti--Cohen symmetry of the SCGF (see Appendix~\ref{sec:GC-sym}).

 Let us focus first on the case $\lambda_\cc^-(f)<\lambda<0$ (the case $0<\lambda<f$, which follows similar lines, is treated in Appendix~\ref{app:rate-function}).
 We make use of the result~(\ref{eq:ULfinalA}) for $U_L(x)$ at leading order in $\epsilon$, to obtain the effective periodic potential $V_\lambda(x)$ from its definition given in Eq.~(\ref{eq:decompfinalUeff}) as:
\begin{equation}
  \label{eq:resVeff}
  V_\lambda(x) =k_1+
   \begin{casesl}
   V(x) - \lambda (1+x) +  \lambda_\cc^-(f)
     & 
    \qquad \text{if} \quad 
    0\leq x \leq x^*_L
    \\[2mm]
    -V(x) + (2f - \lambda) x -  \lambda_\cc^+(f)
    & 
    \qquad \text{if} \quad 
     x^*_L \leq x \leq 1\,.
 \end{casesl}
\end{equation}
%
%
%
\begin{figure}[t]
\begin{center}
  \includegraphics[width=0.45\columnwidth]{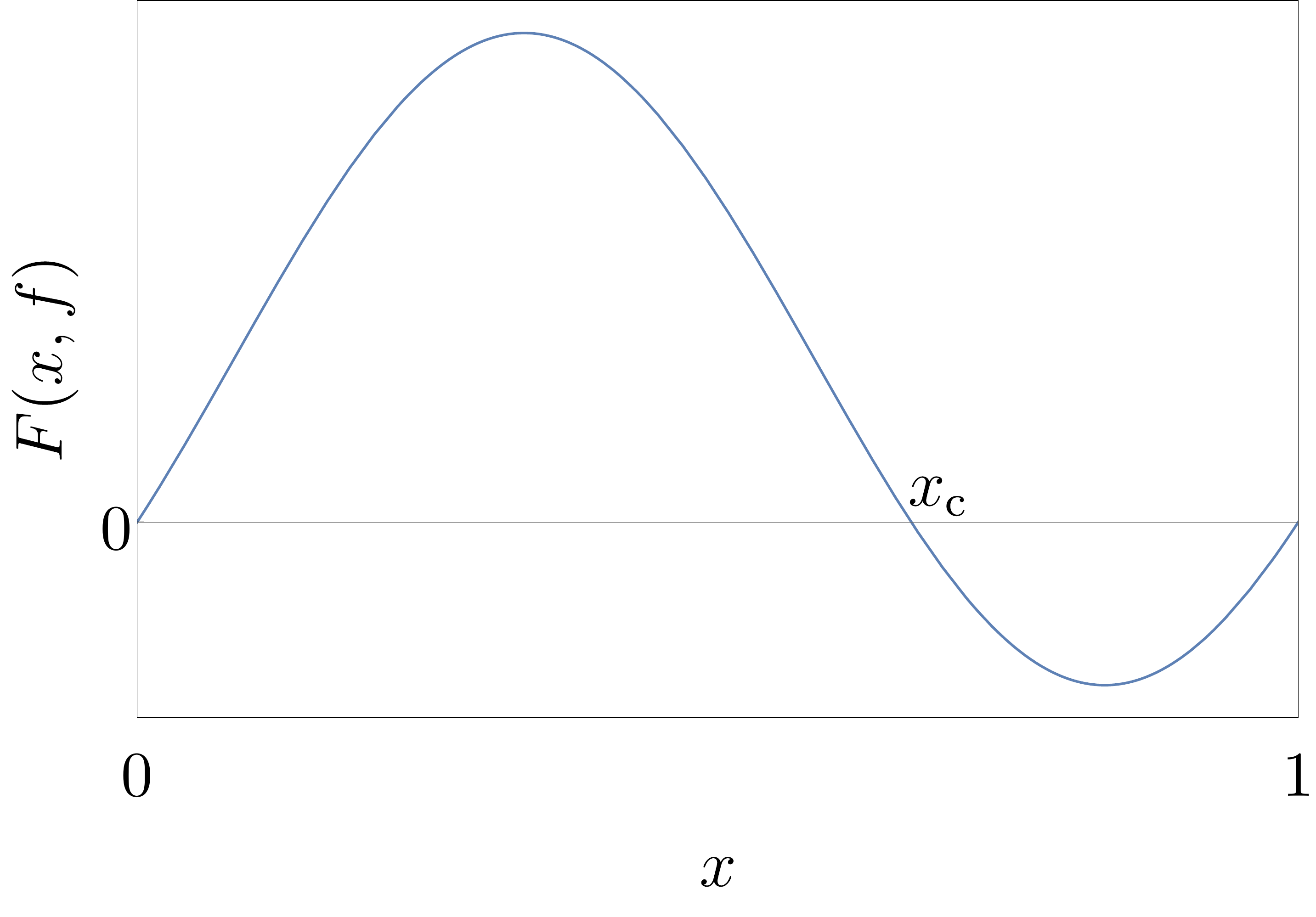}
\qquad
  \includegraphics[width=0.45\columnwidth]{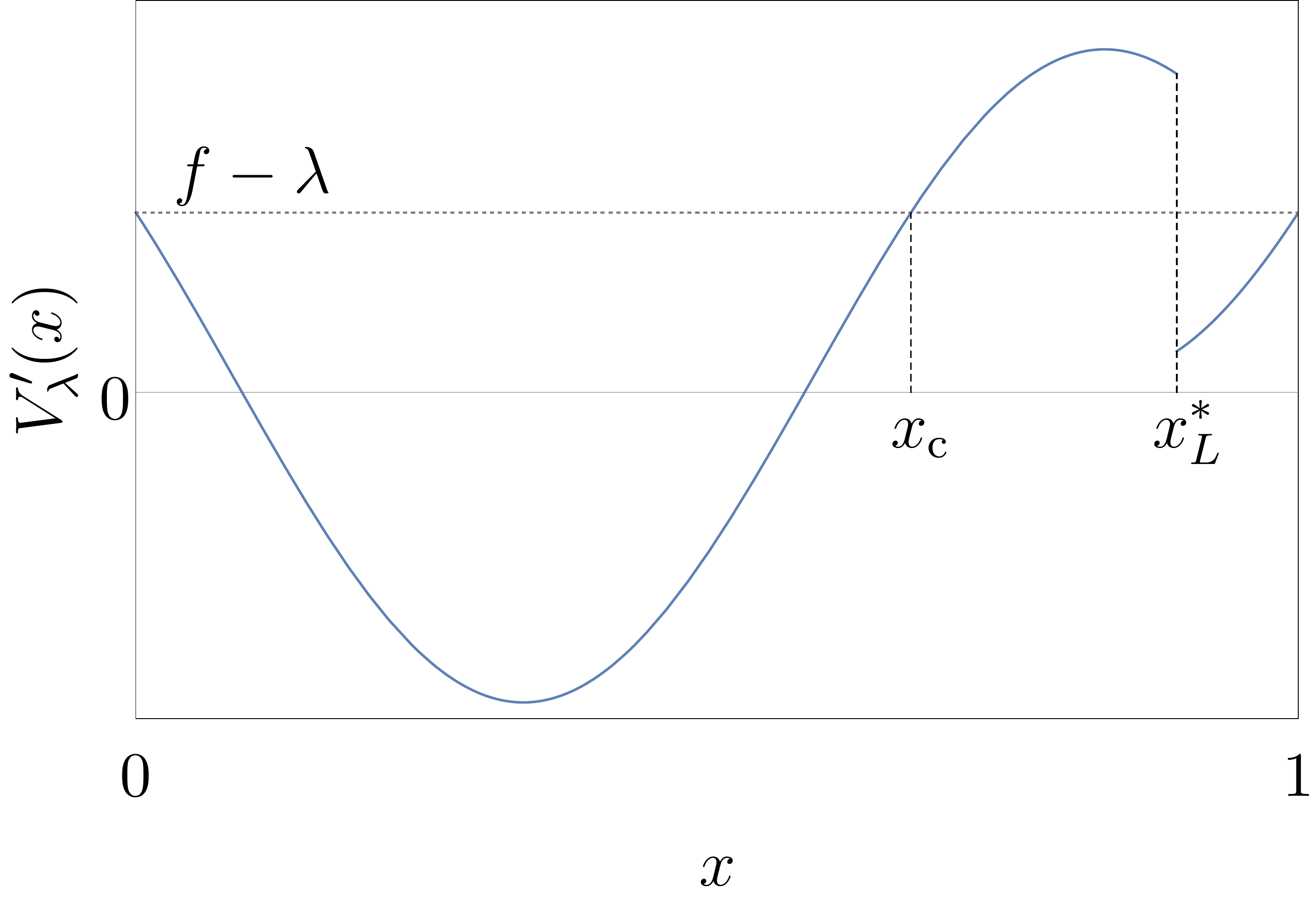}
\end{center}
\caption{Schematic representation of the force, $F(x,f)$ (\textbf{left}) and the derivative of the effective potential, $V'_\lambda(x)$ (\textbf{right}).
\label{fig:scheme_derivative_effective_potential}
}
\end{figure}
%
The first derivative $V'_\lambda(x)$ of the effective potential $V_\lambda(x)$ reads (see Figure~\ref{fig:scheme_derivative_effective_potential} for a schematic illustration for a generic potential)
\begin{equation}
  \label{eq:resVeffder}
  V'_\lambda(x) =
   \begin{casesl}
   -F(x,f)+f-\lambda
     & 
    \qquad \text{if} \quad 
    0\leq x \leq x^*_L
    \\[2mm]
    F(x,f)+f-\lambda
    & 
    \qquad \text{if} \quad 
     x^*_L \leq x \leq 1\,.
 \end{casesl}
\end{equation}
A key property of $V'_\lambda(x)$, to be used below, is that (see Figure~\ref{fig:scheme_derivative_effective_potential})
\begin{eqnarray} \label{eq:sign:Vla:fla:inxcxstar}
V'_\lambda(x) &>& f-\lambda \quad {\rm for} \; x_\cc<x<x^*_L\,, \\
\label{eq:sign:Vla:fla:outxcxstar}
V'_\lambda(x) &<& f-\lambda \quad {\rm for} \; 0\leq x<x_\cc \; {\rm or} \; x^*_L<x<1\,.
\end{eqnarray}
Noting that $V_\lambda(x+z)-V_\lambda(x)=\int_x^{x+z}\dd y\,V'_\lambda(y)$, the optimization problem~(\ref{eq:defXz}) reads as
\begin{equation} \label{eq:opt-prob-1}
 X(z)=    \stackrel[0\leq x\leq 1]{}{\operatorname{argmax}} 
 \int_x^{x+z}\dd y\,V'_\lambda(y)
 \,,
\end{equation}
i.e.,~one has to maximize the algebraic area between $x$ and $x+z$ under the curve $V'_\lambda(x)$. On the other hand, the problem~(\ref{eq:exprPhilambda}) can be also rewritten as
\begin{equation} \label{eq:opt-prob-2}
  \Phi(\lambda,f)=  - 2\min_{z\geq 0}   \zeta(z)\,,
\end{equation}
where
\begin{equation}
\zeta(z)\equiv (f-\lambda)z-\int_{X(z)}^{X(z)+z}\!\! \dd y\,V'_\lambda(y)\,.
\end{equation}
We note that since $f-\lambda>0$ and due to the periodicity of $X(z)$ and $V_\lambda(x)$, the minimum of $\zeta(z)$ is reached for $z\in(0,1)$.

\begin{figure}[t]
\begin{center}
\includegraphics[width=0.45\columnwidth]{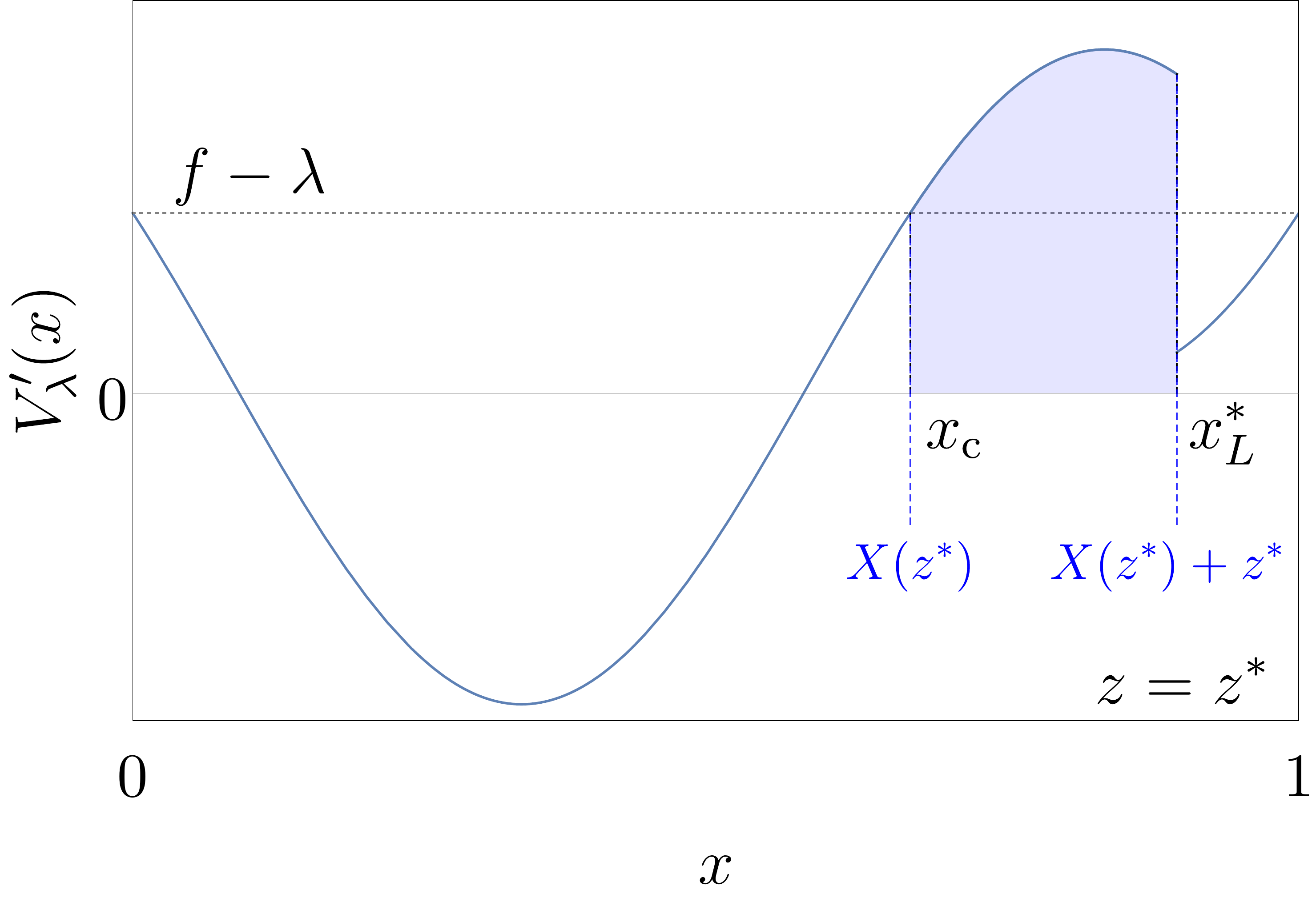}
\quad
\includegraphics[width=0.45\columnwidth]{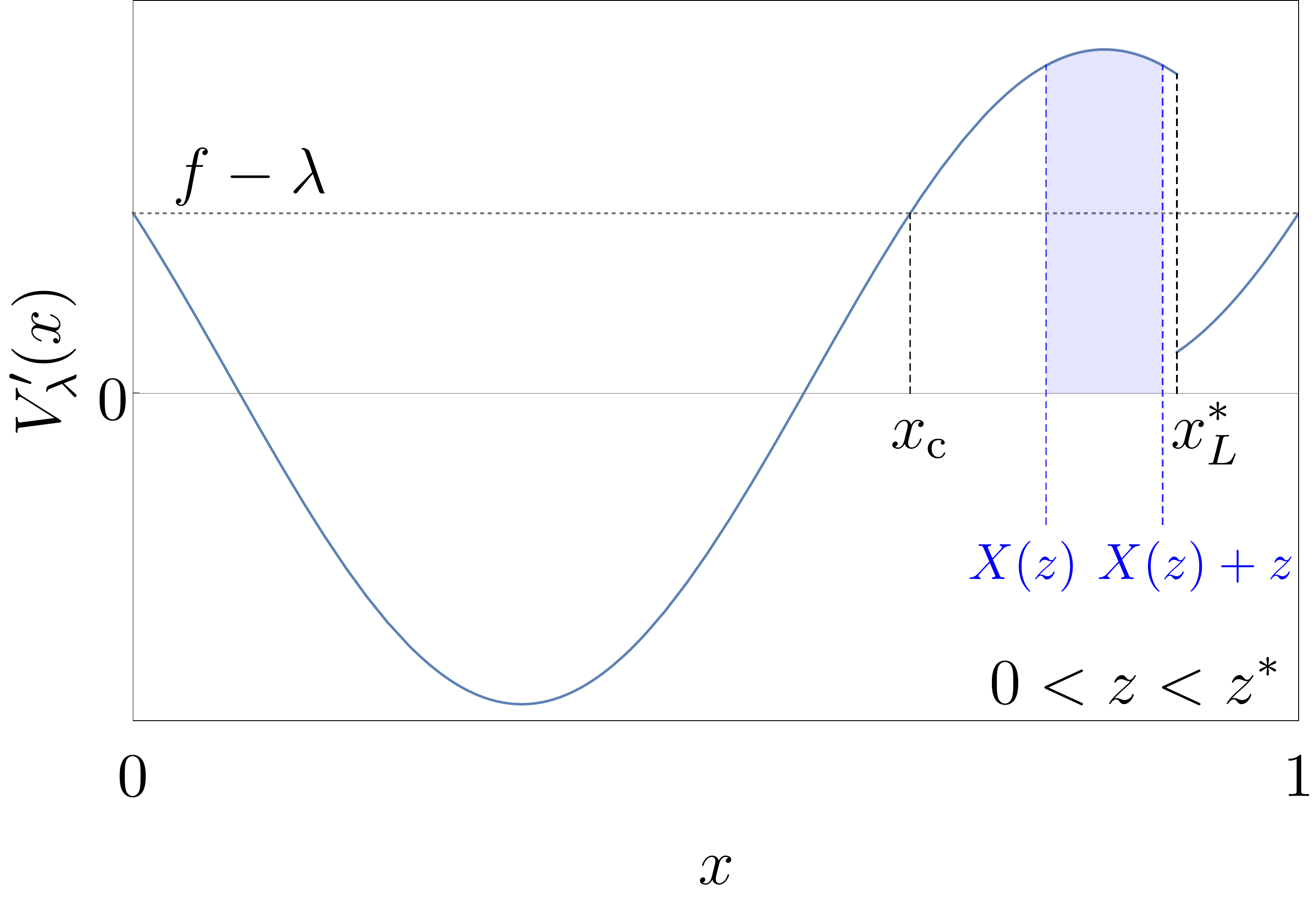}
\\~\\~\\
\includegraphics[width=0.45\columnwidth]{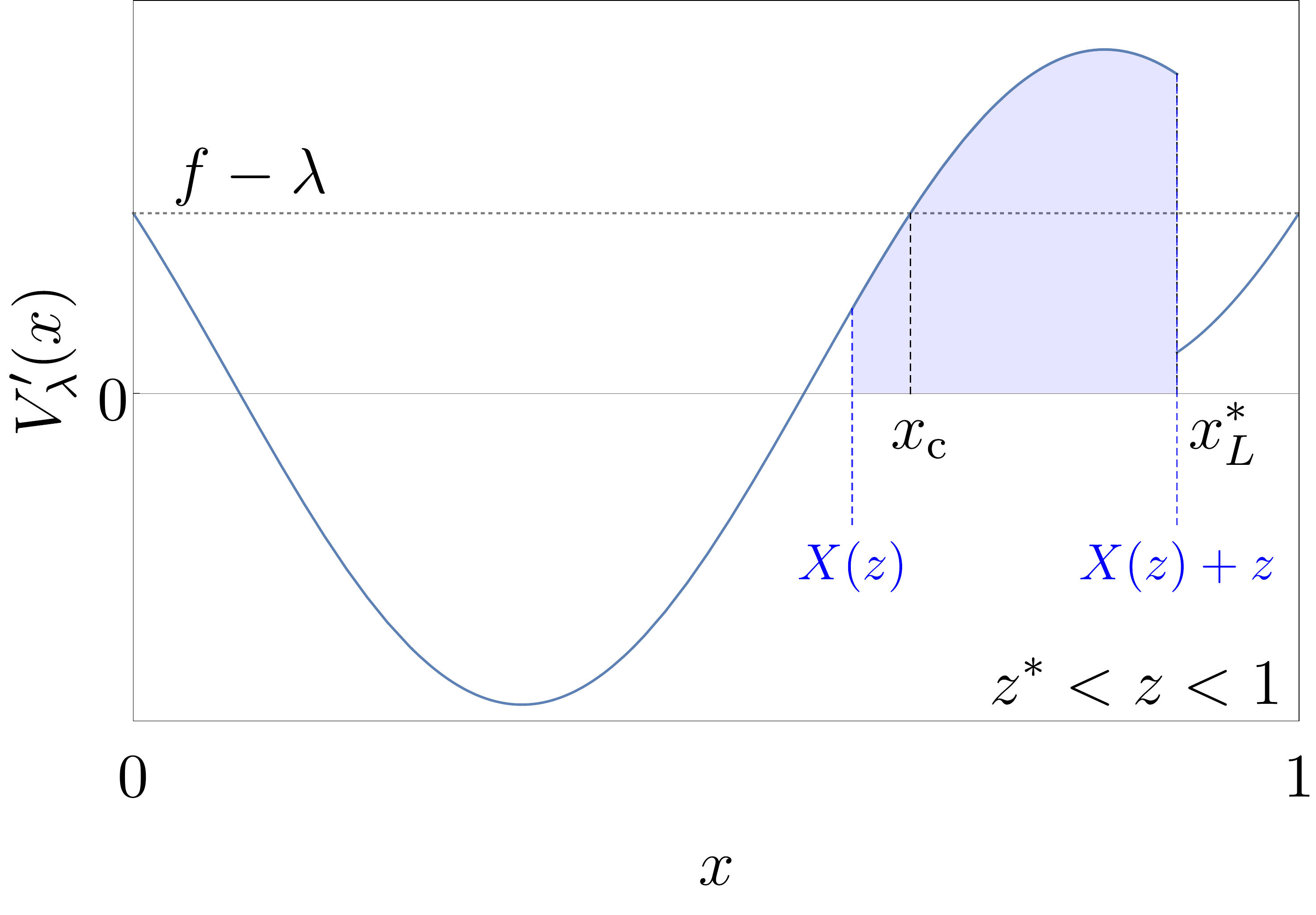}
\quad
\includegraphics[width=0.45\columnwidth]{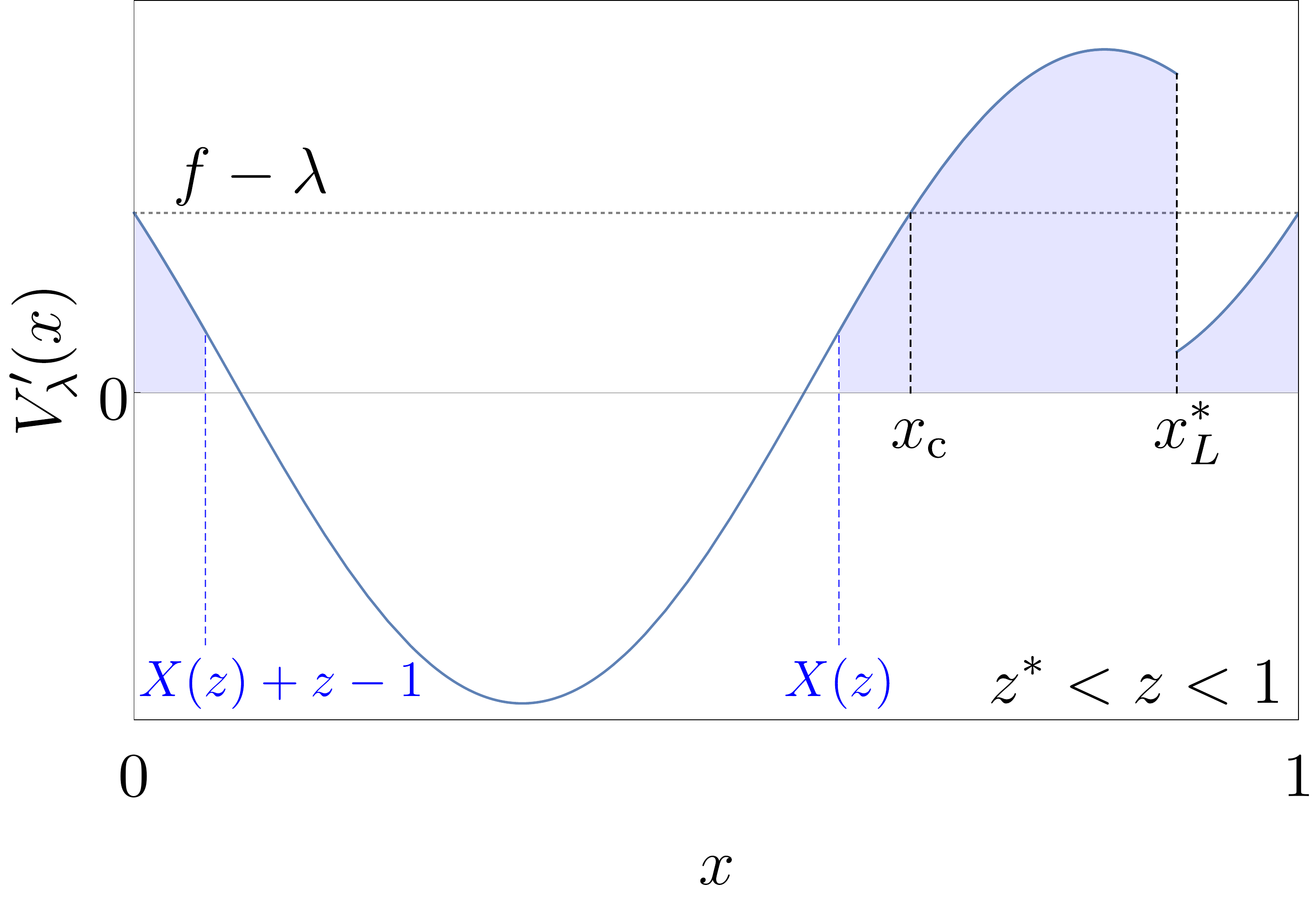}
\end{center}
\caption{Sketch of the graphical argument leading to the key properties of $X(z)$. On each graph, the shaded area indicates the integral of $V'_\lambda(x)$ over the interval $[X(z),X(z)+z]$.
Top left: case $z=z^*$, where $X(z)=x_\cc$.
Top right: case $0<z<z^*$, where $x_\cc<X(z)<x^*_L$ with $X(z)+z\leq x^*_L$.
Bottom: case $z^*<z<1$, illustrated here with $0\leq X(z)<x_\cc$, both for $X(z)+z=x^*_L$ and for $X(z)+z>x^*_L$.
\label{fig:derivative_effective_potential_shaded}
}
\end{figure}

In the following, we describe a graphical argument which helps us to find the solutions of the optimization problems~(\ref{eq:opt-prob-1}) and (\ref{eq:opt-prob-2}).
It turns out that to determine the minimum of $\zeta(z)$, the full determination of $X(z)$ is not required, but one only needs to know some properties of $X(z)$ that can be understood graphically:
\begin{itemize}
 \item If $z=z^*\equiv x^*_L-x_\cc$, then $X(z^*)=x_\cc$ (see top left panel of Fig.~\ref{fig:derivative_effective_potential_shaded});
 
 \item If $0<z<z^*$, then $x_\cc<X(z)<x^*_L$ with $X(z)\leq x^*_L-z$ (see top right panel of Fig.~\ref{fig:derivative_effective_potential_shaded});
 
 \item If $z^*<z<1$, then either $0\leq X(z)<x_\cc$ (see bottom panels of Fig.~\ref{fig:derivative_effective_potential_shaded}) or $x^*_L<X(z)<1$.
 
\end{itemize}
These properties of $X(z)$ mainly rely on the inequalities (\ref{eq:sign:Vla:fla:inxcxstar}) and (\ref{eq:sign:Vla:fla:outxcxstar}).
Let us consider first the case $z=z^*\equiv x^*_L-x_\cc$. Choosing a value $X(z^*)\ne x_\cc$ instead of the optimal value $x_\cc$, one would replace in the integral $\int_{X(z^*)}^{X(z^*)+z^*}V'_\lambda(y) \,dy$ a subinterval where $V'_\lambda(y)>f-\lambda$ by another subinterval where $V'_\lambda(y)<f-\lambda$, leading to smaller value of the integral.
Similar reasonings lead to the conclusion that $X(z)$ lies in the interval $x_\cc<X(z)<x^*_L$ for $0<z<z^*$ and outside this interval for $z^*<z<1$. Note that in the latter case, it seems that in most (if not all) situations, one has $0\leq X(z)<x_\cc$ (at least we could not find examples where $x^*_L<X(z)<1$).

\begin{figure}[t]
\begin{center}
  \includegraphics[width=0.487\columnwidth]{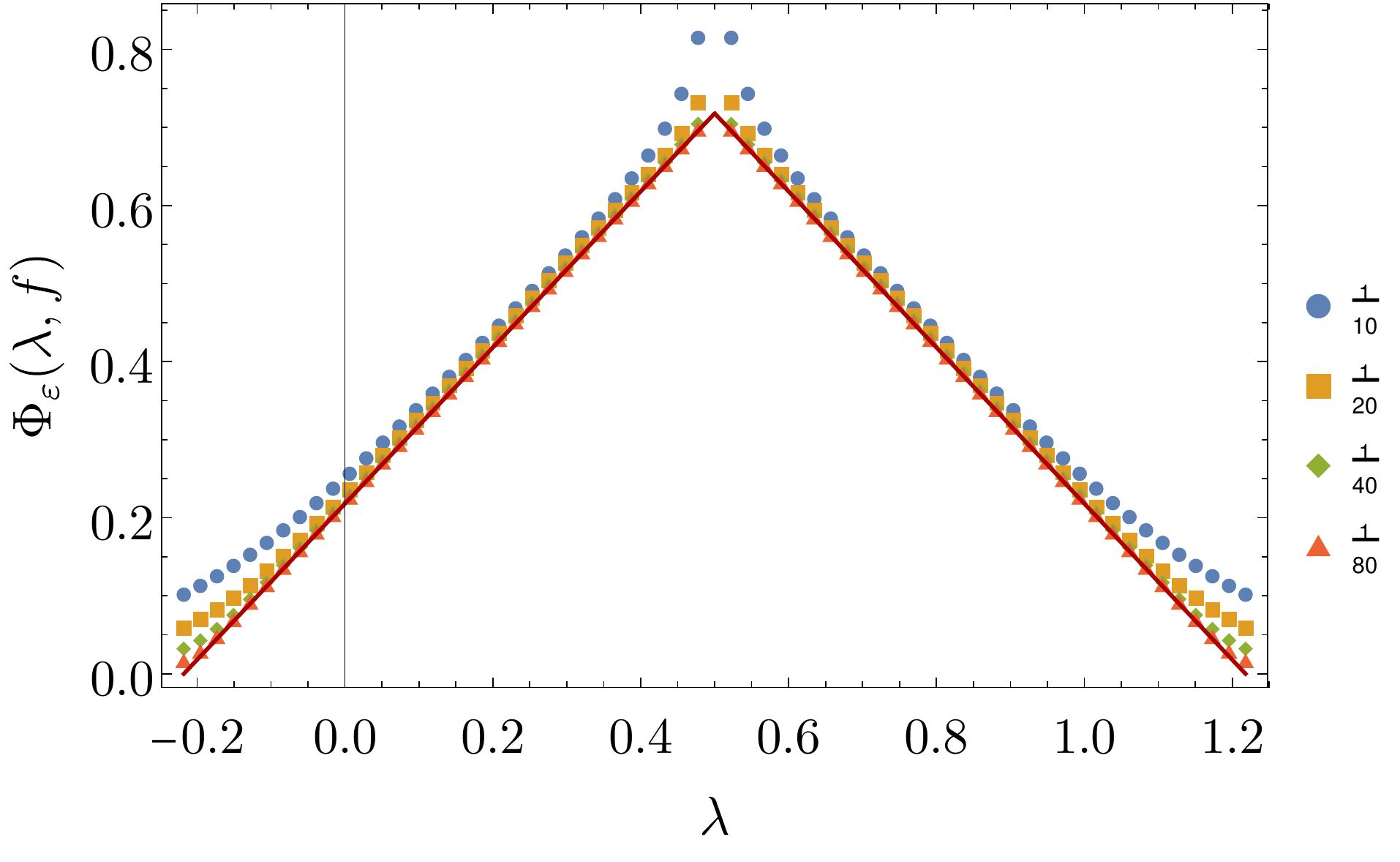}
\hfill
  \includegraphics[width=0.487\columnwidth]{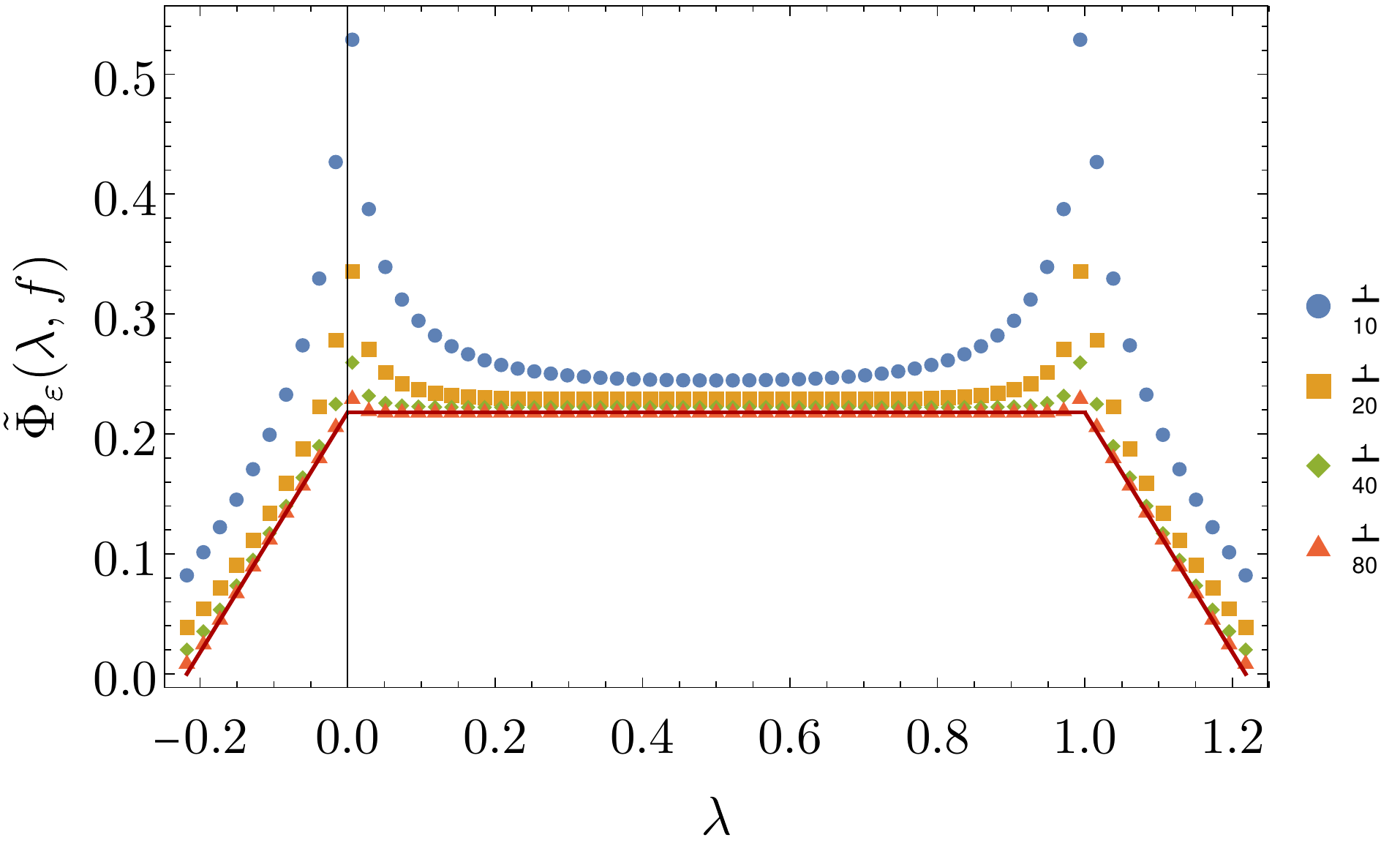}
\end{center}
\caption{
\textbf{(Left)}
Numerical evaluation of $\Phi_\epsilon(\lambda,f) = -\epsilon \log |\partial_\lambda \varphi_\epsilon(\lambda,f)|$
 describing the behavior $\partial_\lambda \varphi_\epsilon(\lambda,f)\asymp  \ee^{-\frac 1\epsilon \Phi(\lambda,f)}$,
 as predicted analytically in the $\epsilon\to 0$ limit (Eq.~(\ref{eq:finPhi}), plain line),
 and as evaluated numerically on a lattice version of the model with $64$ sites and for small but finite values of $\epsilon$ indicated in the legend. The force is the sine model of Eq.~\eqref{eq:sinus:force} with $f=\frac 12$,
and $\lambda$ varies in $[\lambda_\cc^-,\lambda_\cc^+]$.
At finite values of $\epsilon$, one observes boundary layers for $\lambda$ close to $f$ and close to $\lambda_\cc^\pm$,  indicating that the limits $\epsilon\to 0 $ and $\lambda\to \{f, \lambda_\cc^\pm\}$ do not commute. 
\textbf{(Right)} 
Numerical evaluation of $\tilde \Phi_\epsilon(\lambda,f) = -\epsilon \log |\epsilon^{-1} \varphi_\epsilon(\lambda,f)|$
 describing the behavior $\varphi_\epsilon(\lambda,f)\asymp  \ee^{-\frac 1\epsilon \tilde \Phi(\lambda,f)}$ as predicted analytically in the $\epsilon\to 0$ limit (Eq.~(\ref{eq:resultPhitilde}), plain line).
Boundary layers are present in $\lambda=\lambda_\cc^\pm$ (indicating the boundary of the pinned regime) and for $\lambda\in\{0,2f\}$,
corresponding to the switch between one term dominating the other in Eq.~(\ref{eq:phiepsilonPhitilde}).
\label{fig:PhiPhitildecompar}
}
\end{figure}

Using these properties of  $X(z)$, we show in Appendix~\ref{sec:proof-that-zetazz} that $\zeta(z)>\zeta(z^*)$ for all $z\neq z^*$. 
This allows us thus conclude that $\zeta(z)$ has its global minimum at $z = z^*$.
We can now proceed to compute the Arrhenius function $\Phi(\lambda,f)$. According to Eq.~(\ref{eq:opt-prob-2}), we get:
\begin{equation}
 \Phi(\lambda,f) = f\,(x^*_L-x_\cc)-V(x^*_L)+V(x_\cc)\,.
\end{equation}
Taking into account Eq.~(\ref{eq:xstar}) for $x^*_L$, the Arrhenius function can be rewritten as
\begin{equation}
 \Phi(\lambda,f) = \lambda - 2f(1-x_\cc) - 2V(x_\cc)\,.
\end{equation}
Then we note that using the definition (\ref{eq:lambdac:fltfc}) of $\lambda_\cc^-(f)$, we have
 \begin{equation} \label{eq:lambdac:xc}
\lambda_\cc^-(f)=f-\int_0^{x_\cc}\dd y\,F(y,f)+\int_{x_\cc}^{1}\dd y\,F(y,f)=2f(1-x_\cc)+2V(x_\cc)\,.
\end{equation}
Hence, we obtain the final result
\begin{equation}
 \Phi(\lambda,f)=\lambda-\lambda_\cc^-(f)
\end{equation}
for
$\lambda_\cc^-(f)<\lambda<0$.
This expression of the Arrhenius function $\Phi(\lambda,f)$ is also found to be valid for $0<\lambda<f$ (the computation is detailed in Appendix~\ref{app:rate-function}).
In addition, using the Gallavotti--Cohen symmetry  $\Phi(\lambda,f) =   \Phi(2f-\lambda,f)$ (see Appendix~\ref{sec:GC-sym}), one eventually obtains the expression of the Arrhenius function $\Phi(\lambda,f)$
on the full zero-current range $\lambda_\cc^-(f)<\lambda<\lambda_\cc^+(f)$,
\begin{equation} 
 \label{eq:finPhi}
 \Phi(\lambda,f)=f-\lambda_\cc^-(f) -|\lambda-f|\,.
\end{equation}
This is our final result for the function $\Phi$ describing the Arrhenius behavior $\asymp \ee^{-\frac 1\epsilon \Phi(\lambda,f)}$ of $\partial_\lambda\varphi_\epsilon(\lambda,f)$ as $\epsilon\to 0$.
By continuity of the maximal eigenvalue of the tilted operator (which is continuous in $f$), we expect that it is also true for $f=\lambda$.
The derivation of this expression was done for $f>0$. We show in Appendix~\ref{sec:GC-sym} that, thanks to the mirror symmetry, it is also valid for $f<0$.
Again by continuity of the maximal eigenvalue of the tilted operator, we expect that it is also true for $f=0$.
It turns out that the Arrhenius function $\Phi(\lambda,f)$ exhibits a cusp in $\lambda=f$, a property that we further discuss in the following section.
Before that, we derive the Arrhenius function $\tilde \Phi(\lambda,f)$ describing the Arrhenius behavior of the SCGF itself.
Our previous result indicate that a primitive in $\lambda$ of $\partial_\lambda \varphi_\epsilon(\lambda,f)$ takes the form
  $C(\lambda,\epsilon)\, \ee^{-\frac 1\epsilon \Phi(\lambda,f)}$ 
where the function $C(\lambda,\epsilon)$ is exponentially dominated by $\ee^{-\frac 1\epsilon \Phi(\lambda,f)}$ as $\epsilon\to 0$ [i.e.~$\log|C(\lambda,\epsilon|\ll\frac 1\epsilon$].
Hence:
\begin{equation}
  \label{eq:phiepsilonPhitilde}
  \varphi_\epsilon(\lambda,f)
\
 =
\
\int_0^\lambda \dd \bar\lambda\: \partial_{\bar\lambda}\varphi_\epsilon(\bar\lambda,f)
  \stackrel[\epsilon\to 0]{}\approx
  C(\lambda,\epsilon)\, \ee^{-\frac 1\epsilon \Phi(\lambda,f)}
  - 
  C(0,\epsilon)\, \ee^{-\frac 1\epsilon \Phi(0,f)}  
\:,
\end{equation}
since the SCGF is equal to $0$ in $\lambda=0$. We thus have
$
  \tilde\Phi(\lambda,f)
  =
   \min 
  \big\{
    \Phi(\lambda,f) ,  \Phi(0,f)
  \big\}
$
and we find
\begin{equation}
  \label{eq:resultPhitilde}
  \tilde\Phi(\lambda,f)
  =
   \begin{casesl}
     f-\lambda_\cc^-(f) -|\lambda-f|
     & 
    \qquad \text{if} \quad 
    |\lambda - f | \geq |f|
    \\[2mm]
     f-\lambda_\cc^-(f) -|f|
    & 
    \qquad \text{if} \quad 
    |\lambda - f | \leq |f|
\:.
 \end{casesl}
\end{equation}
A comparison of the results of Eqs.~(\ref{eq:finPhi}) and~(\ref{eq:resultPhitilde}) to numerical evaluations of the Arrhenius functions,
performed by diagonalization of the tilted operator of a lattice version of the system at small but finite $\epsilon$,
is presented on Fig.~\ref{fig:PhiPhitildecompar}, showing a good agreement on the example of the sine force of Eq.~(\ref{eq:sinus:force}).
 
\subsection{Interpretation of the form of the Arrhenius function $\Phi(\lambda,f)$ and of its cusp singularity}
\label{sec:interpr-form-rate}

To discuss the physical consequences and interpretation of the form (\ref{eq:finPhi}) of the Arrhenius function in the $\epsilon \to 0$ asymptotics, we recall that 
the derivative
$
\partial_\lambda \varphi_\epsilon \asymp \ee^{-\frac 1\epsilon \Phi}
$
of the SCGF represents (up to prefactors) the average velocity in the biased ensemble, 
$\bar v(\lambda) = \frac 1\epsilon \langle\dot x\rangle_\eff = -\epsilon \, \partial_\lambda \varphi_\epsilon(\lambda,f)$,
but also that when performing the Legendre--Fenchel transform
\begin{equation}
  \label{eq:chensemble}
  \varphi_\epsilon(\lambda,f)
  =
  \max_{a}
  \big\{
   \Pi_\epsilon(a)-\lambda a
  \big\}
\:,
\end{equation}
the maximum is reached in $a=\bar v(\lambda)$.
Thus,
the cusp in the expression (\ref{eq:finPhi}) of the Arrhenius function $\Phi(\lambda,f)$ implies that at this order in $\epsilon$, there is a DPT in $\lambda=f$.
Physically, this point corresponds to the cancellation of the tilt $f_\lambda=f-\lambda$ of the effective potential [see Eq.~(\ref{eq:decompfinalUeff})].
At fixed $f$ and increasing $\lambda$, the transition thus occurs precisely at the point where trajectories switch from escaping towards the left direction to escaping towards the right direction.
A finer analysis of such switch, which would take into account higher-order corrections in $\epsilon$, would detail how the transition from left-moving to right-moving classes of trajectories takes place.
A possible method to do so would be to generalize the WKB and asymptotic matching approach of Proesmans and Derrida~\cite{proesmans_large-deviation_2019} to the case where the dynamics presents metastable states.

However, at the order in $\epsilon$ we have considered, the result~(\ref{eq:finPhi}) we obtained for the Arrhenius function has several interesting consequences.
We first remark that the non-analyticity of the Arrhenius function~$\Phi(\lambda,f)$ in~Eq.~(\ref{eq:finPhi}) implies that the distribution of $\frac 1\tf A(\tf)$ at large $\tf$ develops different regimes, obtained by inverse Legendre transform of~(\ref{eq:chensemble}).
The regime around $\lambda=0$ corresponds to the most typical large deviations [those around the most probable value of $\frac 1\tf A(\tf)$], while the regime corresponding to $\lambda>f$ describes different fluctuations.

We now discuss these regimes, depending on whether $f=0$ or not.
If $f\neq 0$, the expression~(\ref{eq:finPhi}) implies that all scaled cumulants of the current  $\frac 1\tf \langle A(\tf)^n \rangle_c$ are equal, at exponential order in $\epsilon$\footnote{%
Note that in this expression of the cumulants, one should differentiate $n-1$ times w.r.t.~$\lambda$ the expression $\ee^{-\frac 1\epsilon\Phi(\lambda,f)}$ of  $\partial_\lambda \varphi_\epsilon(\lambda,f)$ rather than differentiate $n$ times the expression $\ee^{-\frac 1\epsilon\tilde\Phi(\lambda,f)}$ of $\varphi_\epsilon(\lambda,f)$, because the constant term in Eq.~(\ref{eq:phiepsilonPhitilde}) makes that $\tilde \Phi(\lambda,f)$ is constant for $\lambda\in[0,2f]$, as seen in Eq.~(\ref{eq:resultPhitilde}).
}:
\begin{equation}
  \label{eq:cumulants-arrhenius}
  \lim_{\tf\to\infty}
  \frac 1\tf \langle A(\tf)^n \rangle_c
\
  =
\
 (-\epsilon)^n 
 \frac{\partial^n}{\partial \lambda^n}\Big|_{\lambda=0}
 \varphi_\epsilon(\lambda,f)
\
  \stackrel[\epsilon\to 0]{}\asymp
\
 \ee^{\lambda_{\cc}^{-}(f)/\epsilon}
.
\end{equation}
The equality of all cumulants would show that the distribution of particle current $a\equiv A(\tf)/\tf$ is Poissonian,
provided $a$ is not too far from the average value $\bar v\:$\footnote{The threshold value of $a$ corresponds to $\lambda=f$ in the Legendre transform, where a cusp is observed.}.
Here we can only prove that cumulants are equal at exponential order in~$\epsilon$, as described in~(\ref{eq:cumulants-arrhenius}).
This suggests that the statistics should at least be close to a Poissonian one in the low-noise limit.
It is also possible to show that the prefactors of the exponential in Eq.~(\ref{eq:cumulants-arrhenius}) for the first two cumulants are equal (see Appendix~\ref{sec:deriv-diff-coeff}), which supports the possibility that the Poissonian statistics is also valid beyond the exponential order.
A correspondence to the current statistics in asymmetric random walks, obtained and discussed in Refs~\cite{lebowitz_gallavotticohen-type_1999,speck_large_2012}, could be interestingly probed.

Physically, the Poissonian regime pictures the fact that, at dominant order, the distribution of the time-integrated current is governed by independent escape events that all happen at the same rate.
The occurrence of a Poisson distribution for a continuous variable such as the current can be understood as follows: the motion of the particle occurs along a succession of metastable states, making it almost as discrete. 
The large-time asymptotics implies that obtaining a non-zero value of the time-averaged current is possible only if a large number of discrete jumps in successive metastable states of the tilted potential occur.
In particular, the average current $\bar{v}$ behaves as $\bar v \asymp \ee^{\lambda_{\cc}^{-}(f)/\epsilon}$ for $\epsilon\to 0$.
Interestingly, this behavior has a simple physical interpretation.
Defining the tilted potential $U(x)=V(x)-fx$, such that $U'(x)=-F(x,f)$, the extrema of the tilted potential $U(x)$ are located in $x=0$, $x_\cc$ and $1$; $x_\cc$ corresponds to a local minimum of the tilted potential, whereas $x=0$ and $1$ correspond to local maxima.
It follows that the energy barrier $\Delta U$ to be overcome by a particle moving to right (i.e., along the direction of the force $f$) is
$\Delta U=U(1)-U(x_\cc)= -f(1-x_\cc)-V(x_\cc)$ (we recall that $V(1)=0$).
From Eq.~(\ref{eq:lambdac:xc}), we thus have that $\lambda_{\cc}^{-}(f)=-2\Delta U$. Further recalling that the temperature $T$ is given by $\epsilon=2T$, we obtain the simple Arrhenius law $\bar{v}\asymp\ee^{-\Delta U/T}$ for $T\to 0$, hence providing a clear and simple physical interpretation of our results for $\lambda=0$. 

Coming back to the DPT at $\lambda=f$, the latter tells that, beyond a certain value of the time-integrated current (where the particle is forced to go in the opposite direction to the one naturally imposed by $f$), the distribution changes drastically;
and, because of the non-analyticity, this happens in a way that cannot be seen by a mere expansion of the SCGF in powers of $\lambda$ around $\lambda=0$ (again, at the order in $\epsilon$ we are working).
Note that an analogous phenomenon happens for the fluctuations of current of particles in the WASEP on a ring of $L$ sites~\cite{bodineau05a,bertini_current_2005}: in the large $L$ limit (which is a small-noise limit), the SCGF is quadratic around the origin as  function of $\lambda$, meaning that the fluctuations of the current seem Gaussian (their distribution is Gaussian for a given range of values).
A DPT also occurs for strong enough atypical values of the current, which is manifested as a non-analyticity in the SCGF: the current distribution becomes non-Gaussian for atypical enough values of the current.
For the cumulants, this is manifested at finite $L$ by a dominant order in $L$ which is Gaussian and sub-dominant ones which are non-Gaussian~\cite{PhysRevE.78.021122}.
For our system of interest, one has an analogous phenomenology: current fluctuations are Poissonian on a range close enough to the typical fluctuations, and become non-Poissonian above, in a singular manner.
We expect that this DPT is also manifested in the cumulants of the current at finite temperature: they are Poissonian at dominant order in $\epsilon\to 0$, but present non-Poissonian contributions at higher order.
We illustrate this point for the first and second cumulants in Appendix~\ref{sec:deriv-diff-coeff}, by showing that they are different at any finite $\epsilon$, but they become equivalent in the $\epsilon\to 0$ limit, where both cumulants go to zero.

If $f=0$, the consequences of the DPT, now located in $\lambda=0$ are even more drastic.
The procedure we have followed yields a SCGF which does not even describe expected Gaussian fluctuations for small current fluctuations:
indeed,
for $f=0$ one has
$\lambda_{\cc}^{\pm}(f) =  \pm \lambda_\cc^0$ with $\lambda_\cc^0=\int_0^1 \dd x \, |V'(x)|$,
and with this notation our result~(\ref{eq:finPhi}) for the Arrhenius function reads as $\Phi(\lambda,f)=\lambda_\cc^0-|\lambda|$.
Its derivative in $\lambda=0$ is singular, implying that the Taylor expansion of $\partial_\lambda\phi_\epsilon(\lambda,f)$ in $\lambda=0$ is not well defined (while for instance we expect a finite diffusion coefficient, see Refs.~\cite{reimann_giant_2001,reimann_diffusion_2002} and Appendix~\ref{sec:deriv-diff-coeff}).
We discuss the scaling of such Gaussian fluctuations, in Sec.~\ref{sec:disc-lambd-0} below, specializing to the case of the sine force, and analyzing the non-commutation of the limits $\lambda\to 0$ and $\epsilon\to 0$.
The obtained result~(\ref{eq:finPhi}) can only give us information about current fluctuations which are larger than the Gaussian ones.
We thus expect the phenomenology to be as follows for small but finite $\epsilon$:
(i)~small enough fluctuations of the current are Gaussian (on a range of $\lambda$ that is not captured by our analysis of $\Phi(\lambda,f)$) and correspond to the fact that the particle can escape towards left or right with comparable probability.
(ii)~Larger fluctuations of current 
   (corresponding to $0<|\lambda|<\lambda_\cc^0$) occur again in a Poissonian way, but with a rate that depends on the sign of the current fluctuations we are considering;
   in that regime, the average current (given by $-\epsilon \partial_\lambda \varphi_\epsilon$) is exponentially small with the temperature.
(iii)~Even larger fluctuations of current  (corresponding to $|\lambda|>\lambda_\cc^0$), where now the amplitude of the current is not small with $\epsilon$, are not Poissonian and are described by the SCGF in the propagative regime.

\begin{figure}[t]
\begin{center}
  \includegraphics[height=0.282\columnwidth]{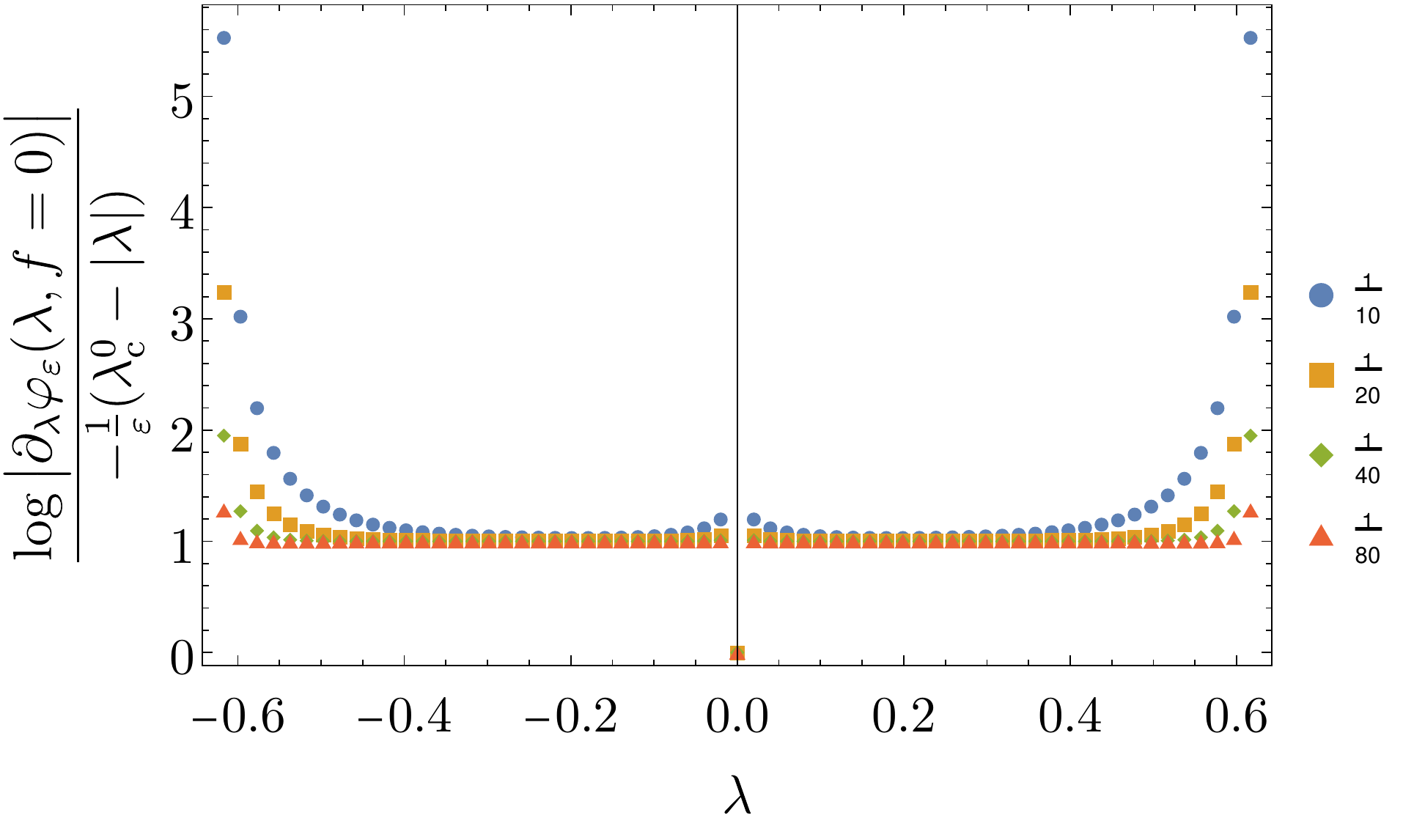}
\hfill
  \includegraphics[height=0.282\columnwidth]{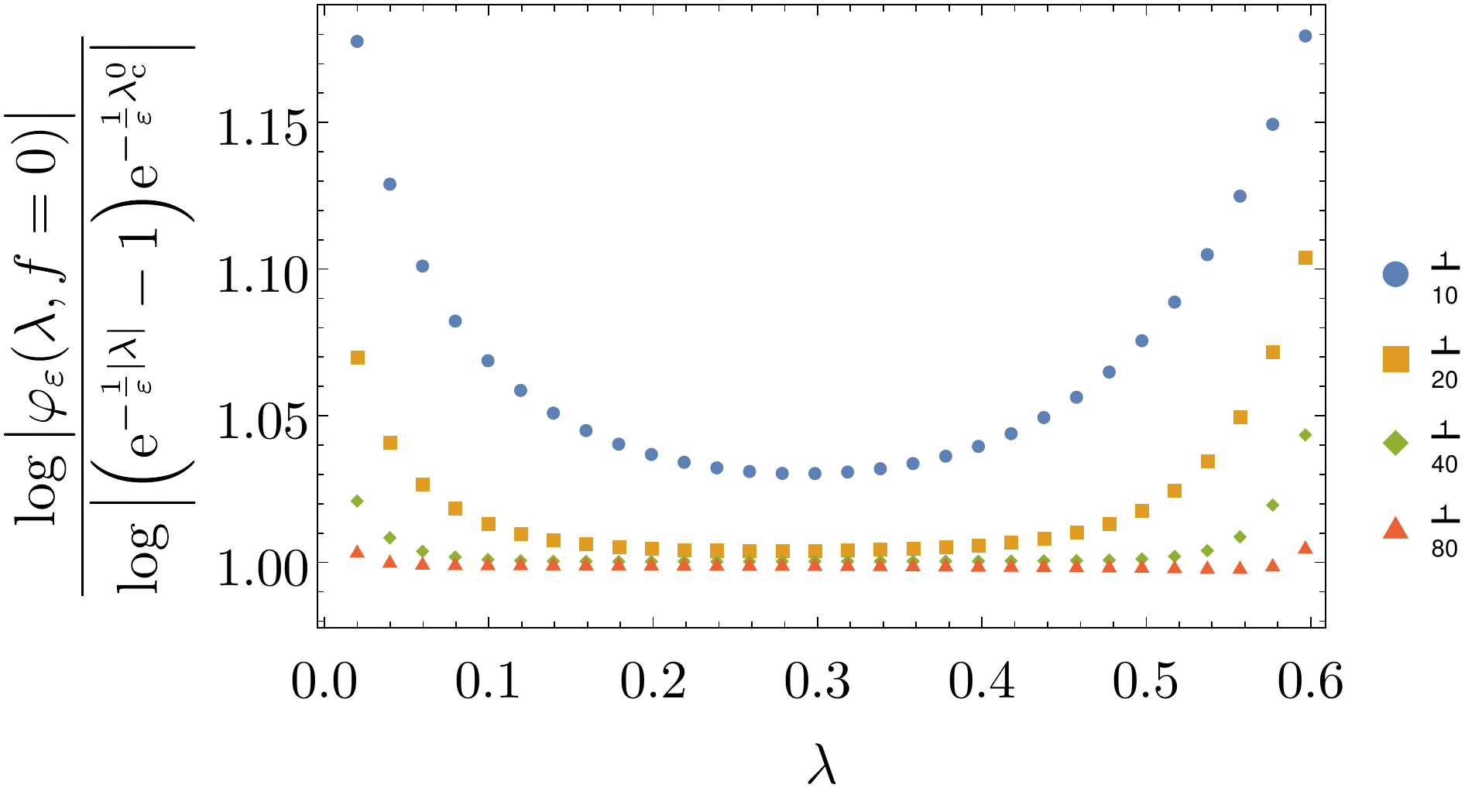}
\end{center}
\caption{
\textbf{(Left)}
Ratio between the logarithms of numerical evaluation of $\partial_\lambda\varphi_\epsilon(\lambda,f=0)$ and the analytical estimate of Eq.~(\ref{eq:exproneovervel}).
One observes boundary layers for $\lambda$ close to $0$ and close  $\pm\lambda_\cc^0$, at finite values of $\epsilon$, indicating that the limits $\epsilon\to 0 $ and $\lambda\to \{0, \pm \lambda_\cc^0\}$ do not commute. 
The numerical estimate is done by diagonalization on a discrete system of $64$ sites at $f=0$ and values of $\epsilon$ indicated in the caption.
\textbf{(Right)} Same but now for $\varphi_\epsilon(\lambda,f=0)$ and the improved analytical estimate Eq.~(\ref{eq:phiepsilontot}).
 It shows that part of the boundary layer for $\lambda$ close to $0$ has been absorbed (while it is not close to $\pm\lambda_\cc^0$). It is compatible with a numerical prefactor $B$ equal to 1 in~(\ref{eq:exproneovervel}).
\label{fig:Phiepsiloncompar}
}
\end{figure}


\subsection{Discussion on the order of limits $\lambda\to 0$ and $\epsilon\to 0$ in the $f=0$ case}
\label{sec:disc-lambd-0}

For $f=0$, the DPT in $\lambda=0$ makes that the cumulant of the average current cannot be determined from $\varphi_\epsilon$ or $\partial_\lambda\varphi_\epsilon$ as $\epsilon\to 0$.
We discuss in this subsection at which scale around $\lambda=0$ a finite value of $\epsilon$ rounds the transition, and allows one to recover the expect Gaussian fluctuations of the current.
We focus here on the case of the sine force, for which 
$
\lambda_{\cc}^{\pm}(0)=\pm\frac{2}{\pi}\equiv \pm\lambda_\cc^0$.
Starting from the expression~(\ref{eq:oneoverV}) of ${\langle\dot x\rangle_\eff}$,
what the computation leading to~(\ref{eq:finPhi}) has shown is actually that
\begin{equation}
  \label{eq:exproneovervel} 
\epsilon  \partial_\lambda \varphi_\epsilon(\lambda,f=0)
  =
  - {\langle\dot x\rangle_\eff} \stackrel[\epsilon\to 0]{}
  =
  \operatorname{sign}(\lambda)\, \epsilon B\, \ee^{-\frac 1\epsilon(\lambda_\cc^0 - |\lambda|)} + o\big( \epsilon\,\ee^{-\frac 1\epsilon(\lambda_\cc^0 - |\lambda|)}  \big) \:.
\end{equation}
The prefactor $B$ depends a priori on $\lambda$ and $\epsilon$, and
to actually compute it, one should integrate the fluctuations around the saddle-point evaluation of Eqs.~(\ref{eq:oneoverV})-(\ref{eq:oneoverVg}),
but this is not an obvious task since the effective periodic potential $V_\lambda(x)$ presents cusps.
We have incorporated in~(\ref{eq:exproneovervel}) a prefactor $\operatorname{sign}(\lambda)$ because for $f=0$, positive (resp.~negative) values of $\lambda$ favor negative (resp.~positive) values of the current.

For the sine force~(\ref{eq:sinus:force}), numerical evidence support the fact that $B$ is independent of $\lambda$ and $\epsilon$, as $\epsilon\to 0$ (see Fig.~\ref{fig:Phiepsiloncompar}).
The rest of this subsection is based on this hypothesis and thus remain more heuristic than the other sections of the paper.
Integrating Eq.~(\ref{eq:exproneovervel}) with the constraint $\varphi_\epsilon(0,0)=0$, one gets
\begin{equation}
  \label{eq:phiepsilontot}
  \varphi_\epsilon(\lambda,0)
  \stackrel[\epsilon\to 0]{}=
  \Big(\ee^{\frac{|\lambda|}{\epsilon}}-1\Big)B\, \ee^{-\frac{\lambda_\cc^0}{\epsilon}} 
  + 
   o\big( \ee^{-\frac 1\epsilon(\lambda_\cc^0 - |\lambda|)}  \big) 
  \:.
\end{equation}
This improved form now goes to $0$ as $\lambda\to 0$, as in the generic expression~(\ref{eq:phiepsilonPhitilde}) and for $\epsilon\to 0$, as shown on Fig.~\ref{fig:Phiepsiloncompar}~(right),  with a numerical prefactor $B$ compatible with a value $1$, that we now heuristically retain.
However, the expression~(\ref{eq:phiepsilontot}) of the SCGF is still non-analytical in the vicinity of $\lambda=0$. In particular, this expression does not describe the Gaussian fluctuations, that occur on a smaller scale that we now discuss.

The diffusion coefficient $D=\lim_{\tf\to\infty}\frac{1}{2\tf}\langle A(\tf)^2\rangle_{\cc}$ 
was derived in Refs.~\cite{reimann_giant_2001,reimann_diffusion_2002} 
(by use of a representation in terms of the moments of the first passage time~\cite{hanggi_reaction-rate_1990}).
In Appendix~\ref{sec:deriv-diff-coeff}, 
we provide for completeness a different derivation, based on a more general approach, and that uses the joint position and current distribution formalism (in the spirit of Derrida's approach for discrete random walks~\cite{derrida_velocity_1983}).
The result reads
\begin{align}
  \label{eq:resDfromAppA}
  D & 
  \stackrel[\phantom{\epsilon\to 0}]{}{=}
  \frac \epsilon 2 
  \frac{1}
  {
    \int_0^1\dd x\: \ee^{-\frac 2\epsilon V(x)}
\:
    \int_0^1\dd x\: \ee^{\frac 2\epsilon V(x)}
  }
\qquad\textnormal{so that}
\\
D &
  \stackrel[\epsilon\to 0]{}{\approx}
 \frac{1}{2\pi} \sqrt{|V''_{\min} V''_{\max}|}
\,
  \ee^{-\frac 2\epsilon\big(V_{\max}-V_{\min}\big)}
\label{eq:dlfhgejkgro}
\:,
\end{align}
with $V_{\max}$ and $V_{\min}$ the maximum and minimum values taken by the potential on its spatial period (and $V''_{\min,\max}$ the value of the second derivative at the position of the extrema).
From the definition~(\ref{eq:defSCGF}) of the SCGF we find that, at fixed $\epsilon$ and for small $\lambda$, the Gaussian fluctuations of velocity are described by
\begin{equation}
  \label{eq:expansionphiepsilon}
  \varphi_\epsilon(\lambda,0) = \frac 12 \frac{\lambda^2}{\epsilon^2} D + O\big((\lambda/\epsilon)^3\big)\:.
\end{equation}
This expansion is valid for small $\lambda/\epsilon$ because we used the parameter $\frac \lambda\epsilon$ in  $\langle \ee^{-\frac\lambda\epsilon A}\rangle$ for the SCGF.

To analyze the importance of the order of limits, it is more convenient to introduce a rescaled counter $\Lambda=\frac\lambda\epsilon$.
Using the values $\lambda_\cc^0=\frac 2\pi$,  $V_{\max}-V_{\min}=\frac 1\pi$ and $|V''_{\min} V''_{\max}| =(2\pi)^2$ for our potential of interest $V(x)=\frac{1}{2\pi}[\cos (2\pi x)-1]$ corresponding to the sine force~(\ref{eq:sinus:force}), we see that
\begin{align}
  \varphi_\epsilon(\epsilon\Lambda,0) = 
  \begin{casesl}
    \Big( \ee^{|\Lambda|}-1\Big) \ee^{-\frac{\lambda_\cc^0}{\epsilon}}
    +  o\big( \ee^{-\frac{\lambda_\cc^0}{\epsilon} }  \big) 
&
\qquad \textnormal{if $\epsilon\to 0$ is taken at fixed $\lambda=\epsilon\Lambda$}
\\[2mm]
    \Big[ \frac 12 \Lambda^2  + O(\Lambda^3)\Big] \ee^{-\frac{\lambda_\cc^0}{\epsilon}}
&
\qquad \textnormal{if $\Lambda\to 0 $ is taken before $\epsilon\to 0$}\:,
  \end{casesl}
\label{eq:resorderPhidev}
\end{align}
from Eqs.~(\ref{eq:phiepsilontot}) and~(\ref{eq:dlfhgejkgro})-(\ref{eq:expansionphiepsilon}) respectively.
These two asymptotic behaviors describe different regimes: the first one is a regime of ``large'' fluctuations where $\Lambda=\lambda/\epsilon\gg 1$, while the second one is valid for values of~$\Lambda$ which remain finite with $\epsilon$.
If one formally expands $\ee^{|\Lambda|}-1$ for $\Lambda\to 0$ (although the expression is not supposed to be valid in this limit),
the order $\Lambda^2$ of the expansion matches that of the $\Lambda\to 0$ expansion (done before taking the $\epsilon\to 0$ limit) shown in Eq.~(\ref{eq:resorderPhidev}).
However in the first case of order of limits, an extra contribution $\propto|\Lambda|$ is present, showing that the order of the two limits cannot be exchanged.
The matching of the $\Lambda^2$ term between the two asymptotics seems to indicate that the crossover regime between the two regimes in~\eqref{eq:resorderPhidev} is smooth.
To summarize, the Gaussian fluctuations happen on a scaling $\lambda=\epsilon\Lambda$ with finite $\Lambda$, i.e.~in a boundary layer around $\lambda=0$ of width $\epsilon$ that vanishes as $\epsilon\to 0$.

%

\section{Conclusion and outlook}

In this work, we have studied the biased Langevin process associated with a generalized time-integrated current, for a particle in a one-dimensional potential with periodic boundary conditions, driven by a uniform non-conservative force. Both the physical driving force $f$ and the statistical bias parameter $\lambda$ allow for the definition of the biased process generating a particle current, but their quantitative effect is different. These two control parameters naturally span a two-dimensional phase diagram within which depinning transitions (i.e., transitions between zero-current and non-zero-current phases) occur.

Focusing on the low-noise limit, we have shown that in this phase diagram, the standard depinning exponent $\frac{1}{2}$ is the exception rather than the rule: it only characterizes the depinning transition as a function of the force $f$, in the absence of statistical bias ($\lambda=0$). At the very same critical point in the phase diagram, we have found that the depinning transition considered as a function of $\lambda$ (with $f$ fixed) instead yields an exponent $\frac{1}{3}$. And even more strikingly, the depinning transition around any other point of the critical lines is characterized by an inverse logarithmic decay of the generalized current, either as a function of the bias $\lambda$ or of the non-conservative force $f$.

We have also studied the thermal rounding effects in the zero-current phase, and characterized the resulting Arrhenius scaling in the low noise limit for both the scaled cumulant generating function and the generalized current. We have found in particular a very simple expression for the effective energy barrier of the biased dynamics, which behaves linearly with $\lambda$, and boils down for $\lambda=0$ to the energy barrier to escape the metastable state of the tilted potential.

As for future work, it may be of interest to push the low-$\epsilon$ expansion to the next order to capture non-Poissonian corrections to the statistics of current. 
A natural picture of the results we have obtained at finite temperature is that the current can be written as a sum of two terms: (i) a number of turns (which is integer and thus a good candidate to present a Poisson statistics)  plus (ii), an extra contribution that is non-integer and that results from thermal fluctuations beyond the Arrhenius scaling.
In this picture, the second contribution would make the integrated current truly non-Poissonian;
it could be interesting to put this image on firm grounds, for instance by studying a possible mapping to the current SCGF in asymmetric random walks~\cite{lebowitz_gallavotticohen-type_1999,speck_large_2012}.
Changing ensemble from the process biased by $\lambda$ to the process conditioned by a fixed value of the current could prove instructive, in particular for the nature of the DPT in the microcanonical ensemble.
Also, the mean velocity and diffusion coefficient were recently computed in an active one-dimensional run-and-tumble model~\cite{doussal_velocity_2020}; an extension of our results to this problem could shed light on the nature of rare events in active matter.
Last, it might be worth exploring other generalizations of the model, by including for instance inertial effects in the Langevin dynamics.

%


\section*{Acknowledgments}
Financial supports from the ``Investissements d'Avenir'' French Government program managed by the French National Research Agency (ANR-16-CONV-0001) and from Excellence Initiative of Aix-Marseille University - A*MIDEX as well as from the grant IDEX-IRS `PHEMIN' of the Université Grenoble Alpes are acknowledged.
VL~acknowledges support by the ERC Starting Grant No. 680275 MALIG, the ANR-18-CE30-0028-01 Grant LABS and the ANR-15-CE40-0020-03 Grant LSD.

\section*{Appendices}
\addcontentsline{toc}{section}{Appendices}

\renewcommand{\thesection}{A}

\section{Gallavotti--Cohen symmetry and mirror symmetry}
\label{sec:GC-sym}

\subsection{Gallavotti--Cohen symmetry}
\label{sec:gall-cohen-symm}

Starting from the expression~(\ref{eq:fp-operator}) of the biased evolution operator $\mathbb W_\lambda $, one decomposes the force as $F(x)=-V'(x)+f$ with $f=\int_0^1\dd x\, F(x)$ and $V(x)$ a periodic function on $[0,1]$, and one defines a diagonal operator $\hat P_{\text{GC}}$ of elements $(\hat P_{\text{GC}})_{xx}=\ee^{-\frac 2\epsilon V(x)}$.
Then, one checks by direct computation that
\begin{equation}
\label{eq:fp-operator-GC}
\fl
\qquad\quad
\hat P_{\text{GC}}^{-1}
\,
\mathbb{W}_\lambda
\,
\hat P_{\text{GC}}
\,
\cdot
\
=
\
\big((F(x)-\lambda')\big)\partial_x\cdot
+\frac{\lambda'}{\epsilon}\left(\frac{\lambda'}{2}-F(x)\right)
+\frac{1}{2}\epsilon\partial_x^2\,\cdot
\
=
\
\mathbb{W}^\dag_{\lambda'}
\!\cdot
\quad
\textnormal{with $\lambda'=2f-\lambda$}
\,,
\end{equation}
where $\dag$ indicates the adjoint operator.
The l.h.s.~of~(\ref{eq:fp-operator-GC}) is a similarity transformation which leaves the spectrum invariant (since $\ee^{-\frac 2\epsilon V(x)}$ is a periodic function of $x$ the vector space is also unchanged).
Since an operator and its adjoint have the same spectrum (hence the same maximal eigenvalue), one finds that
\begin{equation}
  \label{eq:GC}
  \varphi_\epsilon(\lambda,f) =   \varphi_\epsilon(2f-\lambda,f) 
\:.
\end{equation}
This is an instance of Gallavotti--Cohen-type symmetry for the SCGF of the current~\cite{gallavotti_dynamical_1995-1,gallavotti_dynamical_1995,kurchan_fluctuation_1998,lebowitz_gallavotticohen-type_1999}, shown here using a method closely related to that of Kurchan~\cite{kurchan_fluctuation_1998}.

\subsection{Mirror symmetry to change the sign of $f$}
\label{sec:mirrorsym}
Performing the mirror symmetry $X\mapsto 1-X$ on $x$ in the Langevin equation~(\ref{eq:Langevin}), one sees that the mirror variable $x_1(t)=1-x(t)$ of $x(t)$ evolves in a potential $V_1(x_1)\equiv V(1-x_1)$ and a drive $-f$.
Besides, because $\ee^{-\frac \lambda \epsilon \int_0^t\dd t\:\dot x}=\ee^{+\frac \lambda\epsilon\int_0^t\dd t\: \dot x_1}$, we have that, for the SCGF, 
\begin{equation}
  \label{eq:cgfmirror}
  \varphi_\epsilon(\lambda,V,f)=\varphi_\epsilon(-\lambda,V_1,-f)
  \:,
\end{equation}
where we made explicit the dependency in the periodic potential and the drive as an argument of $\varphi_\epsilon$.
This identity allows one to relate the $f>0$ and the $f<0$ domains.
It is easy to see that the quantity $\mathcal V_{\max}$ is invariant by $(V,f)\mapsto(V_1,-f)$ so that, from Eq.~(\ref{eq:lambdacm:def}), $\lambda_\cc^-(V_1,-f)=2f+\lambda_\cc^-(V,f)$.
To show that the expression~(\ref{eq:finPhi}) of the Arrhenius function $\Phi(\lambda,f)$ is valid for any sign of $f$, we finally write%
, for $f<0$:
\begin{equation}
\fl\qquad\quad
  \label{eq:proofPhiminusf}
  \Phi(\lambda,V,f) 
   \stackrel{(\ref{eq:cgfmirror})}{=}
  \Phi(-\lambda,V_1,-f) 
   \stackrel{(\ref{eq:finPhi})}{=}
   -f-\lambda_\cc^-(V_1,-f) -|\lambda-f|
   \stackrel{}{=}
   f-\lambda_\cc^-(V,f) -|\lambda-f|
 \:.
\end{equation}

\renewcommand{\thesection}{B}

\section{Determination of the left eigenvector scaling function $U_L$}
\label{sec:determ-const-k_is}

In this Appendix, we determine the explicit form of the function $U_L(x)$ solution of the extremalization principle~(\ref{eq:minU1U4}).
To determine $k_i^{\min}$ [defined after~(\ref{eq:minU1U4})], we proceed as follows.
First of all, we shift the $x$ variable, without loss of generality, so that $F(x,f)>0$ for $0<x<x_\cc(f)$ and $F(x,f)<0$ for $x_\cc(f)<x<1$.
Here $x_\cc(f)$ is the point $x<x_{\text{m}}$ at which $F(x,f)$ cancels and becomes negative for $x>x_\cc(f)$.
We then evaluate, for each function $U_{L,i}(x)$, the difference $U_{L,i}(1)-U_{L,i}(0)$. Since we are in the zero-current regime, $\lambda$ satisfies
$\lambda_{\cc}^{-}(f) \le \lambda \le \lambda_{\cc}^{+}(f)$.
We recall that we consider $0\leq f<f_\cc$ and we restrict here our computations to the regime $\lambda_{\cc}^{-}(f) \le \lambda < f$ (the complementary regime $f \le \lambda < \lambda_{\cc}^{+}(f)$ is addressed in Appendix~\ref{app:left-eigen}). 
We now distinguish between two cases:

\begin{itemize}
\item \underline{$\lambda_{\cc}^{-}(f) < \lambda <0$ (see Fig.~\ref{fig:UL1234_min} of the Main Text, left)}\,:  Recalling the expressions~(\ref{eq:lambdac:fltfc}) of
  $\lambda_{\cc}^{\pm}(f)$, one finds
\begin{eqnarray} \label{eq:DeltaUL1}
U_{L,1}(1)-U_{L,1}(0) = -\lambda + \lambda_{\cc}^{-}(f) < 0\,\\  \label{eq:DeltaUL2}
U_{L,2}(1)-U_{L,2}(0) = -\lambda + \lambda_{\cc}^{+}(f) > 0\,\\  \label{eq:DeltaUL3}
U_{L,3}(1)-U_{L,3}(0) = -\lambda >0\\  \label{eq:DeltaUL4}
U_{L,4}(1)-U_{L,4}(0) = -\lambda + 2f >0\,.
\end{eqnarray}
As a result, one has $U_{L,1}(0) > U_{L,1}(1)$, whereas $U_{L,i}(0) < U_{L,i}(1)$ for $i\in \{2,3,4\}$.
As we now show, this implies that $U_L(1)=U_{L,1}(1)$.
For all $j\in \{2,3,4\}$, we have
\begin{equation}
U_L(0) = \min{\left\lbrace U_{L,1}(0),U_{L,2}(0),U_{L,3}(0),U_{L,4}(0)\right\rbrace}
\le U_{L,j}(0) < U_{L,j}(1) \,.
\label{eq:minU1U4:0}
\end{equation}
Using the periodicity condition $U_L(0)=U_L(1)$, it follows that $U_L(1)< U_{L,j}(1)$ for all $j\in \{2,3,4\}$.
Hence necessarily $U_L(1)=U_{L,1}(1)$. Then,
the periodicity of $U_L(x)$ in~(\ref{eq:minU1U4}) and the fact that $U_{L,j}(x)$ increases between $0$ and $1$ for $j\in\{2,3,4\}$ impose that the optimal values of the $k_j$'s are determined by $U_{L,j}(0)=U_{L,1}(1)$, for $j\in\{2,3,4\}$ (see Fig.~\ref{fig:UL1234_min}, left).
Using $U_{L,1}(1) = k_1 -\lambda + \lambda_{\cc}^{-}(f)$ and the expressions~(\ref{eq:eigen-2})-(\ref{eq:eigen-4}) one finds that the optimal values of the constants are
\begin{equation}
k_2^{\min} = k_3^{\min} = k_4^{\min} = -\lambda + \lambda_{\cc}^{-}(f) \equiv k_{\min}\,.
\end{equation}
The equality of these constants implies that $U_{L,3}(x)\leq U_{L,2}(x)$ and $U_{L,3}(x)\leq U_{L,4}(x)$ on $[0,1]$ so that the optimization principle~(\ref{eq:minU1U4}) only takes place between $U_{L,1}(x)$ and $U_{L,3}(x)$.
The first inequality $U_{L,3}(x)\leq U_{L,2}(x)$ is easily shown as follows:
\begin{equation}
U_{L,2}(x)-U_{L,3}(x) = -V(x)+f x+\int_0^x \!\!\dd y \, |F(y,f)| = \int_0^x \!\!\dd y \, \big[ F(y,f)+ |F(y,f)|\big] \ge 0 \,.
\end{equation}
The second inequality $U_{L,3}(x)\leq U_{L,4}(x)$ can be obtained by using two different expressions of $U_{L,4}(x)-U_{L,3}(x)$:
\begin{equation}
U_{L,4}(x)-U_{L,3}(x) = \int_0^x \dd y \, F(y,f) =  f - \int_x^1 \dd y \, F(y,f)\,,
\end{equation}
also using $\int_0^1 \dd y \, F(y,f)=f$.
When $x<x_\cc(f)$, $\int_0^x \dd y \, F(y,f) \ge 0$ because the integrand is positive.
For $x>x_\cc(f)$, $\int_x^1 \dd y \, F(y,f) \le 0$
because the integrand is negative; it follows that $f - \int_x^1 \dd y \, F(y,f) \ge 0$ (we recall that $f\ge 0$).

Hence, denoting by $x^*_L$ the intersection point between the two functions $U_{L,1}(x)$ and $U_{L,3}(x)$, we finally obtain:
\begin{equation}
  \label{eq:ULfinalAapp}
  U_L(x) = k_1 + 
  \begin{casesl}
    \lambda_\cc^-(f)-\lambda-\lambda x & \qquad \text{if}\quad 0<x<x^*_L 
    \\
    -2V(x)+(2f-\lambda) x -\lambda_\cc^+(f) & \qquad \text{if}\quad x^*_L <x < 1\,,
  \end{casesl}
\end{equation}
where $x^*_L$ is the solution of the equation
\begin{equation} \label{eq:xstarapp}
  V(x^*_L)=\frac{\lambda}{2}+f\,(x^*_L-1)\,,
\end{equation}
derived by imposing continuity of $U_L(x)$ at $x=x^*_L$.
Note that to obtain Eq.~(\ref{eq:ULfinalAapp}), we used that for $x>x^*_L$ one has $x>x_\cc(f)$, so that $F(x,f)<0$ for $x^*_L <x < 1$ and, from~(\ref{eq:eigen-1})
\begin{eqnarray}
 \nonumber
  U_{L,1}(x)
  &=&k_1-V(x)+(f-\lambda) x - \int_0^1 \dd y\, |F(y,f)| - \int_x^1 \dd y \,\big[ -V'(x) + f\big] \\ \label{eq:intermUL1}
  &=&k_1-2V(x)+(2f-\lambda) x -\lambda_\cc^+(f)\,.
\end{eqnarray}

\item \underline{$0<\lambda <f$ (see Fig.~\ref{fig:UL1234_min} of the Main Text, right)}\,: 
  In this case, $U_{L,1}$ and $U_{L,3}$ decrease their value between $x=0$ and $x=1$ while $U_{L,2}$ and $U_{L,4}$ increase it (as seen from Eqs.~(\ref{eq:DeltaUL1})-(\ref{eq:DeltaUL4})).
Hence Eqs.~(\ref{eq:DeltaUL1}), (\ref{eq:DeltaUL2}) and (\ref{eq:DeltaUL4}) remain valid, while Eq.~(\ref{eq:DeltaUL3}) is replaced by $U_{L,3}(1)-U_{L,3}(0) = -\lambda <0$.
Using the same type of argument as for the previous case, we find that the overall increasing functions  $U_{L,2}$ and $U_{L,4}$ should match at $x=0$, while the overall decreasing functions $U_{L,1}$ and $U_{L,3}$ should match at $x=1$.
The condition $U_{L,1}(1) = U_{L,3}(1)$ leads to $k_3=\lambda_\cc^{-}(f)$, whereas the condition $U_{L,2}(0) = U_{L,4}(0)$ simply yields $k_2=k_4$.
From the periodicity condition of $U_L$, one also has $U_{L,3}(1)=U_{L,4}(0)$, which gives $k_2=k_4=\lambda_\cc^{-}(f)-\lambda$.
One thus obtains:
\begin{equation}
  \label{eq:app:resULothercaseAPP}
  U_L(x) =k_1+
   \begin{casesl}
   -2V(x) + (2f - \lambda) x +  \lambda_\cc^-(f) - \lambda
     & 
    \qquad \text{if} \quad 
    0\leq x \leq x^\dag_L
    \\[2mm]
     - \lambda x +  \lambda_\cc^-(f)
    & 
    \qquad \text{if} \quad 
     x^\dag_L \leq x \leq 1\,,
 \end{casesl}
\end{equation}
where $x^\dag_L$ is determined by imposing $U_L(x)$ to be continuous at $x=x^\dag_L$, i.e.
\begin{equation} \label{eq:app:xdagforULothercase}
  V(x^\dag_L)=f x^\dag_L-\frac{\lambda}{2}\,.
\end{equation}

\end{itemize}
For completeness, a similar derivation of the left eigenvector for $f<\lambda<\lambda_\cc^+(f)$ and of the right eigenvector is given in Appendix~\ref{app:left-eigen}.

\renewcommand{\thesection}{C}

\section{Left and right eigenvectors at leading order in $\epsilon$ in the zero-current phase.}
\label{app:left-eigen}
In this Appendix we complete the derivation of the left eigenvector presented in Sec. \ref{sec:left-right-eigenv-domin} of the Main Text and in Appendix~\ref{sec:determ-const-k_is} for $f<\lambda<\lambda_\cc^+(f)$, and we show the functional form of the right eigenvector at dominant order. Finally, we complete our study by particularizing both solutions to the sinusoidal case $F_{\sin}(x,f)$.

\subsection{Left eigenvector for $f<\lambda<\lambda_\cc^+(f)$}
As we have shown in Sec. \ref{sec:left-right-eigenv-domin} of the Main Text, the left eigenvector can be written as 
\begin{equation} \label{eq:app:linear:combin:L}
L(x)=  \ee^{-U_{L,1}(x)/\epsilon} + \int \dd k_2\, \ee^{-U_{L,2}(x)/\epsilon}
+ \int \dd k_3\, \ee^{-U_{L,3}(x)/\epsilon} + \int \dd k_4\, \ee^{-U_{L,4}(x)/\epsilon}\,,
\end{equation}
where the functions $U_{L,i}$ are given by Eq.~(\ref{eq:eigen-1})-(\ref{eq:eigen-4}) and we have included an integration over the undetermined constants $k_i$. In the limit $\epsilon \to 0$, we assume that $L(x)$ takes the asymptotic form $L(x) \asymp \ee^{-U_L(x)/\epsilon}$ with $U_L(x)$ given by
\begin{equation}
U_L(x)=\min{\left\lbrace U_{L,1}(x),U_{L,2}(x),U_{L,3}(x),U_{L,4}(x)\right\rbrace}\,,
\label{eq:app:minU1U4}
\end{equation}
where one has to minimize also over the constants $k_2$, $k_3$ and $k_4$, compatible with the periodicity constraint $U_L(0)=U_L(1)$. We are now in position to determine the form of the left eigenvector. To do so, we will follow the procedure shown in Appendix~\ref{sec:determ-const-k_is}. We first shift the $x$ variable, without loss of generality, in such a way that $F(x,f)>0$ for $0<x<x_\cc(f)$ and $F(x,f)<0$ for $x_\cc(f)<x<1$. Thus, $x_\cc(f)$ is the point at which $F(x_\cc(f),f)=0$.
Since we are interested in the case $f<\lambda<\lambda_\cc^+(f)$, we now consider two different cases:

\begin{itemize}
\item \underline{$2f < \lambda <\lambda_{\cc}^{+}(f)$}\,: First, to fix the value of the constants $k_i$'s, we study the difference $U_{L,i}(1)-U_{L,i}(0)$. Recalling the expressions~(\ref{eq:lambdac:fltfc}) of  $\lambda_{\cc}^{\pm}(f)$, one finds
\begin{eqnarray} \label{eq:app:DeltaUL1}
U_{L,1}(1)-U_{L,1}(0) = -\lambda + \lambda_{\cc}^{-}(f) < 0\,\\  \label{eq:app:DeltaUL2}
U_{L,2}(1)-U_{L,2}(0) = -\lambda + \lambda_{\cc}^{+}(f) > 0\,\\  \label{eq:app:DeltaUL3}
U_{L,3}(1)-U_{L,3}(0) = -\lambda < 0\\  \label{eq:app:DeltaUL4}
U_{L,4}(1)-U_{L,4}(0) = -\lambda + 2f < 0
\end{eqnarray}
Consequently, one has $U_{L,2}(0) < U_{L,2}(1)$, whereas $U_{L,i}(0) > U_{L,i}(1)$ for $i\in \{1,3,4\}$, which implies that $U_L(0)=U_{L,2}(0)$. Indeed, if we had considered $U_L(0)=U_{L,j}(0)$ for some $j\in \{1,3,4\}$, we would have obtained:
\begin{equation}
U_L(1) = \min{\left\lbrace U_{L,1}(1),U_{L,2}(1),U_{L,3}(1),U_{L,4}(1)\right\rbrace}
\le U_{L,j}(1) < U_{L,j}(0) \,,
\label{eq:app:minU1U4:0}
\end{equation}
and the function $U_L(x)$ would not be a continuous periodic function. Once we have shown that $U_L(0)=U_{L,2}(0)$, we can determine the optimal values of $k_j$'s  by imposing periodicity of the left eigenvector: 
\begin{eqnarray} \label{eq:app:k1}
U_{L,1}(1)=U_{L,2}(0) \quad\rightarrow\quad \bar k_2^{\min}=\lambda^-_\cc(f)-\lambda\,\\  \label{eq:app:k3}
U_{L,3}(1)=U_{L,2}(0) \quad\rightarrow\quad \bar k_3^{\min}=\lambda^-_\cc(f)\,\\  \label{eq:app:k4}
U_{L,4}(1)=U_{L,2}(0) \quad\rightarrow\quad \bar k_4^{\min}=-\lambda^+_\cc(f)\,.
\end{eqnarray}
The values of these constants, together with the form of the solutions~(\ref{eq:eigen-1})-(\ref{eq:eigen-4}) imply that $U_{L,3}(x)\leq U_{L,1}(x)$ and $U_{L,4}(x)\leq U_{L,1}(x)$, $\forall x\in[0,1]$, so 
\begin{equation}
 U_L(x)=\min{\left\lbrace U_{L,2}(x),U_{L,3}(x),U_{L,4}(x)\right\rbrace}\,.
\end{equation}
The first inequality can be proved as:
\begin{eqnarray} \nonumber
 U_{L,1}(x)-U_{L,3}(x)&=&-V(x)+fx-\int_0^x dy|F(y,f)|-\lambda^-_\cc(f)\\
 &=&-\int_x^1 dy\left(F(y,f)-|F(y,f)|\right)\geq0\,,
\end{eqnarray}
while the second one is proven as:
\begin{eqnarray} \nonumber
 U_{L,1}(x)-U_{L,4}(x)&=&V(x)-fx-\int_0^x dy|F(y,f)|+\lambda^+_\cc(f)\\
 &=&\int_x^1 dy\left(F(y,f)+|F(y,f)|\right)\geq0\,.
\end{eqnarray}

As we have shown, $U_{L,2}(x)$ is minimum at $x=0$. One can also see that 
\begin{eqnarray} \label{eq:app:U42}
 U_{L,4}(x)<U_{L,2}(x) \quad\text{for}\,\, x\in[x_{4,2},1],\quad\text{with}\,\,x_{4,2}>x_\cc\\ \label{eq:app:U32}
 U_{L,3}(x)<U_{L,2}(x) \quad\text{for}\,\, x\in[x_{3,2},1],\quad\text{with}\,\,x_{3,2}<x_\cc\\ \label{eq:app:U34}
 U_{L,3}(x)<U_{L,4}(x) \quad\text{for}\,\, x\in[x_{3,4},1],\quad\text{with}\,\,x_{3,4}<x_\cc
\end{eqnarray}
where $x_{3,2}$ and $x_{3,4}$ are the solution of the equations
\begin{eqnarray} \label{eq:app:x32}
 2\left(f x_{3,2}-V(x_{3,2})\right) - \lambda=0\\ \label{eq:app:x34}
 2\left(f x_{3,4}-V(x_{3,4})\right) - 2f     =0\,.
\end{eqnarray}
According to Eqs.~(\ref{eq:app:U42})-(\ref{eq:app:U34}), we observe that $x_{3,2}<x_{4,2}$. In addition, Eqs.~(\ref{eq:app:x32}) and (\ref{eq:app:x34}) show that $x_{3,2}>x_{3,4}$. Therefore, the minimization only takes place between $U_{L,2}(x)$ and $U_{L,3}(x)$, and we obtain:
\begin{equation}
  \label{eq:app:ULfinalA}
  U_L(x) = k_1 + 
  \begin{casesl}
    -2V(x)+(2f-\lambda) x + \lambda^-_\cc(f)-\lambda & \qquad \text{if}\quad 0<x<x^\dag_L
    \\
    -\lambda x + \lambda^-_\cc(f)     & \qquad \text{if}\quad x^\dag_L <x < 1
  \end{casesl}
\end{equation}
with $x^\dag_L=x_{3,2}$. Note that $U_L(x)$ exhibits in this $\lambda$-regime the same functional form as the one found in the regime $0<\lambda<f$ (see Eq.~(\ref{eq:app:resULothercase}) in the Main Text and Eq.~(\ref{eq:app:resULothercaseAPP}) in Appendix~\ref{sec:determ-const-k_is}).

\item \underline{$f<\lambda <2f$}\,:
In contrast to what happens in the previous case, in this regime both $U_{L,1}$ and $U_{L,3}$ are decreasing functions between $x=0$ and $x=1$ while $U_{L,2}$ and $U_{L,4}$ are increasing functions in such an interval, as we can see from Eqs. (\ref{eq:app:DeltaUL1})-(\ref{eq:app:DeltaUL4}). The problem of deriving $U_L$ is then identical to the one already studied in Appendix~\ref{sec:determ-const-k_is} for $0<\lambda<f$, so its form is also given by Eq. (\ref{eq:app:ULfinalA}).

\end{itemize}

\subsection{Right eigenvector for $\lambda_\cc^-(f)<\lambda<\lambda_\cc^+(f)$}
Once the left eigenvector is known, we can derive the form of the right eigenvector at dominant order in $\epsilon$ by using the similarity transformation $\hat P_{\text{GC}}^{-1}\,\mathbb{W}_\lambda\,\hat P_{\text{GC}}\,\cdot=\mathbb{W}_{\lambda'}^\dag\cdot$, found in Appendix~\ref{sec:GC-sym}, where $\hat P_{\text{GC}}$ is the diagonal operator of elements $(\hat P_{\text{GC}})_{xx}=\ee^{-\frac 2\epsilon V(x)}$, $\mathbb{W}_\lambda$ is the Fokker-Planck operator and $\lambda'=2f-\lambda$ (see Eq.~(\ref{eq:fp-operator-GC})). We observe that:
\begin{equation}
 \mathbb{W}_\lambda\,\hat P_{\text{GC}}\,L(\lambda',x)=\hat P_{\text{GC}}\,\left(\mathbb{W}_{\lambda'}^\dag\,L(\lambda',x)\right)=\varphi_\epsilon(\lambda')\,\hat P_{\text{GC}}\,L(\lambda',x)\,,
\end{equation}
where for convenience we have made explicit the dependence of the left eigenvector $L(\lambda,x)$ on $\lambda$. Hence, taking into account the Gallavotti--Cohen symmetry (\ref{eq:GC}), the right eigenvector is of the form:
\begin{equation}
 R(\lambda,x)=\hat P_{\text{GC}}\,L(2f-\lambda,x)\,.
\end{equation}
At leading order, this relation implies:
\begin{equation}
 U_R(\lambda,x)=2V(x)+U_L(2f-\lambda,x)\,,
\end{equation}
where again for convenience we have made explicit the dependence on $\lambda$. We can then determine the functional form of $U_R$ by distinguishing two cases:
\begin{itemize}
 \item \underline{$\lambda_{\cc}^{-}(f) < \lambda <2f$}\,:
 \begin{equation}
  \label{eq:app:URfinalA}
  U_R(x) = k_1 + 
  \begin{casesl}
    \lambda x+ \lambda - \lambda^+_\cc(f) & \qquad \text{if}\quad 0<x<x^*_R
    \\
    2 V(x)-(2f-\lambda)x + \lambda^-_\cc(f)    & \qquad \text{if}\quad x^*_R <x < 1\,,
  \end{casesl}
\end{equation}
with $x^*_R$ the intersection point between the two branches.

 \item \underline{$2f < \lambda <\lambda_{\cc}^{+}(f)$}\,:
 \begin{equation}
  \label{eq:app:URfinalB}
  U_R(x) = k_1 + 
  \begin{casesl}
    2 V(x)-(2f-\lambda)x + \lambda - \lambda^+_\cc(f) & \qquad \text{if}\quad 0<x<x^\dag_R
    \\
    \lambda x - \lambda^+_\cc(f)    & \qquad \text{if}\quad x^\dag_R <x < 1\,,
  \end{casesl}
\end{equation}
with $x^\dag_R$ again the intersection point between the two branches.
\end{itemize}
Interestingly, in the non-driven case $f=0$ one can easily find that:
\begin{equation}
 x^*_R = 1 - x^*_L\,,\quad x^\dag_R = 1 - x^\dag_L\,.
\end{equation}

\subsection{Explicit expression for the left and right eigenvectors for the sinusoidal case}
\label{sect:app:left-right-eigen}

In order to illustrate the different results obtained for the left and right eigenvectors at dominant order, we now particularize to the case of a sinusoidal force of the form:
\begin{equation}
 F_\Ss(x,f)=\sin{2\pi x_f}+f\,,
\end{equation}
where $x_f=x-\arcsin{f}/(2\pi)$. Note that we have shifted the $x$ variable in order to satisfy that $F_\Ss(x,f)>0$ for $x\in[0,x_\cc]$ and $F_\Ss(x,f)<0$ for $x\in[x_\cc,1]$, with $x_\cc=\arcsin{f}/\pi+1/2$. In this case, $U_L(x)$ takes the following form.

\begin{figure}[t]
\begin{center}
  \includegraphics[height=4.7cm]{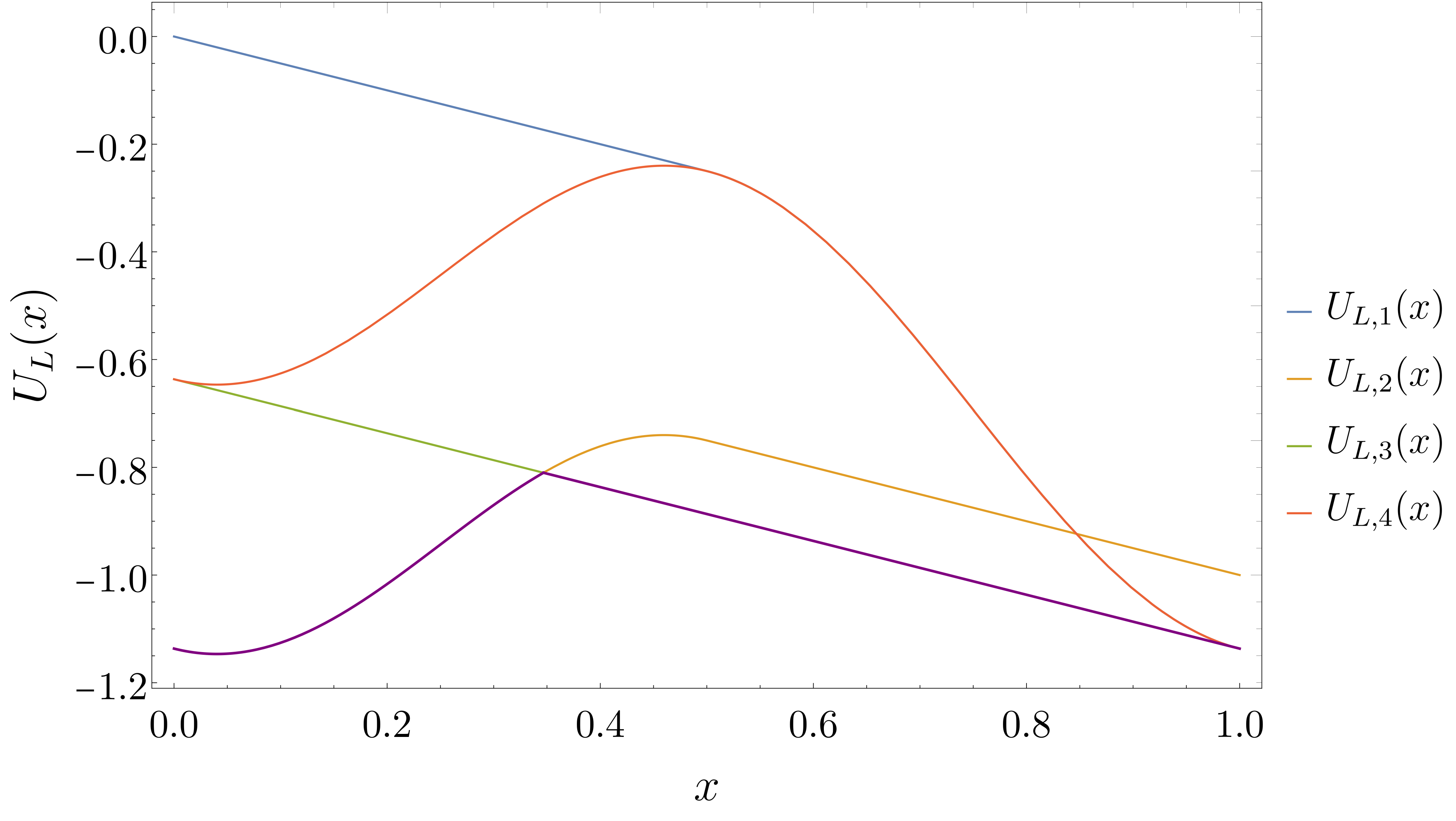}\quad\quad
  \includegraphics[height=4.7cm]{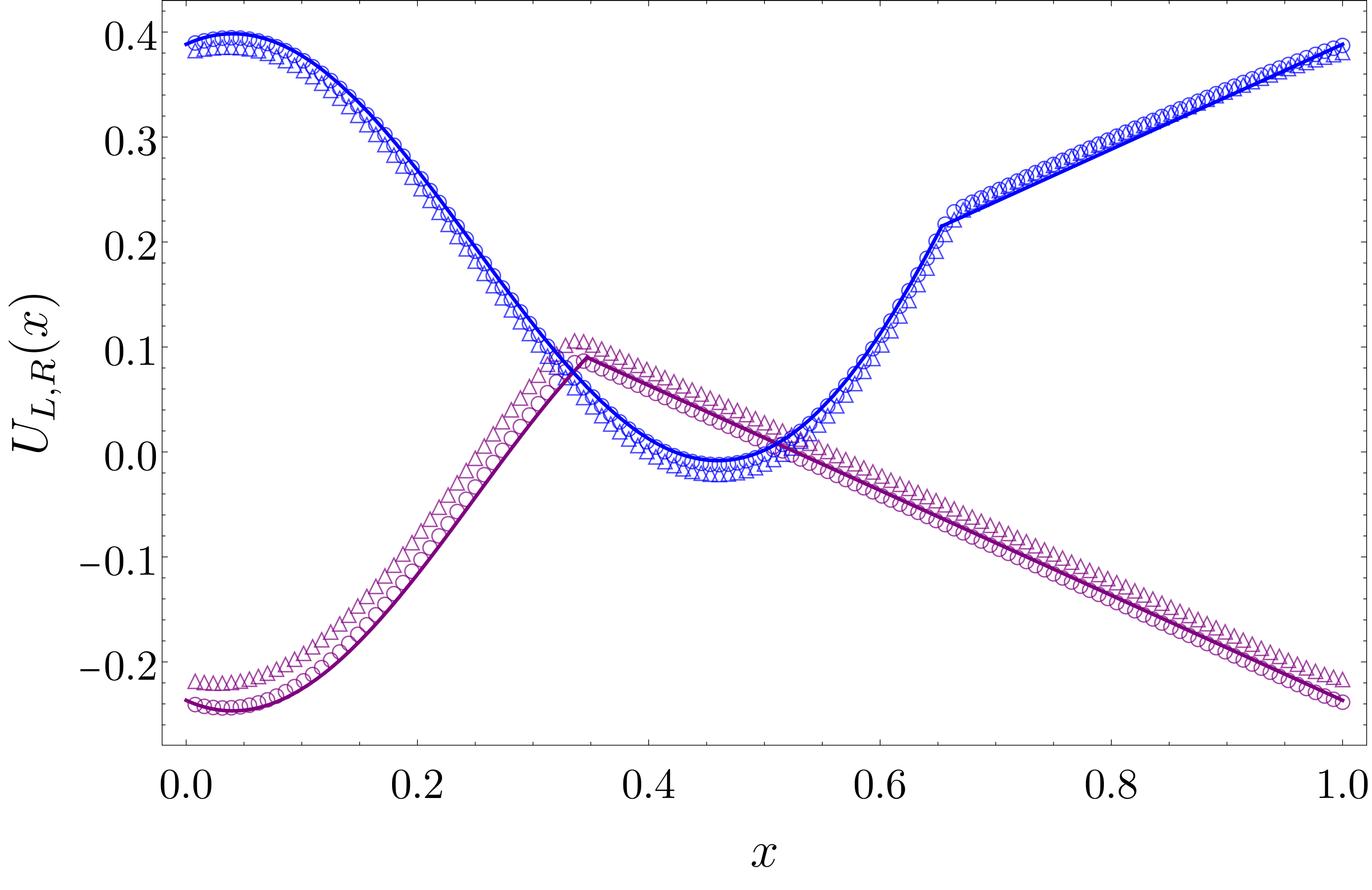}
\end{center}
\caption{\textbf{(Left)} The different solutions $U_{L,i}(x)$ with $i=1,...,4$ entering into play in the minimization problem for dominant order of the left, for $k_1=0$, $f=0$ and $\lambda=\frac{1}{2}$ for the sine force of Eq.~(\ref{eq:sinus:force}). 
The purple envelop encodes the final form of both $U_L(x)$.
\textbf{(Right)} Comparison between the analytical solution of $U_L(x)$ (purple line) and $U_R(x)$ (blue line), see Eqs.~(\ref{eq:app:UL-particular-B}) and (\ref{eq:app:UR-particular-B}), and numerical results for a discretized system with $128$ sites and noise amplitude $\epsilon=2^{-7}$ (triangles) and $\epsilon=2^{-8}$ (circles).
\label{fig:ULcompar}
}
\end{figure}

\begin{itemize}
 \item \underline{$\lambda_\cc^-(f)<\lambda <0$}\,:
 \begin{equation}
  U_L(x)= k_1 + 
  \begin{casesl}
    -\lambda(1+x) + \lambda_\cc^-(f) & \qquad \text{if}\quad 0<x<x^*_L
    \\
    -\frac{\cos{2\pi x_f}-\sqrt{1-f^2}}{\pi}+(2f-\lambda)x -\lambda_\cc^+(f)    & \qquad \text{if}\quad x^*_L <x < 1
  \end{casesl}
\end{equation}
where $x^*_L>x_\cc$ is the solution of the equation:
\begin{equation}
 -\frac{\cos{\left(2\pi x^*_L - \arcsin{f}\right)}-\sqrt{1-f^2}}{\pi}+2 f x^*_L+\lambda-2f=0\,,
\end{equation}
and with $\lambda_\cc^-(f)$ given by Eq.~(\ref{eq:lambdac:fltfc:sin}).

 \item \underline{$0<\lambda <\lambda_\cc^+(f)$}\,
  \begin{equation} \label{eq:app:UL-particular-B}
  U_L(x)= k_1 + 
  \begin{casesl}
    -\frac{\cos{2\pi x_f}-\sqrt{1-f^2}}{\pi}+(2f-\lambda)x - \lambda + \lambda_\cc^-(f) & \qquad \text{if}\quad 0<x<x^\dag_L\\
    -\lambda x + \lambda_\cc^-(f)    & \qquad \text{if}\quad x^\dag_L <x < 1
  \end{casesl}
\end{equation}
where $x^\dag_L<x_\cc$ is the solution of the equation:
\begin{equation}
 -\frac{\cos{\left(2\pi x^\dag_L - \arcsin{f}\right)}-\sqrt{1-f^2}}{\pi}+2 f x^\dag_L-\lambda=0\,.
\end{equation}
\end{itemize}
On the other hand, the leading contribution to the right eigenvector is:
\begin{itemize}
 \item \underline{$\lambda_\cc^-(f)<\lambda <2f$}\,:
 \begin{equation}
  U_R(x)= k_1 + 
  \begin{casesl}
    \lambda x + \lambda - \lambda_\cc^+(f) & \qquad \text{if}\quad 0<x<x^*_R
    \\
    \frac{\cos{2\pi x_f}-\sqrt{1-f^2}}{\pi}-(2f-\lambda) x + \lambda_\cc^-(f)     & \qquad \text{if}\quad x^*_R <x < 1
  \end{casesl}
\end{equation}
where $x^*_R<x_\cc$ is the solution of the equation:
\begin{equation}
 \frac{\cos{\left(2\pi x^*_R - \arcsin{f}\right)}-\sqrt{1-f^2}}{\pi}+2 f (1-x^*_R)-\lambda=0\,.
\end{equation}

 \item \underline{$2f<\lambda <\lambda_\cc^+(f)$}\,
  \begin{equation} \label{eq:app:UR-particular-B}
  U_R(x)= k_1 + 
  \begin{casesl}
    \frac{\cos{2\pi x_f}-\sqrt{1-f^2}}{\pi}-(2f-\lambda)x+\lambda-\lambda_\cc^+(f) & \qquad \text{if}\quad 0<x<x^\dag_R\\
    \lambda x -\lambda_\cc^+(f)     & \qquad \text{if}\quad x^\dag_R <x < 1
  \end{casesl}
\end{equation}
where $x^\dag_R>x_\cc$ is the solution of the equation:
\begin{equation}
 \frac{\cos{\left(2\pi x^\dag_R - \arcsin{f}\right)}-\sqrt{1-f^2}}{\pi}-2 f x^\dag_R+\lambda=0\,.
\end{equation}
\end{itemize}
In Fig.~\ref{fig:ULcompar} (left) we show the analytical results of the dominant order of the left eigenvector for the non-driven case $f=0$. In addition, numerical results obtained by direct diagonalization of a discretized version of the biased operator (and small values of $\epsilon$) are also shown in Fig.~\ref{fig:ULcompar} (right), where we observe they are in good agreement with the asymptotic prediction.

\renewcommand{\thesection}{D}

\section{Proof that $\zeta(z)>\zeta(z^*)$ for $z\neq z^*$}
\label{sec:proof-that-zetazz}

In this Appendix, we provide the technical proof used in Sec.~\ref{sec:determ-therm-effects} of the Main Text that $\zeta(z)>\zeta(z^*)$ for $z\neq z^*$. We first recall that, giving the form of the effective potential $V_\lambda(x)$ (see Eq.~(\ref{eq:resVeff})), we can deduce that:
\begin{eqnarray} \label{eq:app:sign:Vla:fla:inxcxstar}
V'_\lambda(x) &>& f-\lambda \quad {\rm for} \; x_\cc<x<x^*_L\,, \\
\label{eq:app:sign:Vla:fla:outxcxstar}
V'_\lambda(x) &<& f-\lambda \quad {\rm for} \; 0\leq x<x_\cc \; {\rm or} \; x^*_L<x<1\,.
\end{eqnarray}
We then distinguish two situations:

\begin{itemize}
 \item \underline{$0<z<z^*$}: In this case, we can write, defining the domain $I_1 \equiv [x_\cc,X(z)]\cup[X(z)+z,x^*_L]$,
 \begin{eqnarray} \nonumber
  \zeta(z^*)-\zeta(z)&=&(f-\lambda)(z^*-z)-\int_{x_\cc}^{x^*_L}\dd y\,V'_\lambda(y)+\int_{X(z)}^{X(z)+z}\dd y\,V'_\lambda(y)\\
&=& -\int_{I_1} \dd y\,\left[V'_\lambda(y)-(f-\lambda)\right] < 0,
 \end{eqnarray}
using the inequality (\ref{eq:app:sign:Vla:fla:inxcxstar}). It follows that $\zeta(z)>\zeta(z^*)$ for $0<z<z^*$.

 \item \underline{$z^*<z<1$}: In this interval of $z$, we need to study separately the two possibilities $0\leq X(z)<x_\cc$ and $x^*_L<X(z)<1$. We consider first the case $0\leq X(z)<x_\cc$.
If we assume that $X(z)+z\leq1$, we have, defining $I_2 \equiv [X(z),x_\cc]\cup[x^*_L,X(z)+z]$
  \begin{eqnarray} \nonumber
  \zeta(z)-\zeta(z^* )&=& (f-\lambda)(z-z^*)-\int_{X(z)}^{X(z)+z}\dd y\,V'_\lambda(y)+\int_{x_\cc}^{x^*_L}\dd y\,V'_\lambda(y)\\
  &=& -\int_{I_2} \dd y\,\left[V'_\lambda(y)-(f-\lambda)\right] > 0,
\end{eqnarray}
using the inequality (\ref{eq:app:sign:Vla:fla:outxcxstar}), whence $\zeta(z)>\zeta(z^*)$.

On the other hand, if $X(z)+z>1$, taking into account the periodicity of $V_\lambda(x)$ we can write
 \begin{eqnarray} \nonumber
  V_\lambda(X(z)+z)-V_\lambda(X(z))&=&V_\lambda(X(z)+z-1)-V_\lambda(X(z))\\
  &=&\int_0^{X(z)+z-1}\!\!\!\!\!\!\dd y\,V'_\lambda(y)+\int_{X(z)}^1\dd y\,V'_\lambda(y)\,,
 \end{eqnarray}
 where we note that $V_\lambda(0)=V_\lambda(1)=0$.
Thus,
\begin{eqnarray} \nonumber
 \zeta(z)-\zeta(z^*)&=&(f-\lambda)(z-z^*)-\int_0^{X(z)+z-1}\!\!\!\!\!\!\!\!\!\!\dd y\,V'_\lambda(y)-\int_{X(z)}^1\!\!\!\!\dd y\,V'_\lambda(y)+\int_{x_\cc}^{x^*_L}\!\!\dd y\,V'_\lambda(y)\\
&=& -\int_{I_3} \dd y\,\left[V'_\lambda(y)-(f-\lambda)\right] > 0,
\end{eqnarray}
where we have defined the domain $I_3 \equiv [0,X(z)+z-1]\cup[X(z),x_\cc]\cup[x^*_L,1]$, and used the inequality (\ref{eq:app:sign:Vla:fla:outxcxstar}).
So we also find in this case that $\zeta(z)>\zeta(z^*)$.

Finally, we mention for completeness the case  $x^*_L<X(z)<1$, although we could not find an example where this situation occurs.
In this case we would have that $1+x^*_L\leq X(z)+z<X(z)+1$ (see Fig.~\ref{fig:scheme_derivative_effective_potential}). Using again the periodicity of $V'_\lambda(x)$, we get after reorganizing the terms
\begin{equation}
 \zeta(z)-\zeta(z^*) = -\int_{I_4} \dd y\,\left[V'_\lambda(y)-(f-\lambda)\right] > 0
\end{equation}
from Eq.~(\ref{eq:app:sign:Vla:fla:outxcxstar}), with $I_4 \equiv [0,x_\cc]\cup[x^*_L,X(z)+z-1]\cup[X(z),1]$, leading again to the result $\zeta(z)>\zeta(z^*)$.

\end{itemize}
Thus, we have shown that for all $z \ne z^*$, $\zeta(z)>\zeta(z^*)$.

\renewcommand{\thesection}{E}

\section{Determination of the Arrhenius scaling in the regime $0<\lambda<f$}
\label{app:rate-function}
\begin{figure}[t]
\begin{center}
\includegraphics[width=0.45\columnwidth]{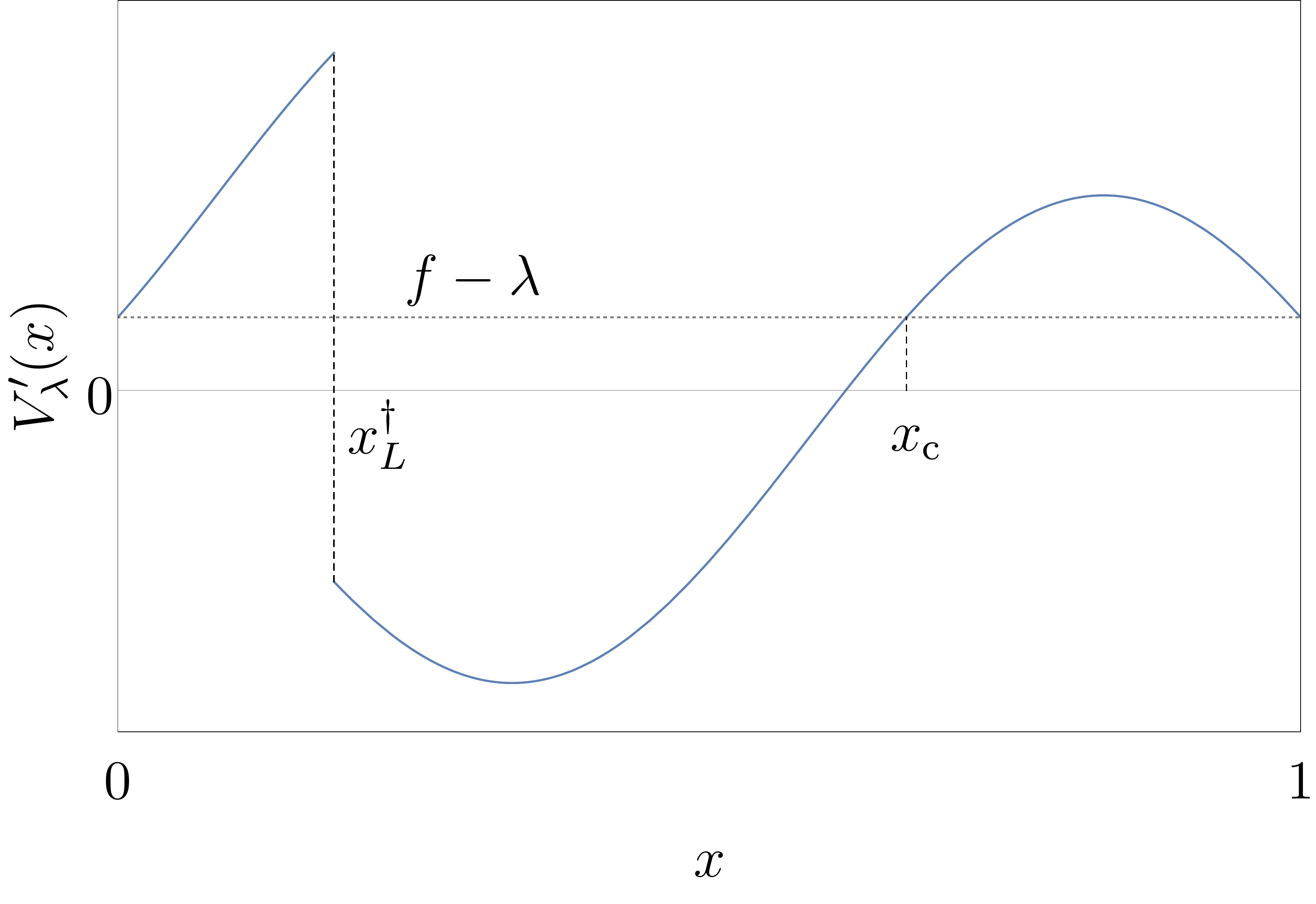}
  \quad
\includegraphics[width=0.45\columnwidth]{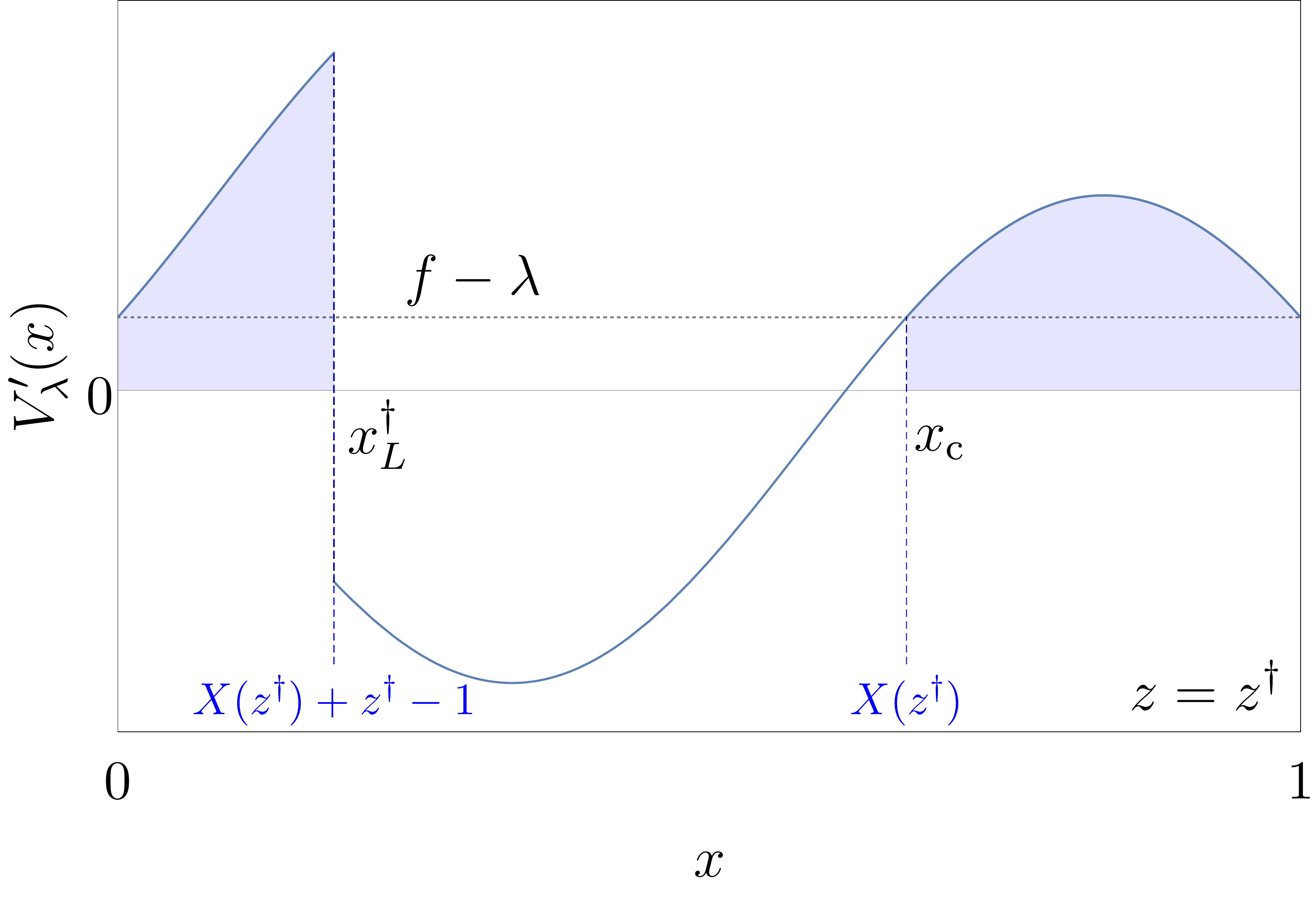}
\\~\\~\\
\includegraphics[width=0.45\columnwidth]{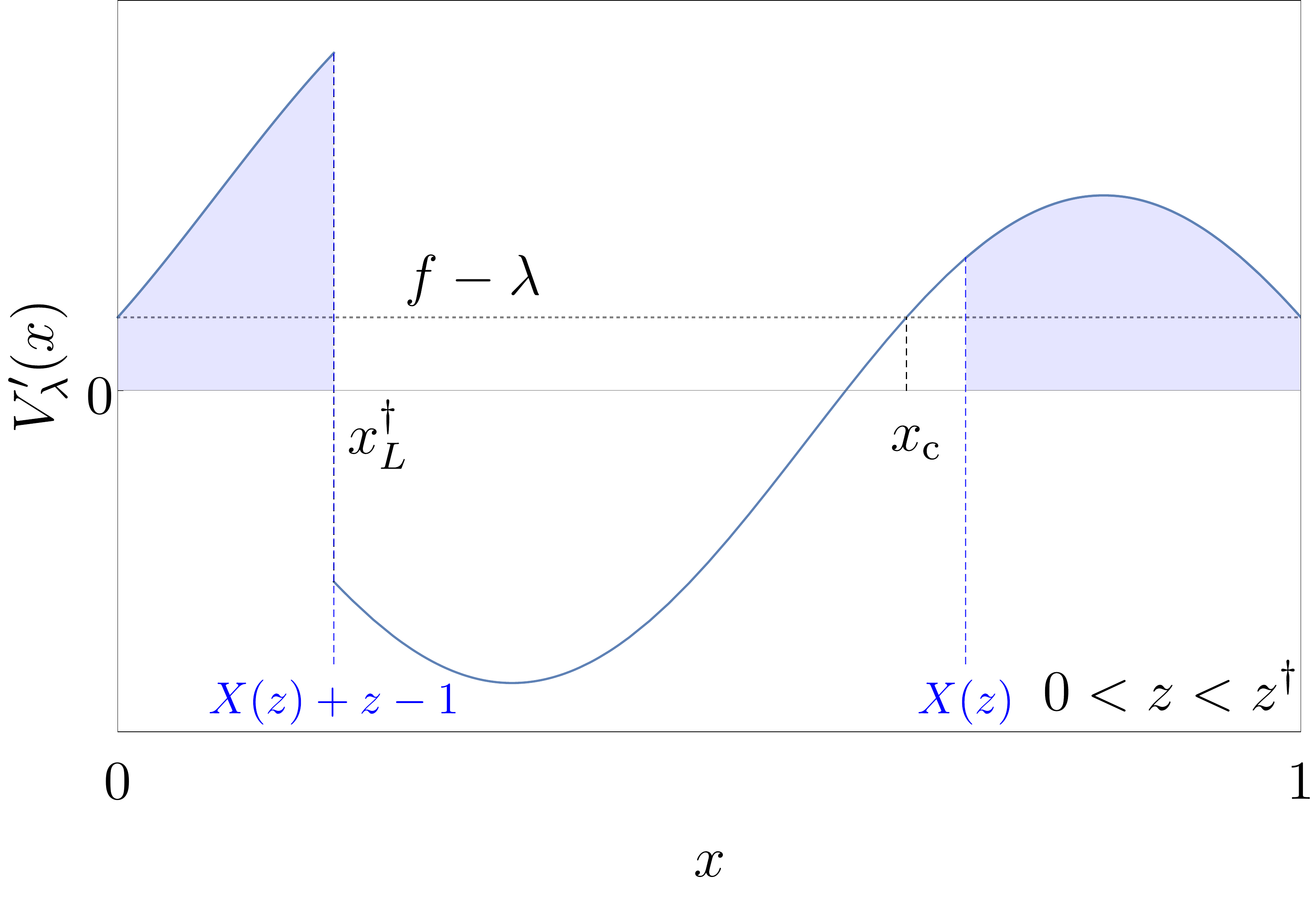}
  \quad
\includegraphics[width=0.45\columnwidth]{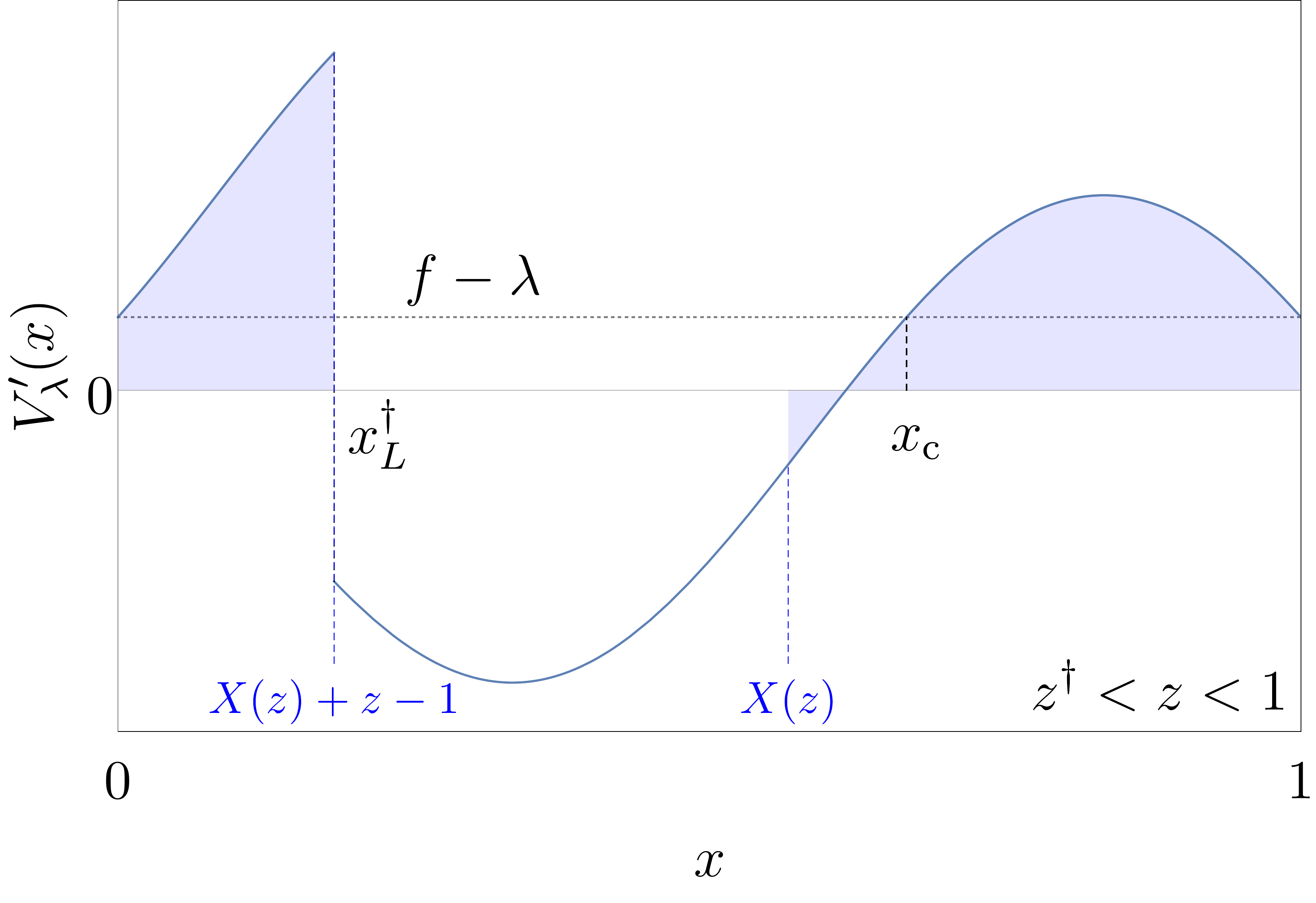}
\end{center}
\caption{Top left: Schematic representation of the derivative of the effective potential, $V'_\lambda(x)$ for $0<\lambda<f$.
The three other panels display a sketch of the graphical argument leading to the key properties of $X(z)$. On each graph, the shaded area indicates the integral of $V'_\lambda(x)$ over the interval $[X(z),X(z)+z]$.
Top right: case $z=z^\dag$, where $X(z)=x_\cc$.
Bottom left: case $0<z<z^\dag$, where $x_\cc<X(z)<1$.
Bottom right: case $z^\dag<z<1$, illustrated here with $x^\dag_L\leq X(z)<x_\cc$.
\label{fig:app:scheme_derivative_effective_potential}
}
\end{figure}

In this Appendix we derive the form of the Arrhenius function $\Phi(\lambda,f)$ associated with the $\epsilon \to 0$ Arrhenius scaling $\partial_{\lambda} \varphi_{\epsilon}(\lambda,f) \asymp \ee^{-\Phi(\lambda,f)/\epsilon }$ for the derivative of the SCGF when $0<\lambda<f$. 
This complements the case $\lambda_\cc^-(f)<\lambda<0$ treated in Sec.~\ref{sec:determ-therm-effects} of the Main Text.
To do so, we recall the formalism already described in Sec.~\ref{sec:strat-obta-arrh}, where we show that the Arrhenius function can be determined by solving the two optimization problems
\begin{eqnarray} \label{eq:app:opt-prob-1}
 X(z) &=&    \stackrel[0\leq x\leq 1]{}{\operatorname{argmax}} \left[\int_x^{x+z}dy\,V'_\lambda(y)\right]\\ \label{eq:app:opt-prob-2}
   \Phi(\lambda,f) &=&  - 2\min_{0\leq z<1}  [\zeta(z)]\,,
\end{eqnarray}
with $\zeta(z)$ defined as
\begin{equation}
\zeta(z) \equiv (f-\lambda)z-\int_{X(z)}^{X(z)+z}dy\,V'_\lambda(y)\,.
\end{equation}
The effective potential $V_\lambda(x)=V(x)+U_L(x)$ is periodic, and in this expression $U_L(x)$ is the leading contribution of the left eigenvector in the limit $\epsilon\rightarrow0$. Using the expression \eqref{eq:app:resULothercase} of $U_L(x)$, we get
\begin{equation}
  \label{eq:app:resVeff}
  V_\lambda(x) =k_1+
   \begin{casesl}
   -V(x) + (2f - \lambda) x +  \lambda_\cc^-(f) - \lambda
     & 
    \qquad \text{if} \quad 
    0\leq x \leq x^\dag_L
    \\[2mm]
    V(x) - \lambda x +  \lambda_\cc^-(f)
    & 
    \qquad \text{if} \quad 
     x^\dag_L \leq x \leq 1\,,
 \end{casesl}
\end{equation}
where $x^\dag_L$ is determined by imposing $V_\lambda(x)$ to be continuous at $x=x^\dag_L$, i.e.
\begin{equation} \label{eq:app:xdag}
  V(x^\dag_L)=f x^\dag_L-\frac{\lambda}{2}\,.
\end{equation}
It follows that the derivative $V'_\lambda(x)$ of the effective potential is given by
\begin{equation}
  \label{eq:resVeffder2}
  V'_\lambda(x) =
   \begin{casesl}
   F(x,f)+f-\lambda
     & 
    \qquad \text{if} \quad 
    0\leq x \leq x^\dag_L
    \\[2mm]
    -F(x,f)+f-\lambda
    & 
    \qquad \text{if} \quad 
     x^\dag_L \leq x \leq 1\,.
 \end{casesl}
\end{equation}
The shape of $V'_\lambda(x)$ is illustrated on Fig.~\ref{fig:app:scheme_derivative_effective_potential} for a sine force field.
As seen on Fig.~\ref{fig:app:scheme_derivative_effective_potential}, $V'_\lambda(x)$ satisfies the following inequalities:
\begin{eqnarray} \label{eq:sign:Vla:fla:inxcxdag}
V'_\lambda(x) &<& f-\lambda \quad {\rm for} \; x_L^\dag<x<x_\cc\,, \\
\label{eq:sign:Vla:fla:outxcxdag}
V'_\lambda(x) &>& f-\lambda \quad {\rm for} \; 0\leq x<x_L^\dag \; {\rm or} \; x_\cc<x<1\,.
\end{eqnarray}
These inequalities play a key role in what follows.

Since $f-\lambda>0$ and due to the periodicity of $X(z)$ and $V_\lambda(x)$, the minimum in Eq.~(\ref{eq:app:opt-prob-2}) is reached for $z\in[0,1)$. Following a similar approach to the one presented in Sec.~\ref{sec:determ-therm-effects} of the Main Text, we partly use a graphical argument to find a solution of the optimization problems~(\ref{eq:app:opt-prob-1}) and (\ref{eq:app:opt-prob-2}).
To optimize the function $\zeta(z)$, we need to know some properties of the function $X(z)$ that provides
the maximum of the area under the curve $V'_\lambda(x)$ between $x$ and $x+z$, for $0\leq z<1$. Graphically, one can deduce from Fig.~\ref{fig:app:scheme_derivative_effective_potential} the following properties of $X(z)$:
\begin{itemize}
 \item for $z=z^\dag\equiv x^\dag_L+1-x_\cc$, one finds $X(z^\dag)=x_\cc$.
 
 \item for $0<z<z^\dag$, one has either $0<X(z)<x^\dag_L$ or $x_\cc<X(z)<1$.
 
 \item for $z^\dag<z<1$, $X(z)$ satisfies $x^\dag_L\leq X(z)<x_\cc$.
 
\end{itemize}
In the following, we then show that $\zeta(z)>\zeta(z^\dag)$ for all $z\neq z^\dag$. We distinguish two situations (see Fig.~\ref{fig:app:scheme_derivative_effective_potential}):

\begin{itemize}
 \item \underline{$0<z<z^\dag$}: In this case, one has to study separately the regimes $0<X(z)<x^\dag_L$ and $x_\cc<X(z)<1$. For $0<X(z)<x^\dag_L$, we graphically observe that $X(z)+z\leq x^\dag_L$ (see Fig.~\ref{fig:app:scheme_derivative_effective_potential}). Hence, by using the periodicity of  $V_\lambda(x)$ we get
\begin{eqnarray} \nonumber
  \zeta(z^\dag) - \zeta(z) &=& (f-\lambda)(z^\dag -z)-\int_{x_\cc}^{1}dy\,V'_\lambda(y)-\int_{0}^{x^\dag_L}dy\,V'_\lambda(y)+\int_{X(z)}^{X(z)+z}\!\!\!\!\!\!dy\,V'_\lambda(y)\\
&=& -\int_{I_5} \dd y\,\left[V'_\lambda(y)-(f-\lambda)\right] < 0\,,
 \end{eqnarray}
where we have defined the domain $I_5 \equiv [0,X(z)]\cup[X(z)+z,x_L^\dag]\cup[x_\cc,1]$, and used the inequality (\ref{eq:sign:Vla:fla:outxcxdag}).

On the other hand, when $x_\cc<X(z)<1$, one can deduce that $X(z)+z\leq 1+x^\dag_L$. Then, assuming that $X(z)+z>1$ and using again the periodicity of $V_\lambda(x)$, we have
   \begin{eqnarray} \nonumber
  \zeta(z^\dag)-\zeta(z) &=& (f-\lambda)(z^\dag-z)-\int_{x_\cc}^{x^\dag_L+1}dy\,V'_\lambda(y)+\int_{X(z)}^{X(z)+z}dy\,V'_\lambda(y)\\
&=& -\int_{I_6} \dd y\,\left[V'_\lambda(y)-(f-\lambda)\right] < 0\,,
 \end{eqnarray}
with $I_6 \equiv [X(z)+z-1,x_L^\dag]\cup[x_\cc,X(z)]$, and thanks to the inequality (\ref{eq:sign:Vla:fla:outxcxdag}).
Finally, if we now assume $X(z)+z<1$ we can write
    \begin{eqnarray} \nonumber
  \zeta(z^\dag)-\zeta(z)&=&(f-\lambda)(z^\dag-z)-\int_{x_\cc}^{x^\dag_L+1}dy\,V'_\lambda(y)+\int_{X(z)}^{X(z)+z}dy\,V'_\lambda(y)\\
&=& -\int_{I_7} \dd y\,\left[V'_\lambda(y)-(f-\lambda)\right] < 0\,,
 \end{eqnarray}
having defined the domain $I_7 \equiv [0,x_L^\dag]\cup[x_\cc,X(z)]\cup[X(z)+z,1]$, and using again Eq.~(\ref{eq:sign:Vla:fla:outxcxdag}).
We have thus shown that $\zeta(z)>\zeta(z^\dag)$ for all $z$ in the interval $(0,z^\dag)$;

  \item \underline{$z^\dag<z<1$}: Given the form of $V_\lambda(x)$, we can see that the condition $x^\dag_L+1\leq X(z)+z<X(z)+1$ is satisfied in this regime (see Fig.~\ref{fig:app:scheme_derivative_effective_potential}). Then, we obtain
    \begin{eqnarray} \nonumber
  \zeta(z)-\zeta(z^\dag)&=&(f-\lambda)(z-z^\dag)-\int_{X(z)}^{X(z)+z}dy\,V'_\lambda(y)+\int_{x_\cc}^{x^\dag_L+1}dy\,V'_\lambda(y)\\
&=& -\int_{I_8} \dd y\,\left[V'_\lambda(y)-(f-\lambda)\right] > 0\,,
 \end{eqnarray}
where $I_8 \equiv [x_L^\dag,X(z)+z-1]\cup[X(z),x_\cc]$, and using the inequality (\ref{eq:sign:Vla:fla:inxcxdag}).
Hence $\zeta(z)>\zeta(z^\dag)$ for $z^\dag<z<1$.

 \end{itemize}
 
Now that we have established that the global minimum of $\zeta(z)$ is located at $z^\dag$, we determine now the Arrhenius function $\Phi(\lambda,f)$. From Eq.~(\ref{eq:app:opt-prob-2}) we obtain
\begin{equation}
 \Phi(\lambda,f)=f(x^\dag_L+1-x_\cc)-\lambda-V(x^\dag_L)+V(x_\cc)\,.
\end{equation}
Using Eqs.~(\ref{eq:app:xdag}) and (\ref{eq:lambdac:xc}), we eventually find
\begin{equation}
 \Phi(\lambda,f) = \lambda-\lambda_\cc^-(f)
\end{equation}
for $0<\lambda<f$, as announced in Sec.~\ref{sec:determ-therm-effects} of the Main Text.

\renewcommand{\thesection}{F}

\section{Derivation of the diffusion coefficient}
\label{sec:deriv-diff-coeff}

If the average velocity $\bar v$ of the driven particle in 1D has been known for a long time~\cite{stratonovich_oscillator_1965}, it is only relatively recently~\cite{reimann_giant_2001,reimann_diffusion_2002} that the diffusion coefficient~$D$ has been computed for $f\neq 0$,
using exact representations in terms of the moments of a first passage time~(see e.g.~\cite{hanggi_reaction-rate_1990}).
To compute $D$, we rather adapt the Langevin dynamics method of Derrida~\cite{derrida_velocity_1983}, who studied a similar question for a particle hopping on a discrete ring of finite size.
We translate his approach into the language of joint probability densities, which allows us to make the link with the LDF framework.
This approach is also more general (for instance, it allowed recently for a study of a similar problem in the context of active particles~\cite{doussal_velocity_2020}).

\subsection{Settings}
\label{sec:settingsD}

We recall that we consider an additive observable $A(\tf)$ of the form~(\ref{eq:generalobservable}) with $h(x(t))= 0$ and $g(x(t))=1$; i.e.~$A(\tf)$ represents the position counted algebraically
\begin{equation}
  \label{eq:}
  A(\tf)= \int_0^{\tf} \dd t\: \dot x(t)
\end{equation}
on a time window of duration $\tf$.
The average velocity $v$ and the diffusion coefficient $D$ are 
\begin{equation}
  \label{eq:defvdefD}
  \bar v = \lim_{\tf\to\infty} \frac 1 \tf \langle A(\tf) \rangle 
\ ,
\qquad\quad
  D = \lim_{\tf\to\infty} \frac 1 {2\tf} \langle A(\tf)^2 \rangle_c
  \:.
\end{equation}

The time evolution of the joint probability $\mathring P(x,A,\tf)$ and of its Laplace transform $\hat P_\lambda (x,\tf) = \int \dd A\, \ee^{-\frac\lambda\epsilon A} \mathring P(x,A,\tf)$ are governed by the operators
\begin{align}
\label{eq:defWcirc}
\mathring {\mathbb W } \cdot
&= 
  -\partial_x\big((F(x)-\epsilon \partial_A)\cdot\big)
  +\partial_A\Big(\frac{\epsilon}{2}\partial_A\cdot-F(x)\cdot\Big)
  +\frac{1}{2}\epsilon\partial_x^2\cdot
\\ 
\label{eq:defWlambda} 
\mathbb{W}_\lambda\cdot
&=
  -\partial_x\big((F(x)-\lambda)\cdot\big)
  +\frac{\lambda}{\epsilon}\Big(\frac{\lambda}{2}-F(x)\Big)\cdot
  +\frac{1}{2}\epsilon\partial_x^2\cdot
\ ,
\end{align}
in the sense that $\partial_\tf \mathring P = \mathring {\mathbb W}  \mathring P$ and $\partial_\tf \hat P_\lambda =  {\mathbb W_\lambda}  \hat P_\lambda$.
The two operators are related by the correspondence $\partial_A \leftrightarrow \frac\lambda\epsilon$.
This can be easily proved by multiplying the evolution equation for $\mathring P(x,A,\tf)$ by a factor $\ee^{-\frac\lambda\epsilon A}$, integrating over $A$ [we recall that the domain of $A$ is the whole real line], and performing an integration by parts.
We now introduce
\begin{align}
  \label{eq:defPQ0}
P(x,\tf) &= \int \dd A\: \mathring P(x,A,\tf) = \hat P_{\lambda=0}(x,\tf)
\\
  \label{eq:defPQ1}
Q(x,\tf) &= \int \dd A\ A\, \mathring P(x,A,\tf) = - \epsilon \partial_\lambda \hat P_{\lambda}(x,\tf)\big|_{\lambda=0}
\:,
\end{align}
which are periodic functions of $x$.
To determine their evolution equation, one notes that $P(x,\tf)$ is the usual probability density that verifies the Fokker--Planck equation, while 
$ \partial_\tf Q = \mathbb W Q - \epsilon \partial_\lambda \mathbb W_\lambda\big|_{\!_{\lambda=0^{\vphantom{|}}}}\!\!\!\!\!\!\! P$, implying
\begin{align}
  \label{eq:evolP}
  \partial_t P
&=
  -\partial_x\Big( F P -\frac \epsilon 2 \partial_x P \Big)
\\
  \label{eq:evolQ}
  \partial_\tf Q
&=
  -\partial_x\Big( F Q -\frac \epsilon 2 \partial_x Q \Big)
  -\epsilon \partial_x P + F P
\:,
\end{align}
where for simplicity we do not indicate the arguments of the functions.
The introduction of the two functions $P(x,\tf)$ and $Q(x,\tf)$, that satisfy the coupled equations~(\ref{eq:evolP}) and~(\ref{eq:evolQ}), allows one to determine the mean velocity and the diffusion coefficient by solving these equations instead of determining the full dependency in $A$ or in $\lambda$ of $\mathring P(x,A,\tf)$ or of $\hat P_\lambda(x,\tf)$.

\subsection{Relation between the functions $P$ and $Q$ and the diffusion coefficient $D$}
\label{sec:relat-betw-funct}

From~(\ref{eq:defPQ1}), the average value of $A(\tf)$ reads $\langle A(\tf)\rangle  = \int \dd x\;\dd A\: A\, \mathring P(x,A,\tf) = \int \dd x\: Q(x,\tf)$ and from~(\ref{eq:evolQ}) it verifies
\begin{equation}
  \label{eq:evol_aveA}
   \partial_\tf  \langle A(\tf) \rangle
  = \int \dd x\: F(x) P(x,\tf)
  \:,
\end{equation}
where and thereafter $\int \dd x$ denotes $\int_0^1 \dd x$, and we used the periodicity in $x$ of the functions $F$ and $Q$.
Similarly, using the periodicity in $x$ we see from~(\ref{eq:defWcirc}) that the second moment of $A(\tf)$ verifies
\begin{align} \nonumber
   \partial_\tf  \langle A(\tf)^2 \rangle
&
  = \int \dd x\; \dd A\: 
  A^2\: 
  \partial_A\Big(\frac{\epsilon}{2}\partial_A\mathring P(x,A,\tf)-F(x)\,\mathring P(x,A,\tf)\Big)
\\
  \nonumber
&
  = \int \dd x\; \dd A\: 
  \Big(
   \epsilon \mathring P(x,A,\tf)
+
   2 F(x) A\,\mathring P(x,A,\tf)
\Big)
\\
  \label{eq:evol_aveA2c}
&
  = \epsilon + 2 \int \dd x\: F(x) Q(x,\tf)
\:.
\end{align}
At large time, one expects that~\cite{derrida_velocity_1983}
\begin{equation}
  \label{eq:PQlarget}
  P(x,\tf) \stackrel[\tf \to\infty]{}{\to} P_0(x)
\:,
\qquad\quad
  Q(x,\tf) \stackrel[\tf \to\infty]{}{=} Q_0(x) + \tf Q_1(x) + o(\tf^0)
\:,
\end{equation}
where corrections to these asymptotic behaviors are exponentially decreasing in time.
For $P(x,\tf)$ this follows from the Fokker--Planck evolution, and $P_0(x)$ is the steady state distribution, normalized to $\int \dd x\:P_0=1$.
For $Q(x,\tf)$ we now show that the large-time behavior~(\ref{eq:PQlarget}) is compatible with the evolution equation~(\ref{eq:evolQ}).
Inserting~(\ref{eq:PQlarget}) into Eqs.~(\ref{eq:evolP})-(\ref{eq:evolQ}), we indeed obtain:
\begin{align}
  \label{eq:evolP0}
  0
 &=
  -\partial_x\Big( F P_0 -\frac \epsilon 2 \partial_x P_0 \Big)
\\
  \label{eq:evolQ0Q1a}
0
&=
    -\partial_x\Big( F Q_1 -\frac \epsilon 2 \partial_x Q_1 \Big)
\\
  \label{eq:evolQ0Q1b}
Q_1
&=
  -\partial_x\Big( F Q_0 -\frac \epsilon 2 \partial_x Q_0 \Big)
  -\epsilon \partial_x P_0 + F P_0
\:.
\end{align}
Since $P_0(x)$ and $Q_1(x)$ verify the same linear equation with the same periodic boundary condition, they are proportional.
The normalization of $Q_1$ is fixed by integrating~\eqref{eq:evolQ0Q1b} on $[0,1]$, which imposes $\int \dd x\: Q_1(x)=\int \dd x\: F(x) P_0(x)$, hence finally:
\begin{equation}
  \label{eq:resP_0}
  Q_1(x)
  =
  P_0(x) \ \int \dd x' F(x') P_0(x')
\:.
\end{equation}

We now determine from these large-time asymptotics the behavior of the first and second moment of $A(\tf)$.
From the identity $\langle A(\tf)\rangle  =  \int \dd x\: Q(x,\tf)$  and Eq.~(\ref{eq:PQlarget}) we have that at large time
\begin{align} \nonumber
  \langle A(\tf) \rangle 
&
  \stackrel[\tf \to\infty]{\phantom{(\ref{eq:resP_0})}}{=}
\
  \int \dd x\:  Q_0(x)
+
  \tf    \int \dd x\:  Q_1(x) + o(\tf^0)
\
\\
\
&
  \stackrel[\phantom{\tf \to\infty}]{(\ref{eq:resP_0})}{=}
\
  \int \dd x\:  Q_0(x)
+
  \tf    \int \dd x\: F(x) P_0(x) + o(\tf^0)
\:,
   \label{eq:Aavelargetime}
\end{align}
and thus for the average velocity [see also Eq.~(\ref{eq:evol_aveA}) which yields the same result]:
\begin{equation}
  \label{eq:vinf}
  \bar v 
  = \lim_{\tf\to\infty} \frac 1 \tf \langle A(\tf) \rangle
  = \lim_{\tf\to\infty} \partial_\tf \langle A(\tf) \rangle
  = \int \dd x\: F(x) P_0(x) 
\:.
\end{equation}
Similarly, from Eqs.~(\ref{eq:evol_aveA2c}) and~(\ref{eq:PQlarget}):
\begin{equation}
  \label{eq:A2inf}
  \partial_\tf \langle A(\tf)^2 \rangle 
  \stackrel[\tf \to\infty]{}{=}
  \epsilon
   +
   2 \int \dd x\: F(x) Q_0(x)
   +
   2 \tf \int \dd x\: F(x) Q_1(x) + o(\tf^0)
\:.
\end{equation}
Inserting in this equation the form~(\ref{eq:resP_0}) of $Q_1(x)$, one observes that for the diffusion coefficient $D$, the term proportional to $\tf$ in the expression of the second moment is canceled by the one coming from the first moment:
\begin{align}
  \nonumber
  D & 
      \stackrel[]{\phantom{(\ref{eq:Aavelargetime})}}{=}
      \lim_{\tf\to\infty} \frac{1}{2\tf} \Big[ \langle A(\tf)^2\rangle - \langle A(\tf) \rangle ^2 \Big]
      = \lim_{\tf\to\infty} \frac 12 \partial_\tf \Big[ \langle A(\tf)^2\rangle - \langle A(\tf) \rangle ^2 \Big]
\\  
  \label{eq:Dexpr2}
   & 
      \stackrel[]{\phantom{(\ref{eq:Aavelargetime})}}{=}
     \lim_{\tf\to\infty} \Big[ \frac 12 \partial_\tf  \langle A(\tf)^2\rangle - \langle A(\tf) \rangle\, \partial_\tf \langle A(\tf) \rangle \Big]
\\  
  \label{eq:Dexpr3}
D
   & 
      \stackrel[]{{(\ref{eq:Aavelargetime})}}{=}
  \frac \epsilon 2
   +
     \int \dd x\: F(x) Q_0(x)
   -
     \Big( \int \dd x\:  Q_0(x)      \Big)
     \Big( \int \dd x\: F(x) P_0(x)  \Big)
\:.
\end{align}
We remark that to obtain this expression it was essential to determine the subleading contribution $\propto \tf^0$ of $\langle A(\tf)\rangle$ in~(\ref{eq:Aavelargetime}).
We see from the expression~(\ref{eq:Dexpr3}) that the diffusion coefficient is fully determined by the knowledge of the functions $P_0$ and $Q_0$, solutions of the equations~(\ref{eq:evolP0}) and~(\ref{eq:evolQ0Q1b}),
which we now solve at equilibrium ($f=0$, Sec.~\ref{sec:diff-coeff-an}) and  out of equilibrium ($f\neq 0$, Sections~\ref{sec:diff-coeff-caseneq} and~\ref{sec:simpl-expr-diff}).

\subsection{Diffusion coefficient for an equilibrium (reversible) dynamics}
\label{sec:diff-coeff-an}

If the force $F(x)$ derives from a periodic potential $V(x)$ as $F(x)=-V'(x)$, the steady state is the Boltzmann distribution at temperature $\epsilon/2$ 
that verifies $F P_0 = \frac \epsilon 2 \partial_x P_0 $
and reads:
\begin{equation}
  \label{eq:P0eq}
  P_0(x) = \frac 1 Z \ee^{-\frac 2\epsilon V(x)}
\ ,
\qquad\quad
  Z = \int \dd x \: \ee^{-\frac 2\epsilon V(x)}
\:.
\end{equation}
From~(\ref{eq:resP_0}), one has $Q_1(x)=0$ (as expected, at equilibrium, the average velocity $\bar v$ is zero, see Eq.~(\ref{eq:vinf})).
To determine the diffusion coefficient from~(\ref{eq:Dexpr3}), there remains to solve the equation~(\ref{eq:evolQ0Q1b}), which can be rewritten
\begin{align}
0
  &=
  -\partial_x\Big( 
    F Q_0 -\frac \epsilon 2 \partial_x Q_0 
    + \frac \epsilon 2 P_0
\Big)
\:.
\end{align}
Thus there is a constant $C_1$ such that
\begin{align}
\label{eq:eqC1} 
   F Q_0
   = C_1
    + \frac \epsilon 2 \partial_x Q_0 
    - \frac \epsilon 2 P_0
\:.
\end{align}
Inserting this result into~(\ref{eq:Dexpr3}), one finds by periodicity:
\begin{equation}
  \label{eq:DC1}
  D = C_1
\end{equation}
(the term $\epsilon$ in Eq.~(\ref{eq:Dexpr3}) is indeed canceled).
To determine the constant $C_1$ without directly solving for $Q_0(x)$, one sets
$
  Q_0(x)
  =
  R_0(x) \ee^{-\frac 2\epsilon V(x)}
$
with $R_0(x)$ a periodic function on $[0,1]$.
From~(\ref{eq:eqC1}) one finds:
\begin{equation}
  \label{eq:defR0}
  \partial _x R_0 = 
  \frac 1 Z -\frac 2\epsilon C_1 \ee^{\frac 2\epsilon V(x)}
\end{equation}
and the expression of $C_1$ is finally determined by integrating~(\ref{eq:defR0}) on $[0,1]$:
\begin{equation}
  \label{eq:C1}
  C_1 = 
  \frac \epsilon 2 \frac 1 Z
  \frac{1}{\int \dd x \: \ee^{\frac{2}{\epsilon} V(x)}}
  \:.
\end{equation}
The final expression of the equilibrium diffusion coefficient is determined from~(\ref{eq:DC1}) and from the expression of the partition function $Z$:
\begin{equation}
  \label{eq:resDfinalanyepsi}
  D
  =
  \frac\epsilon 2
  \frac 1  {
    \Big(\int \dd x \: \ee^{\frac{2}{\epsilon} V(x)}\Big)
    \Big(\int \dd x \: \ee^{-\frac{2}{\epsilon} V(x)}\Big)
  }
  \:.
\end{equation}
This result, first obtained in~\cite{lifson_selfdiffusion_1962}, can also be found using the fluctuation-dissipation relation (applying for instance the results of §11.3.1 in Risken's book~\cite{risken_fokker-planck_1996}),
valid at equilibrium when the force derives from a potential.

\subsection{Average velocity $\bar v$  in the case of a generic force}
\label{sec:vel-caseneq}

One writes the periodic force as deriving from a non-periodic tilted potential $U(x)$ such that
\begin{equation}
  \label{eq:FUf}
  F(x) = - U'(x)
  \;,
\qquad \quad
  U(1)-U(0) = - \int \dd x \: F(x) = -f
  \:,
\end{equation}
where $f$ is the tilt of the potential. We also decompose $U(x)=V(x)-fx$ where $V(x)$ is a periodic function of $x$ on $[0,1]$.
Then the steady state $P_0(x)$ can be written as follows, using periodicity~\cite{risken_fokker-planck_1996,proesmans_large-deviation_2019}:
\begin{align}
  \label{eq:exprP0noneq}
  P_0(x)
&=
  \frac{1}{Z_{\!f}}
  \ee^{-\frac 2\epsilon U(x)}
\left(
  \int_0^x \dd y\: \ee^{\frac 2\epsilon U(y)}
+
  \ee^{\frac 2 \epsilon f}
  \int_x^1 \dd y\: \ee^{\frac 2\epsilon U(y)}
\right)
\:,
\end{align}
where $Z_{\!f}$ is the normalization constant ensuring $\int \dd x\, P_0(x)=1$.
From~(\ref{eq:vinf}), the average current $\bar v$ is~\cite{risken_fokker-planck_1996,proesmans_large-deviation_2019}:
\begin{align}
\nonumber
  \bar v 
&=
  \int \dd x \: F(x) P_0(x)
\\[-2mm]
&=
  \frac \epsilon 2  
  \frac{1}{Z_{\!f}}
  \int \dd x \:
  \left[
    \partial_x\big( \ee^{-\frac 2\epsilon U(x)}\big)
\left(
  \int_0^x \dd y\: \ee^{\frac 2\epsilon U(y)}
+
            \ee^{\frac 2 \epsilon f}
  \int_x^1 \dd y\: \ee^{\frac 2\epsilon U(y)}
\right)
  \right]
    \label{eq:vbarnoneq1}
\\
\label{eq:vbarfinal01}
\bar v
&=
  \frac \epsilon 2  
  \frac{1}{Z_{\!f}}
  \left[
  \ee^{\frac 2 \epsilon f}
   -1
  \right]
\end{align}
where we integrated by parts and used the definition~(\ref{eq:FUf}) of the tilt.

Such expressions of the steady state~(\ref{eq:exprP0noneq}) and of the average velocity~\eqref{eq:vbarfinal01} are interesting because they only involve integrals on $[0,1]$, but their use is limited because the expression of the normalization constant $Z_{\!f}$ is rather cumbersome.
In the spirit of~\cite{le_doussal_creep_1995,scheidl_mobility_1995}, 
it is also interesting to use a different representation of the steady state $P_0$ and of the average velocity $\bar v$, valid for $f>0$\,%
\footnote{%
The case $f<0$ is obtained by mirror symmetry $x\mapsto 1-x$, see Appendix~\ref{sec:GC-sym}.
},
that provides a clearer physical picture and proves useful to determine the diffusion coefficient $D$.
We write
\begin{align}
  \label{eq:defP0P0tilde}
  P_0(x)
  &=
    \frac{1}{Z} \ee^{-\frac 2\epsilon U(x)} \tilde P_0(x)
\\
\label{eq:defP0tile}
 \tilde P_0(x)
  &=
    \int_0^{+\infty} \dd y \: \ee^{\frac 2\epsilon U(y+x)}
   =
    \int_x^{+\infty} \dd y \: \ee^{\frac 2\epsilon U(y)}
\:,
\end{align}
where the integrals converge since we have assumed $f>0$, and $Z$ is determined by imposing the normalization $\int \dd x\, P_0(x)=1$.
We first notice that this probability is periodic, as required, since
\begin{equation}
  \label{eq:computP0P1}
  \frac{P_0(x+1)}{P_0(x)}
  = 
  \frac{\ee^{-\frac{2}{\epsilon}[V(x)-(x+1)f]}}{\ee^{-\frac{2}{\epsilon}[V(x)-xf]}}
  \frac
  {\int_0^{+\infty} \dd y \: \ee^{\frac 2\epsilon [V(y+x)-(y+x+1)f]}}
  {\int_0^{+\infty} \dd y \: \ee^{\frac 2\epsilon [V(y+x)-(y+x)f]}}
  =
  \ee^{\frac2\epsilon f}
  \ee^{-\frac2\epsilon f}
  =1
\:.
\end{equation}
Thanks to this periodicity, the corresponding average velocity, given by Eq.~(\ref{eq:vinf}), reads
\begin{align}
  \nonumber
  \bar v 
  &= 
  - \int \dd x\: U'(x) P_0(x)
  =
  \frac \epsilon 2
  \frac 1Z 
  \int \dd x\: \partial_x\big(\ee^{-\frac 2\epsilon U(x)}  \big) \tilde P_0(x)
\\
\nonumber
  &=
  \frac \epsilon 2
  \bigg\{
   \big[P_0(x)\big]_0^1
   - 
   \frac 1 Z
   \ee^{-\frac 2\epsilon U(x)}
   \tilde P_0'(x)
  \bigg\}
\\
\label{eq:vofZP0tilde}
  &=
  \frac \epsilon 2
  \frac 1 Z
\:,
\end{align}
where we used that, from Eq.~(\ref{eq:defP0tile}), $\tilde P_0'(x)=-\ee^{\frac 2\epsilon U(x)}$.
Last, to check that the expression~(\ref{eq:defP0P0tilde})-(\ref{eq:defP0tile}) of the steady state is indeed the periodic solution of the Fokker--Planck equation~(\ref{eq:evolP0}), one computes
\begin{equation}
  \label{eq:demoFPP0P0tilde}
  -\frac \epsilon 2 \partial_x P_0(x)
\:
  = 
\:
  U'(x) P_0(x)
  -
  \frac \epsilon 2
  \frac 1Z 
  \ee^{-\frac 2\epsilon U(x)}
  \tilde P_0'(x)
  \stackrel{\eqref{eq:vofZP0tilde}}=
  U'(x) P_0(x) + \bar v
\:,
\end{equation}
which shows that
$ 
 -\frac \epsilon 2 \partial_x P_0(x) + F(x) P_0(x)
$
is a constant (equal to the average current), and by differentiation w.r.t.~$x$ that indeed $P_0(x)$ is the periodic solution of Eq.~(\ref{eq:evolP0}).

\smallskip

The interest of the representation~(\ref{eq:defP0P0tilde})-(\ref{eq:defP0tile}) of the steady state (a periodic method of variation of constants) is that the constant $Z$ and thus the average velocity~(\ref{eq:vofZP0tilde}) take the simple form
\begin{align}
  \label{eq:Zcontmirror}
  Z 
  &= 
  \int_{0}^{+\infty} \dd y \int_0^1 \dd x \:\ee^{\frac 2\epsilon [U(y+x)-U(x)]}
\\
\label{eq:vofZP0tildefinal}
  \bar v 
  &=
  \frac \epsilon 2
  \frac 1
  {\int_{0}^{+\infty} \dd y \int_0^1 \dd x \:\ee^{\frac 2\epsilon [U(y+x)-U(x)]}}
\:,
\end{align}
that we used in the main text (Eqs.~(\ref{eq:oneoverV})-(\ref{eq:oneoverVg})) in order to determine the SCGF of the current in the low-temperature limit.
Decomposing the first integral of~(\ref{eq:vofZP0tildefinal}) as a union of the intervals $[n,n+1]$ for $n\in\mathbb N$ and summing over $n$, 
one recovers the previous expression~(\ref{eq:vbarfinal01}) of the velocity, as easily checked.

\subsection{Diffusion coefficient $D$ in the case of a generic force}
\label{sec:diff-coeff-caseneq}

We now wish to determine the expression of $D$ from Eq.~(\ref{eq:Dexpr3}).
As we remarked,
the steady-state  ``constant of motion'' implied by Eq.~(\ref{eq:evolP0}) is precisely the average current:
\begin{align}
\label{eq:propFP0}
F(x) P_0(x) -\frac \epsilon 2 \partial_x P_0(x)= \bar v
\:.
\end{align}
(This can also be verified from Eqs.~(\ref{eq:exprP0noneq}) and~(\ref{eq:vbarfinal01})).
The expression of $Q_1(x)$ inferred from~(\ref{eq:resP_0}) is thus:
\begin{equation}
Q_1(x) = \bar v \, P_0(x)
\label{eq:Q1solgeneric}
\:.
\end{equation}
It thus remains to determine properties of the function $Q_0$, by solving Eq.~(\ref{eq:evolQ0Q1b}),
which we rewrite as
\begin{align}
    \label{eq:evolQ0Q1bbis}
\bar v \, P_0
&=
  -\partial_x\Big( F Q_0 -\frac \epsilon 2 \partial_x Q_0 + \frac \epsilon 2 P_0 \Big)
\underbrace{  - \frac \epsilon 2 \partial_x P_0 + F P_0 }_
{
  \
  \stackrel[]{\eqref{eq:propFP0}}{=}
  \
  \bar v
}
\:.
\end{align}
This implies that there exists a constant $C_2$ such that
\begin{equation}
  \label{eq:eqQ0C2}
  F Q_0 -\frac \epsilon 2 \partial_x Q_0 
  =
  C_2 -  \frac \epsilon 2 P_0
  + 
  \bar v\, (x-\Pi_0)
\:,
\qquad\quad
\text{with}
\quad
\Pi_0(x) = \int_0^x \dd x'\: P_0(x')
\:.
\end{equation}
Integrating this identity on $[0,1]$ and using the periodicity of $Q_0$ one obtains:
\begin{equation}
  \label{eq:intFQ0}
  \int \dd x\: F(x) Q_0(x)
  =
  C_2 -  \frac \epsilon 2 
  + 
  \bar v\, \Big\langle x-\frac 12\Big\rangle_0
\:,
\qquad\quad
\text{with}
\quad
\big\langle ...\big\rangle_0 \equiv  \int \dd x\,...\, P_0(x)
\:.
\end{equation}
We used the equality $\int \dd x\: \Pi_0(x) = \langle 1-x\rangle_0$, obtained by integration by part.
Inserting~(\ref{eq:vbarnoneq1}) and~(\ref{eq:intFQ0}) in the expression~(\ref{eq:Dexpr3}) of the diffusion coefficient, one obtains:
\begin{equation}
  \label{eq:Dinterm2}
  D = 
  C_2 + \bar v \, \Big\langle x-\frac 12\Big\rangle_0 - \bar v \int \dd x\: Q_0(x)
\:.
\end{equation}

In contrast to the equilibrium case, one now needs to solve for $Q_0(x)$ in order to evaluate the last missing integral, $\int \dd x\: Q_0(x)$.
To do so, we define a periodic function $R(x)$ which is the r.h.s.~of the Eq.~(\ref{eq:eqQ0C2}) 
\begin{equation}
  \label{eq:defRx}
  R(x) = 
  C_2 -  \frac \epsilon 2 P_0(x)
  + 
  \bar v\, (x-\Pi_0(x))
\:,
\end{equation}
so that $Q_0(x)$ verifies the equation
\begin{equation}
  \label{eq:sqQ0R0}
    F Q_0 -\frac \epsilon 2 \partial_x Q_0 
    = R
\:.
\end{equation}
We now show that similarly to the expression~(\ref{eq:defP0P0tilde})-(\ref{eq:defP0tile}) of $P_0$, the periodic solution of~\eqref{eq:sqQ0R0} reads
\begin{align}
  \label{eq:defQ0Q0tilde}
  Q_0(x)
  &=
    \frac 2 \epsilon
    \ee^{-\frac 2\epsilon U(x)} \tilde Q_0(x)
\\
\label{eq:defQ0tile}
 \tilde Q_0(x)
  &=
    \int_0^{+\infty} \dd y \: \ee^{\frac 2\epsilon U(y+x)} R(y+x)
    =
    \int_x^{+\infty} \dd y \: \ee^{\frac 2\epsilon U(y)} R(y)
\:.
\end{align}
The integrals converge since $R$ is periodic and bounded and we assumed $f>0$.
A computation analogous to Eq.~(\ref{eq:computP0P1}) shows that the expression~(\ref{eq:defQ0Q0tilde})-(\ref{eq:defQ0tile}) of $Q_0(x)$ is a continuous periodic function.
Besides, since $\tilde Q_0'(x)=-\ee^{\frac 2\epsilon U(x)} R(x)$, one has
\begin{equation}
  \label{eq:demoFPmodifQ0Q0tilde}
  -\frac \epsilon 2 \partial_x Q_0(x)
\:
  = 
\:
  U'(x) P_0(x)
  -
  \ee^{-\frac 2\epsilon U(x)}
  \tilde Q_0'(x)
\:
  = 
\:
  -F(x) P_0(x) + R(x)
\:,
\end{equation}
which shows that indeed the expression (\ref{eq:defQ0Q0tilde})-(\ref{eq:defQ0tile}) of $Q_0$ is the periodic solution to Eq.~(\ref{eq:sqQ0R0}).
Also, from the expression~(\ref{eq:vofZP0tildefinal}) of $\bar v$,
we remark that the constant $C_2$ in the expression~(\ref{eq:Dinterm2}) of $D$
is cancelled by the contribution $\propto C_2$ coming from $R(t)$ in $\bar v \int \dd x\: Q_0(x)$.
 Hence, $C_2$ can safely be set to 0.
Collecting the previous results, the expression of the diffusion coefficient (valid for $f>0$) is:
\begin{equation}
  \label{eq:Dprefinal}
  D = 
  \bar v \, \Big\langle x-\frac 12\Big\rangle_0 
  - \frac 1 Z 
  \int_0^1 \dd x\:
  \int_x^{+\infty} \dd y \:
    \ee^{\frac 2\epsilon [U(y)-U(x)]} 
   \Big\{
   \bar v\, \big(y-\Pi_0(y)\big)
   -
   \frac \epsilon 2 P_0(y)
   \Big\} 
\:.
\end{equation}

\subsection{Simplification of the expression of the diffusion coefficient $D$}
\label{sec:simpl-expr-diff}

The obtained expression~(\ref{eq:Dprefinal}) is rather cumbersome. To simplify it we introduce the function
\begin{equation}
  \label{eq:defUpsilon}
  \Upsilon(x) = \int_x^{+\infty}  \dd w \: \ee^{\frac 2\epsilon U(w)}
\:,
\end{equation}
which, as previously, is well defined for $f>0$. It satisfies $\Upsilon'(x)=-\ee^{\frac 2\epsilon U(x)}$ and $P_0(x)=\frac 1Z \ee^{-\frac 2\epsilon U(x)}\Upsilon(x)$.
\noindent
We decompose $D=D_0+D_1+D_2$ with
\begin{align}
\label{eq:defD0}
  D_0
  & = 
  \bar v \, \Big\langle x-\frac 12\Big\rangle_0 
\\
\label{eq:defD1}
D_1
  & =
 - \frac{\bar v}{Z} 
  \int_0^1 \dd x\:
  \int_x^{+\infty} \dd y \:
    \ee^{\frac 2\epsilon [U(y)-U(x)]} 
    \,
    \big(y-\Pi_0(y)\big)
\\
\label{eq:defD2}
D_2
  & =
  \frac \epsilon 2 
  \frac 1 Z 
  \int_0^1 \dd x\:
  \int_x^{+\infty} \dd y \:
    \ee^{\frac 2\epsilon [U(y)-U(x)]} 
    \,
    P_0(y)
\:.
\end{align}
We have that
\begin{align}
\label{eq:D2_reform}
  D_2
  & =
  \frac{\bar v}{Z}
  \int_0^1 \dd x\:
  \int_x^{+\infty} \dd y \:
    \ee^{-\frac 2\epsilon U(x)} 
    \,
    \Upsilon(y)
\end{align}
and, integrating by parts:
\begin{align}
\nonumber
\fl\qquad\qquad\,
  D_1
  & =
  \frac{\bar v}{Z}
  \int_0^1 \dd x\:
    \ee^{-\frac 2\epsilon U(x)} 
  \int_x^{+\infty} \dd y \:
     \Upsilon'(y)
    \,
  \int_0^y \dd z\: \big(1-P_0(z)\big)
\\
\nonumber
\fl\qquad\qquad\,
  & =
  \frac{\bar v}{Z}
  \int_0^1 \dd x\:
    \ee^{-\frac 2\epsilon U(x)} 
\bigg\{
    \bigg[ \Upsilon(y)
    \int_0^y \dd z\: \big(1-P_0(z)\big)
    \bigg]_{y=x}^{y=+\infty}
    -
  \int_x^{+\infty} \dd y \:
     \Upsilon(y)
     \big(1-P_0(y)\big)
\bigg\}
\\
\fl\qquad\qquad\,
  & =
    -
  \frac{\bar v}{Z}
  \int_0^1 \dd x\:
    \bigg\{
    \ee^{-\frac 2\epsilon U(x)} 
    \Upsilon(x)
    \int_0^x \dd z\: \big(1-P_0(z)\big)
    +
    \ee^{-\frac 2\epsilon U(x)} 
    \int_x^{+\infty} \dd y \:
     \Upsilon(y)
     \big(1-P_0(y)\big)
    \bigg\}
\:,
\end{align}
so that, by compensation with the expression~(\ref{eq:D2_reform}) of $D_2$:
\begin{align}
\label{eq:D1plusD2}
\fl\qquad\,
D_1+D_2
  & =
    -
  \frac{\bar v}{Z}
  \int_0^1 \dd x\:
    \bigg\{
    \ee^{-\frac 2\epsilon U(x)} 
    \Upsilon(x)
    \int_0^x \dd z\: \big(1-P_0(z)\big)
    -
    \ee^{-\frac 2\epsilon U(x)} 
    \int_x^{+\infty} \dd y \:
     \Upsilon(y)
     P_0(y)
    \bigg\}
\:.
\end{align}
Then, using $\ee^{-\frac 2\epsilon U(x)}\Upsilon(x)= Z P_0(x)$, the first integral in~(\ref{eq:D1plusD2}) reads
\begin{align}
\fl\qquad\,
      -
  \frac{\bar v}{Z}
  \int_0^1 \dd x\:
    \ee^{-\frac 2\epsilon U(x)} 
    \Upsilon(x)
    \int_0^x \dd z\: \big(1-P_0(z)\big)
&=
 - \bar v \, \langle x \rangle _0
 +
  \bar v
\underbrace{
  \int_0^1 \dd x\:
    \Pi_0'(x)
    \Pi_0(x)
}_{
= \frac 12 \big[\Pi_0(x)^2\big]_0^1=\frac 12
}
= - \bar v \, \Big\langle x-\frac 12\Big\rangle_0
\:,
\end{align}
which compensates exactly with $D_0$ (see Eq.~(\ref{eq:defD0})).
Finally, from Eq.~(\ref{eq:D1plusD2}) and from the expression~(\ref{eq:defUpsilon}) of the function~$\Upsilon$, the resulting expression of the diffusion coefficient, valid for $f>0$, is:
\begin{align}
  \nonumber
D
&
\
=
\
  \frac{\bar v}{Z}
  \int_0^1 \dd x\:
    \ee^{-\frac 2\epsilon U(x)} 
    \int_x^{+\infty}\!\!\!\! \dd y \:
     \Upsilon(y)
     P_0(y)
 \\
 &
   \label{eq:Dfinal2}
\
=
\
  \bar v
  \int_0^1 \dd x
    \int_x^{+\infty}\!\!\!\! \dd y \:
    \ee^{-\frac 2\epsilon [U(x)-U(y)]} 
    \,
    P_0(y)^2
\:.
\end{align}
This result is valid for any $\epsilon>0$.
Although these expressions are simpler than Eq.~(\ref{eq:Dprefinal}), we could not find a more direct derivation of them within our approach.
They are equivalent to the ones obtained in Refs.~\cite{reimann_giant_2001,reimann_diffusion_2002}.

\subsection{Low-temperature asymptotics for $\bar v$ and $D$ in the pinned regime}
\label{sec:low-temp-asympt}

We have seen in Sec.~\ref{sec:interpr-form-rate} that as $\epsilon\to 0$ in the pinned regime $0<f<f_\cc$, all scaled cumulants of $A(\tf)$ scale logarithmically in the same way, as $\asymp \ee^{\lambda_c^-(f)/\epsilon}$, corresponding to a Poissonian regime of fluctuations in some range of current fluctuations.
We also discussed that finite-temperature corrections to the behavior of the cumulants encode the DPT observed in the Arrhenius function $\Phi(\lambda,f)$.
The exact expressions~(\ref{eq:vofZP0tildefinal}) of $\bar v$ and~(\ref{eq:Dfinal2}) of $D$, valid for all $\epsilon>0$, are different, showing that at finite $\epsilon$ the current fluctuations are not Poissonian.
In order to illustrate these aspects of the distribution of $A(\tf)$, we now analyze the small-$\epsilon$ behavior of $\bar v$ and~$D$.

For $0<f<f_\cc$, using a  $\epsilon\to 0$ saddle-point evaluation of the integrals in the expression~(\ref{eq:vofZP0tildefinal})  of $\bar v$ in order to estimate its exponential and sub-exponential behavior, one finds:
\begin{equation}
  \label{eq:vbarsmallepsilon}
  \bar v
  \stackrel[\epsilon \to 0]{}{\approx}
  \frac{1}{2\pi}
  \sqrt{|U''_{\text{max}} U''_{\text{min}}|}
  \,
  \ee^{-\frac 2\epsilon (U_{\text{max}}- U_{\text{min}})}
\ ,
\qquad \quad
\text{for}~ 0<f<f_\cc
\:.
\end{equation}
Here $U_{\text{min}}$ is the minimum value of the tilted potential $U(x)$ for $x\in[0,1]$ and $U_{\text{max}}$ the value of the local maximum of $U(x)$ that is located immediately to its \emph{right} (and similarly for the second derivatives).
One should beware that the  $\epsilon\to 0$ limit does not commute with the  $f\to 0$ limit (for $f=0$, one has $\bar v=0$), nor with the  $f\uparrow f_\cc$ limit.
This is because at $f>0$, the $\epsilon\to 0$ asymptotics means that the barrier $U_{\text{max}}- U_{\text{min}}$ is 
(i) strictly smaller than the barrier to the \emph{left} direction (but the barriers to the left and the right are equal for $f=0$); 
and (ii) much larger than $\epsilon$ (but the barrier is equal to 0 at $f=f_\cc$).

 For the diffusion coefficient, one uses the expression~(\ref{eq:Dfinal2}) valid for $f>0$
and the form~(\ref{eq:defP0P0tilde})-(\ref{eq:defP0tile}) of the steady state $P_0$ to write that
 \begin{align}
 \nonumber
  D
&  \stackrel[\phantom{\epsilon \to 0}]{\phantom{(\ref{eq:vofZP0tilde})}}{=}
  \frac{\bar v}{Z^2}
  \int_0^1 \dd x
    \int_x^{+\infty}\!\!\ \dd y \:
    \ee^{-\frac 2\epsilon [U(x)-U(y)]} 
    \,
\Big\{
    \ee^{-\frac 2\epsilon U(y)}
    \int_y^{+\infty} \dd z\: 
    \ee^{\frac 2\epsilon U(z)}
\Big\}
^2
\\
\label{eq:fullintegralsforD}
&  \stackrel[\phantom{\epsilon \to 0}]{(\ref{eq:vofZP0tilde})}{=}
  \Big(\frac 2\epsilon\Big)^2 \,
  \bar v^3
  \int_0^1 \dd x\:
    \int_x^{+\infty}\!\!\!\! \dd y 
    \int_y^{+\infty}\!\!\!\! \dd z 
    \int_y^{+\infty}\!\!\!\! \dd z' 
    \ee^{-\frac 2\epsilon [U(x)+U(y)-U(z)-U(z')]} 
\:.
\\
&  \stackrel[\epsilon \to 0]{\phantom{(\ref{eq:vofZP0tilde})}}{\asymp}
  \bar v^3
\
  \exp
  \Big\{
     \frac 2\epsilon
     \max_{}
     \big[U(z)+U(z')-U(x)-U(y)\big]
  \Big\}
\end{align}
where the maximum is evaluated for $x\in[0,1]$ and $y,z,z'\in \mathbb R$ with the constraint $x\leq y \leq z,z'$.
It is reached for $x=y=x_{\min}$ and $z=z'=x_{\max}$ and takes the value $2(U_{\max}-U_{\min})$.
To conclude, using Eq.~(\ref{eq:vbarsmallepsilon}), the logarithmic equivalents of $\bar v$ and $D$ are the same:
\begin{align}
  \label{eq:vbarsmallepsilonlog}
\;
  \bar v
\stackrel[\epsilon \to 0]{}{\asymp}  
  \ee^{-\frac 2\epsilon (U_{\text{max}}- U_{\text{min}})}
\ ,
\qquad \quad
\text{for}~ 0<f<f_\cc
\\
  \label{eq:Dsmallepsilonlog}
  D
\stackrel[\epsilon \to 0]{}{\asymp}  
  \ee^{-\frac 2\epsilon (U_{\text{max}}- U_{\text{min}})}
\ ,
\qquad \quad
\text{for}~ 0<f<f_\cc
\:,
\end{align}
which illustrates the Poisson structure in the distribution of $A(t)$ at small noise, that we determined in the main text.
The prefactors, obtained by a saddle-point analysis of Eqs.~(\ref{eq:vofZP0tildefinal}) and~(\ref{eq:fullintegralsforD}), are also the same:
\begin{align}
  \label{eq:vbarsmallepsilonbis}
  \bar v
&
  \stackrel[\epsilon \to 0]{}{\approx}
  \frac{1}{2\pi}
  \sqrt{|U''_{\text{max}} U''_{\text{min}}|}
  \,
  \ee^{-\frac 2\epsilon (U_{\text{max}}- U_{\text{min}})}
\ ,
\qquad \quad
\text{for}~ 0<f<f_\cc
\\
  \label{eq:Dsmallepsilonprefactor}
D
&
  \stackrel[\epsilon \to 0]{}{\approx}
  \frac{1}{2\pi}
  \sqrt{|U''_{\text{max}} U''_{\text{min}}|}
  \,
  \ee^{-\frac 2\epsilon (U_{\text{max}}- U_{\text{min}})}
\ ,
\phantom{2}
\qquad \quad
\text{for}~ 0<f<f_\cc
\:,
\end{align}
illustrating that to observe a departure from the purely Poissonian behavior, 
higher-order corrections in powers of $\epsilon$ are needed.

\section*{References}

\addcontentsline{toc}{section}{References}
\bibliographystyle{plain_url}

\bibliography{depinning-conditioned}

\end{document}